\documentclass[aps,prd,reprint,twocolumn,superscriptaddress,showpacs]{revtex4-2}

\usepackage{amsfonts}
\usepackage{amsmath}
\usepackage{graphicx}
\usepackage{bm}
\usepackage{amssymb}
\usepackage{dcolumn}
\usepackage{color}
\usepackage{multirow}
\usepackage{booktabs}
\usepackage[colorlinks,
linkcolor=blue,
anchorcolor=blue,
citecolor=blue,
urlcolor=blue,
]{hyperref}
\begin{document}
\title{T-carbon: experiments, properties, potential applications and derivatives}

\author{Xin-Wei Yi}
\email{These authors make equal contribution to this work.}
\affiliation{School of Physical Sciences, University of Chinese Academy of Sciences, Beijing 100049, China}

\author{Zhen Zhang}
\email{These authors make equal contribution to this work.}
\affiliation{School of Physical Sciences, University of Chinese Academy of Sciences, Beijing 100049, China}

\author{Zheng-Wei Liao}
\email{These authors make equal contribution to this work.}
\affiliation{School of Physical Sciences, University of Chinese Academy of Sciences, Beijing 100049, China}

\author{Xue-Juan Dong}
\affiliation{School of Physical Sciences, University of Chinese Academy of Sciences, Beijing 100049, China}

\author{Jing-Yang You}
\email{phyjyy@nus.edu.sg}
\affiliation{Department of Physics, Faculty of Science, National University of Singapore, 117551, Singapore}

\author{Gang Su}
\email{gsu@ucas.ac.cn}
\affiliation{Kavli Institute for Theoretical Sciences, and CAS Center for Excellence in Topological Quantum Computation, University of Chinese Academy of Sciences, Beijing 100190, China}
\affiliation{School of Physical Sciences, University of Chinese Academy of Sciences, Beijing 100049, China}

\date{\today}
\begin{abstract}
T-carbon is a novel carbon allotrope with many appealing properties. Since the proposal of T-carbon, there has been a lot of intensive studies devoted to its physical, chemical, optical, magnetic, thermoelectrical  and topological properties and possible applications in diverse areas in recent years. In this review, we provide a comprehensive review on the advances of the experiments, intriguing properties and various potential applications of T-carbon in energy storage, optoelectronics, thermoelectrics, topological states, etc. As intercalation and doping are effective methods to modify the electronic structure of materials or even to convert into sparkly different new structures or phases, we also discuss different atom doped T-carbon, which exhibit more intriguing properties and lead to promising potential applications in solar cells, photocatalysis, magnetism, superconductivity and so on. In addition, it is interesting to mention that the hydrogenated T-carbon molecules were found to be possibly related to the physical origin of the UV extinction feature in interstellar medium that has been the half-century long unsolved puzzle. A number of novel derivative structures either derived or inspired from T-carbon are included as well. Finally, we give prospects and outlook for future directions of study on T-carbon and related structures.
\end{abstract}
\maketitle
\tableofcontents

\begin{figure*}[!!!htp]
	\includegraphics[scale=0.35,angle=0]{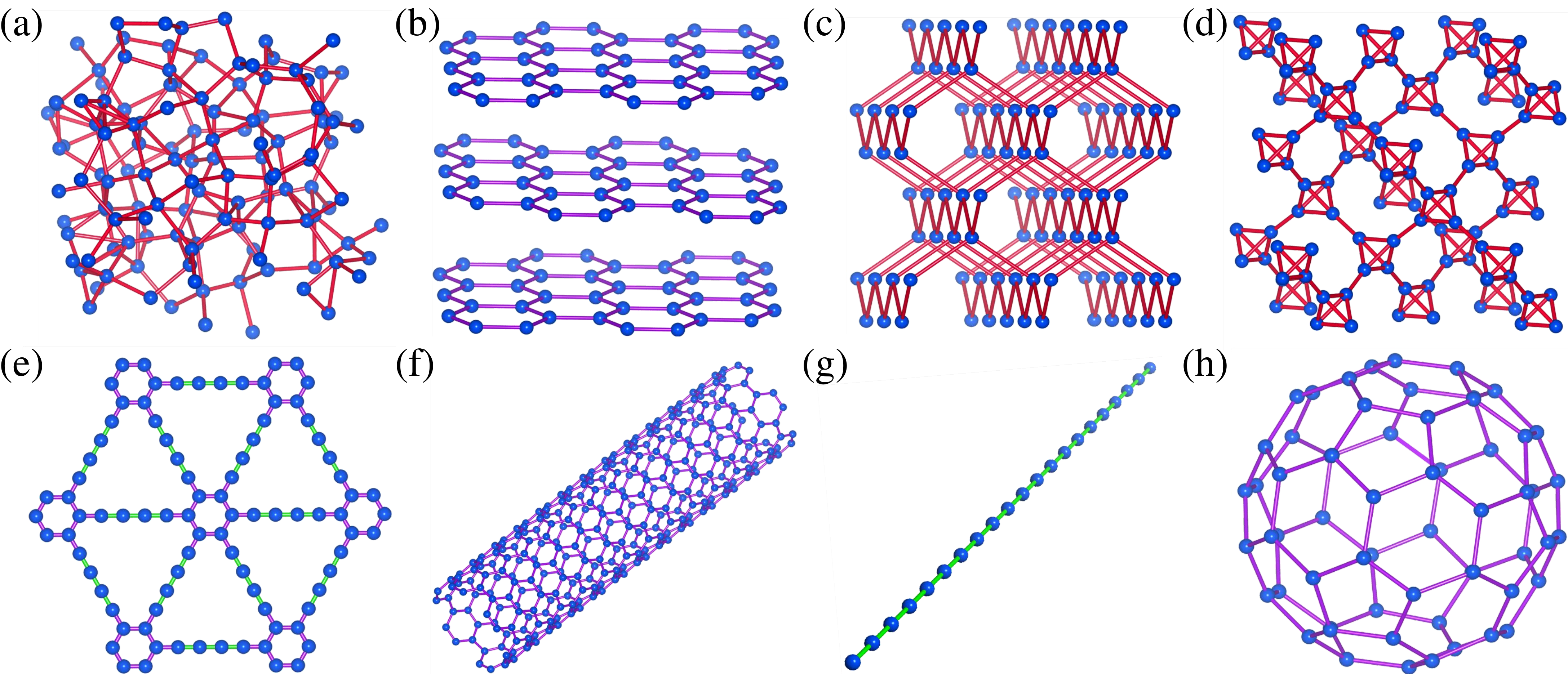}
	\caption{Carbon allotropes: (a) amorphous carbon, (b) graphite, (c) diamond, (d) T-carbon, (e) graphdiyne, (f) carbon nanotube, (g) carbyne, and (f) C$_{60}$ fullerene.}
	\label{fig1-1}
\end{figure*}

\section{Carbon allotropes}
Carbon is an extremely versatile element with highest mass content in earth's crust. Each carbon atom has four valence electrons with 2$s^2p^2$. As carbon atom can form $sp$, $sp^2$ and $sp^3$ hybrid chemical bonds, it has strong ability to bind itself with other elements to form a variety of allotropes or compounds with different chemical bonds, leading to the colorful world. It is thought that 10$\%$ to 25$\%$ of interstellar carbon in the form of polycyclic aromatic hydrocarbons (PAHs) that may explain the origin of life. Very recently, one group has discovered the PAHs  in the interstellar cloud via spectral matched filtering for the first time~\cite{McGuire2021}, where the abundance of PAHs is found to far exceed the expected target. Studying these PAHs and other similar molecules can help human better understand how life began in space. Besides the best known and studied graphite and diamond, since 1980s, several new carbon allotropes, including fullerenes~\cite{Schultz1965, Kroto1985}, carbon nanotubes~\cite{Iijima1993, Bethune1993}, and graphene~\cite{Novoselov2004}, etc., have been synthesized, which inspired numerous attempts to find new carbon structures in the past decades. Due to the abundant physical and chemical properties, the research of carbon materials has always been a hot topic. According to the statistics of SACADA (Samara Carbon Allotrope Database), there are currently more than 500 carbon allotropes that have been proposed~\cite{Hoffmann2016}. In the following, we will briefly discuss several typical carbon allotropes that have been extensively studied, as shown in Fig.~\ref{fig1-1}.

\subsection{Diamond}
Diamond has two different structures: cubic and hexagonal with all $sp^3$ hybrid chemical bond. Diamond is an insulator with a band gap of 5.5 eV and has high thermal conductivity. It is the hardest material that occurs naturally, and thus is widely used in industry, such as arts and crafts, cutting tools, and so on. It is also a precious gem.

Nanodiamond has many new mechanical and optical properties, such as high specific surface area and adjustable surface structure, non-toxicity and biocompatibility, which are very suitable for the preparation of specific drugs, drug delivery, electrochemistry, and new materials~\cite{Mochalin2011}. As a green catalyst, metal-free nanodiamond also shows great potential in the remediation of organic pollutants. Some scientists have designed nanodiamond-based catalysts with specific active sites and clear oxidation mechanism, and established a very promising catalytic oxidation system for the green remediation of actual polluted water~\cite{Shao2018}.

Diamond can be produced naturally or synthetically. Common synthetic techniques include CVD, high temperature and high pressure technology, chemical synthesis, and so on. Recently, an experiment realized the chemical induced transformation from AB-stacked bilayer graphene grown by CVD to fluorinated single-layer diamond~\cite{Bakharev2019}.

Conductive diamond for electrochemical research can be obtained by doping diamond, and doped diamond is usually synthesized by various CVD technologies. Among the doped diamond, boron doped diamond (BDD) is the most popular one, which can achieve high conductivity through high concentration doping. Conductive diamond has wide electrochemical potential window, low and stable background current, fast electron transfer rate, long-term response, chemical inertness, biocompatibility, and rich surface reactivity~\cite{Yang2019}. With these properties, conductive diamond films have been applied in many different fields, such as chemical and biochemical sensing, anode and cathode electrosynthesis of different compounds, electrocatalytic reduction of carbon dioxide, and manufacture of batteries and supercapacitors~\cite{Yang2019}. In addition, it is worth noting that BBD has been widely studied as an anode for electrochemical oxidation of organic wastewater~\cite{He2019}.

Nitrogen vacancy (NV) is a kind of point defect that fixes a nitrogen atom near the vacancy of diamond. Electron spin-based NV in diamond can be used as magnetometer to explore the static and dynamic magnetic structures and electronic transport phenomena~\cite{Casola2018}. In recent years, the analysis of related theories and experiments has been in forefront of the field of magnetometry~\cite{Casola2018}. Besides, the NV and group-IV color centers in diamond are used as main quantum emitters in the field of quantum information~\cite{Thiering2018,Kalb2017,Trusheim2020}.

The high thermal conductivity, low expansion coefficient, and low anisotropy of diamond make it a promising heat-dissipating material and thermal accelerator for thermal management composites. Recently, 3D porous diamond foam with 3D connected diamond network was obtained by the hot filament assisted CVD technology~\cite{Zhang2019a}. Adding diamond foam into latent heat storage devices can achieve high thermal response and good structural stability.

It is well known that diamond has brittle fracture, which hinders its application in certain areas. Recently, the existence of ultra-large, reversible elastic deformation in nanoscale single-crystalline and polycrystalline diamond needles has been demonstrated~\cite{Banerjee2018}. This surprising high elastic deformation is attributed to the lack of defects in the nanoscale needle-like structure and relatively smooth surface~\cite{Banerjee2018}.

\begin{table*}
	\renewcommand\arraystretch{1.15}
	\caption{The equilibrium density ($\rho$), Hardness, bulk modulus ($B$), cohesive energy (E$_{coh}$), energy gap ($E_g$), electrical conductivity ($\sigma$) and thermal conductivity ($\kappa$) at zero pressure of several typical carbon allotropes.}
	{\centering
		\begin{tabular}{lp{3cm}<{\centering}p{1.6cm}<{\centering}p{1.8cm}<{\centering}p{1.6cm}<{\centering}p{1.6cm}<{\centering}p{1.6cm}<{\centering}p{1.6cm}<{\centering}p{1.6cm}<{\centering}}
			\hline
			\hline
	& $\rho$ (g/ m$^3$) &Hardness (GPa)  &$B$ (10$^2$ GPa)  & E$_{coh}$  (eV/atom) & $E_g$ (eV) &$\sigma$ ($\Omega^{-1}$cm$^{-1}$) & $\kappa$ (Wm$^{-1}$K$^{-1}$) & Reference\\
	\hline
c-diamond    &3.52   &96$\pm$5 &4.43     &7.37        &5.45       &10$^{-18}$ &200     & \cite{Jones1993} \\
h-diamond    &3.52   &92.8     &4.30          &7.732     &3.13  &-  &-  &\cite{shengxianlei2011}\\
Graphite       &2.27   &-   &2.86-3.19    &7.374 &-0.04   &2.5$\times$10$^4$ &2000 &  \cite{Cahn1983}\\
Amorphous carbon &1.4-1.8   &0.2 &-   &- &0.4-0.7 &10$^{-2}$-10$^{-3}$    &2.2  &\cite{Bullen2000}\\
Carbon nanotubes &0.01   &- &1-8  &- &- &10$^{5}$-10$^{6}$ &3500     & \cite{Bachtold1998,Pop2006,Wang2014,Li2009a}\\
Fullerene (C$_{60}$) & 1.65 & 30 &7.0-9.0 & - &1.8 &10$^{-14}$ &0.4 &  \cite{Rabenau1993,Tea1993,Chernogorova2014,Bhakta2020,Braun2018,Ivetic2002}\\
M-carbon &3.45   &83.1 &4.31 &7.636   &3.56    &- &-  &  \cite{Li2009}\\
Bct-C$_4$ &3.35 &92.2 &4.09 &7.533 &2.47 &- &- &  \cite{Umemoto2010}\\
T-carbon &1.50 &$<$10 &1.69 &6.573 &2.25 &- &- &  \cite{shengxianlei2011,Chen2011}\\ 
			\hline
			\hline
	\end{tabular}}
	\label{Table2-1}
\end{table*}

\subsection{Graphite and Graphene}
Graphite is a three-dimensional (3D) van der Waals bulk carbon material with $sp^2$ hybridization, which is stacked in terms of ABAB... [Fig~\ref{fig1-1} (b)]. The adjacent graphene sheets in graphite can slide easily with each other, which makes graphite soft and lubricating. The special electronic distribution in graphite makes it thermally and electrically conductive. Graphite is the most stable carbon allotrope under ambient temperature and pressure and is inert with almost all materials. Owing to the special structure and excellent properties, graphite is widely used in industry and science. 

Due to the large interspace between graphene layers, graphite can host many atoms and molecules. Based on the specific structure of graphite, researchers have devoted to the modification of graphite. There are three main forms of modification, including graphite oxide (GrO), graphite intercalation compounds (GICs) and expanded graphite~\cite{Sengupta2011}. GrO is oxidized functionalized graphite, which is insulating and hydrophilic. The distance between neighboring layers of GrO is larger than that of graphite, making it easier for ions with a large radius such as Na$^{+}$ to pass through, which increases the possibility to use it as the anode of the battery. The reduced GrO was reported to have potential applications in the anode of lithium- (LIB) and sodium-ion batteries (NIB) due to the excellent electrochemical performance and large specific capacity, which can reach 917 mAh$\cdot$g$^{-1}$ in LIB~\cite{Sun2017}. GrO can also be used for the cheap production of graphene~\cite{Pei2012}. There are many ways to prepare GrO, including Brodie, Staudenmaier, Hummers, modified Hummers, and improved Hummers methods~\cite{PradoLavinLopez2016}. Among them, the modified Hummers method has realized the massive production of GrO with low-cost expanded graphite~\cite{Sun2013}. 

GICs, obtained by intercalating metal oxides, metal chlorides, bromides, fluorides, oxyhalides, acidic oxides, Lewis acids, etc. into the interlayer of graphite, exhibit excellent physical and chemical properties. For instance, several metal-intercalated compounds of graphite, such as KC$_8$, CaC$_6$ and YbC$_6$, are reported to be superconductors~\cite{Smith2015}. Several GICs exhibit better physical and chemical properties than the pristine graphite and can be used as electrodes~\cite{Xu2017a}. GICs has been used as anode materials of LIB in commercialization for many years. Additionally, graphite as anodes for NIB~\cite{Jache2014} and K-ion batteries (KIB)~\cite{Jian2015} has high reversible charge storage capacity. The different host-guest interactions between different alkali metal elements and graphite have been discussed in detail~\cite{Li2019}. It was reported that the limited cycle time and inferior stability of KIB can be solved by constructing an organic-rich passivation layer on the graphite anodes~\cite{Fan2019}. High-performance NIB can be obtained by adjusting the relative stability of ternary graphite intercalation compounds and the solvent activity in electrolytes~\cite{Xu2019}. Graphite can be used not only as the anode material by inserting cations, but also as the cathode material by inserted anions. The structure, characterization methods, mechanical and electrochemical properties of graphite cathode have been widely investigated~\cite{Zhang2019}. The study on graphite cathodes mainly focus on Al-ion and dual-ion batteries (also called “dual-graphite batteries”), where GICs are used as both anode and cathode. The advantages of dual-ion batteries are high energy density, environment-fridenly, low cost and switchable polarity~\cite{Xu2017a,Zhang2019}. The further research and commercialization of the related directions is very promising.

In 2004, single-layer graphite was obtained by mechanical exfoliation, named graphene~\cite{Novoselov2004}. Since then, there has been an upsurge in the study of two-dimensional (2D) materials dominated by graphene. In graphene, carbon atoms are hybridized with $sp^2$ bond. Graphene is a Dirac semimetal with the electron mobility of 2.5$\times$10$^5$ cm$^2$$\cdot$V$^{-1}$$\cdot$s$^{-1}$~\cite{Mayorov2011} and the thermal conductivity of 3000-5000 W/mK~\cite{Balandin2008}. Since the energy dispersion of graphene near the Fermi energy is linear, the carriers exhibit the characteristics of the Dirac fermion. Owing to the intriguing optical, electrical and mechanical properties, graphene has important applications in materials science, micro nano processing, energy, biomedicine and drug delivery, and is considered a revolutionary material. 

Graphene with high electron mobility is gradually used in transistors, which can work at low voltage and room temperature~\cite{Tiwari2020}. Due to the large surface-to-volume ratio and high electrical conductivity etc., the application of graphene sensors has been widely studied in recent years, including biosensors, diagnostics, DNA sensors and gas sensors~\cite{Tiwari2020,Nag2018}. Graphene can improve the photovoltaic and electrical properties of perovskite solar cells, which has already been used as conductive electrode, carrier transporting material and stabilizer material for perovskite solar cells~\cite{Lim2017}. Spin transport in graphene was confirmed in 2007~\cite{Tombros2007}. Since then, the emergence of graphene based heterostructures has been discussed~\cite{Avsar2020}. It seems that low-power all-spin-logic devices based on 2D van der Waals heterostructures will be seen in a few years~\cite{Avsar2020}. Both graphene oxide and reduced graphene oxide contain active oxygen groups, which brings new properties to the materials. They are widely applied as the main fillers of nanocomposites, and can be used in electrochemical biosensors, solar cells, supercapacitors~\cite{Lawal2019}. 

To realize the industrialization of graphene requires its rapid and economic mass production. This is also the main research hotspot of graphene synthesis technology in recent years. The synthesis methods of graphene are mainly divided into two categories, namely constructive and destructive methods. The former is mainly to delaminate graphite into graphene, including mechanical exfoliation, arc discharge, liquid-phase exfoliation (LPE), oxidative exfoliation-reduction, etc., while the latter uses atomic level precursors to construct graphene, such as chemical vapor deposition (CVD), epitaxial growth, substrate-free gas-phase synthesis, etc.~\cite{Lee2019}. Among them, oxidative exfoliation-reduction, CVD and LPE have been shown to possess high industrial application in the future~\cite{Lee2019}.

Gao $et$ $al.$ used scanning tunneling microscope (STM) to manipulate graphene, and realized that graphene can be folded and unfolded repeatedly in any direction, with the maximum twist angle of 60$^{\circ}$~\cite{Chen2019}. The folded single-crystal graphene nanoislands can form a special chiral tube edge, which has one-dimensional electronic characteristics similar to carbon nanotubes. The atomic-level manipulation of graphene provides a new possibility to construct quantum devices with carbon nanostructures.

Two recent experiments provide evidence for 2D electron fluid in graphene~\cite{Berdyugin2019,Gallagher2019}, revealing the flow behavior of quantum fluid that cannot be observed in water. On the basis of relevant research, we can design new thermoelectric materials, which can convert electric current into thermal current (and vice versa).

Simple single-layer graphene (SLG) has made many remarkable achievements. Due to the interlayer interaction, multilayer graphene (MLG) stacks usually show other novel properties different from SLG~\cite{Mogera2020}. If the graphene layer in MLG is not stacked in the form of “AB” in graphite, but with an angular twist, it is called twisted graphene (tBLG). Due to the overlap of two Dirac cones, DOS of tBLG shows the van Hove singularity~\cite{Yan2012}. When the rotation angle is small, the electronic properties~\cite{Luican2011} and Raman spectra~\cite{Kim2012} of tBLG are different from those of SLG, whereas when the rotation angle is large, they are basically the same as SLG. In addition, the optical properties~\cite{Carozo2013} and moiré patterns~\cite{Brihuega2012} of superlattice structure are also dramatically dependent on the rotation angle. 

In 2018, unconventional superconductivity was discovered in magic-angle $\sim1.1^{\circ}$ twisted bilayer graphene (TBG) superlattices~\cite{Cao2018}. This amazing property was realized by appropriately charging the carrier density using an external gate. The superconducting temperatures in the two devices were measured to be 0.5 and 1.7 K, respectively. Since then, breakthrough research of magic-angle TBG has emerged in an endless stream. There have been many studies to explain the novel superconducting properties~\cite{Dodaro2018,Guinea2018,Tarnopolsky2019,Lian2019,Wu2018,Ray2019,Gonzalez2019}. Recently, two different groups~\cite{RodanLegrain2020,Vries2020} have independently established Josephson junctions with magic-angle TBG, thus forming a controllable superconducting valve. It is reported that Moiré superlattices based on magic-angle TBG can be controlled by unconventional ferroelectricity~\cite{Zheng2020} and magnetism~\cite{Polshyn2020}, which will bring new directions for the innovation of new generation of quantum materials and electronic devices. Additionally, the superconductivity of TBG was extended to a trilayer graphene structure~\cite{Hao2021}. The superconductivity of the sandwich magic-angle TBG is believed to be from the strong electron couplings.

\subsection{Graphdiyne}
Graphyne is a 2D planar network structure formed by inserting one alkyne bond between the benzene rings of graphene, which was first proposed in 1987~\cite{Baughman1987}. Another structure, named graphdiyne, can be obtained by inserting a diacetylene bond (two alkyne bonds) between the benzene rings of graphene~\cite{Haley1997}, which was first synthesized experimentally in 2010~\cite{Li2010}. Graphdiyne has the hexagonal structure with space group of P6/mmm, and its lattice constant is 9.48 \AA~\cite{Long2011}. It is a narrow band gap semiconductor with a band gap of 0.46 eV~\cite{Long2011}. At present, the main synthesis techniques of graphdiyne include dry chemistry~\cite{Liu2017a,Qian2015,Wang2018,Zuo2017} and wet chemistry~\cite{Li2010,Matsuoka2017,Gao2018,Zhou2015a,Zhou2018}. As graphdiyne is a 2D porous structure, it has a broad application in separation of mixed gases. In 2011, Smith $et$ $al.$ theoretically studied the ability of graphdiyne to separate H$_2$ from CH$_4$ and CO~\cite{Jiao2011}, and found that the near center adsorption diffusion process of H$_2$ has the lowest energy, which therefore exhibits the best permeability, 10$^4$ times faster than that of porous graphene. One year later, Buehler $et$ $al.$ studied this issue in depth through molecular dynamics simulations~\cite{Cranford2012}, and found that these gases need a certain pressure to release through graphdiyne, and at atmospheric pressure, only H$_2$ can pass through.

At room temperature, the intrinsic electron mobility of graphdiyne is 10$^5$ cm$^2\cdot$V$^{-1}\cdot$s$^{-1}$, while its hole mobility is an order of magnitude lower than the electron mobility. The high carrier mobility, suitable band gap and special structure make it have great potential in photocatalysis. Wang $et$ $al.$ synthesized TiO$_2$/graphdiyne composites by hydrothermal method, and found that compared with TiO$_2$/graphene composites, TiO$_2$/graphdiyne can better catalyze the degradation of methylene blue~\cite{Wang2011}. Then, by first-principles calculations, they showed that the TiO$_2$/graphdiyne composite has more positive charges than TiO$_2$/graphene, easier to capture electrons, and has stronger photocatalytic effect~\cite{Yang2013a}. In addition, Venugopal $et$ $al.$ designed and synthesized a ZnO/graphdiyne composite, using ZnO instead of TiO$_2$, with similar photocatalytic performance~\cite{Thangavel2015}. Similarly, the nice performance of graphdiyne has also promoted the development of photoelectrochemistry and electrolytic water decomposition catalysts~\cite{Li2016a,Han2018,Zhang2018a,Xue2016,Xue2016a,Li2017,Xue2017}. 

Because graphdiyne combines the characteristics of 2D materials and porous materials, it exhibits good electronic transport properties and can be used in energy storage devices. Theoretical calculations show that the atomic density of graphdiyne is reduced due to the existence of diacetylene bonds, and therefore it can form LiC$_3$ by the surface adsorption of lithium. The theoretical specific capacity of graphdiyne is about 744 mAh$\cdot$g$^{-1}$~\cite{Shekar2013,Sun2012}, which is twice that of graphite~\cite{Dahn1991,Toyoura2010}. Graphdiyne~\cite{Huang2015,Zhang2015} and its derivatives, such as N-graphdiyne~\cite{Zhang2016}, Cl-graphdiyne~\cite{Wang2017}, H-graphdiyne~\cite{He2017}, B-graphdiyne~\cite{Wang2018a} participated in the preparation of lithium battery, which has the characteristics of large capacity, good cycle stability and high Coulomb efficiency. At the same time, graphdiyne can also be used as a good material to assemble solar cells~\cite{Xiao2015,Kuang2015,Liu2020}, which can greatly improve the efficiency of solar cells. In addition, graphdiyne has great potential in hydrophobic materials~\cite{Gao2015,Liu2017b} and temperature sensors~\cite{Yan2018}.

\subsection{Carbon nanotubes}
Carbon nanotubes (CNTs) are 1D systems with carbon atoms bonding each other by entirely $sp^2$ hybridization. As early as 1952, Radushkevich and Lukianovich published a clear image showing carbon nanotubes with a diameter of 50 nm~\cite{Radushkevich1952,Monthioux2006}. The rediscovery of CNTs in 1991 by lijima~\cite{Wang2007} set off an upsurge in the study of CNTs~\cite{Saito1997,Rueckes2000}. The CNTs include single-walled (SWCNTs) and multi-walled carbon nanotubes (MWCNTs). MWCNTs contain double-walled (DWCNTs) and triple-walled carbon nanotubes (TWCNTs)~\cite{Fujisawa2016} and so on. The longest CNTs are known to be more than 0.5 m~\cite{Zhang2013}, and the shortest are cycloparaphenylene~\cite{Jasti2008}. The structure of CNTs is determined by the diameter and chiral angle (helicity), which are usually described by two Hamada indexes (n, m). The synthesis methods of CNTs include arc discharge~\cite{Journet1997}, laser ablation~\cite{Rinzler1998}, microwave heating~\cite{Kharissova2009}, hydrocarbons CVD~\cite{Li1996} and so on.

The diversity of diameters and chiral angles leads to a wide variety of structures, accompanied by a wealth of properties, such as electronic, thermal and mechanical properties~\cite{Saito1998}. The density of CNTs was observed to be 1.6 g$\cdot$cm$^{-3}$~\cite{Sugime2013}. The Young’s modulus of SWCNTs is 1-5 TPa, while that of MWCNTs is 0.2-0.95 TPa. CNTs can behave as both metal and semiconductor, which is determined by Hamada indexe (n, m). For SWCNTs, if n-m can be divided by 3, then CNTs are considered as a metal, otherwise, CNTs act as semiconductors. Metallic SWCNTs can be used to study the single-electron charging, Luttinger liquid, quantum interference, etc~\cite{Dekker1999,Liang2001}, whereas semiconducting CNTs have been studied in transistors, sensory devices, etc. Superconductivity was observed in SWCNTs with diameters of 1.4 and 0.4 nm at about 0.55 K~\cite{Kociak2001} and 15 K~\cite{Tang2001}, respectively. Moreover, the phonon propagation in CNTs is anisotropic, and the thermal conductivity of a single MWCNT along the tube direction is about 3000 W/mK at room temperature, which is much larger than that of diamond and graphite ($\sim$2000 W/mK). Therefore, CNTs are expected to be a good thermal conductor along the tube direction, while they are thermal insulating transverse to the tube axis. The thermal stability of CNTs is maintained at 2800 $^{\circ}$C in vacuum and 750 $^{\circ}$C in air. 

CNTs have a wide range of applications. MWCNTs as electrically conductive components in polymer composites are the first to realize commercial application. CNTs can also be used as supercapacitor electrodes, because the capacitance of SWCNT and MWCNT electrodes are 180 and 102 F/g, respectively~\cite{Niu1997}. In addition, CNTs can also be used for hydrogen storage, field emission devices~\cite{Heer1995,Rinzler1995}, sensors and probes~\cite{Zaporotskova2016}, etc. N-doped CNTs have good stability and can be implemented as efficient metal-free catalysts for nitrobenzene hydrogenation~\cite{Gong2009,Xiong2020}.

\subsection{Carbyne}
Carbon chain is a 1D carbon allotrope formed by multiple carbon atoms arranged in a straight line, which has been studied since 1885~\cite{Baeyer1885}. According to the chemical bonds connected, it can be divided into two types: 1) carbon atoms are alternately connected by single and triple bonds, i.e. polyyne~\cite{Shun2006}; 2) carbon atoms are continuously arranged by double bond, i.e. cumulene~\cite{Januszewski2014}. In general, with the increase of the length of these carbon chains, the stability will become worse, and it is easy to transform into other structures. An ideal 1D carbon chain is infinitely long, and its properties do not change with the length, so it is called carbyne. Carbyne was first detected in 1961~\cite{Korshak1961}, and was reproduced in 1978~\cite{WHITTAKER1978,WHITTAKER1978a}. In 2016, Shi $et$ $al.$ synthesized a stable carbon chain composed of more than 6000 carbon atoms by using the carbon nanotube template method~\cite{Shi2016}. In 2018, Heeg $et$ $al.$ studied this ultra-long carbon chain by near-field Raman spectroscopy, and confirmed that the Raman spectrum peak is no longer affected by the length when the length exceeds 30 nm, thus experimentally confirming the existence of carbyne~\cite{Heeg2018}. At present, how to control the synthesis of carbon chain with different length, as well as the synthesis of stable carbyne, is still a concern.

Theoretical calculations show that the mechanical strength of carbyne is higher than that of diamond~\cite{Liu2013a}, which is expected to be used in mechanical reinforcement materials. Meanwhile, carbon fiber compounds composed of carbyne may be stronger than existing carbon fiber compounds, and are expected to make progress in aerospace and many fields~\cite{Heimann1983}. In addition, the band gap of carbyne can be adjusted with length and external strain, and the minimum band gap is 1.8 eV. This means that carbyne is a typical semiconductor, which has great potential in the synthesis of semiconductor materials with a single atom width~\cite{Shi2017}. The nonlinear optical properties of carbyne also contribute to the fabrication of frequency conversion devices~\cite{Ma2016}. It also has potential in energy storage~\cite{Slepkov2004,Shiraishi2005} and hydrogen storage~\cite{Sorokin2011}.

\subsection{Fullerenes}
Fullerene is another kind of carbon allotrope, whose molecules are connected by carbon atoms through double or single bonds to form a closed curved surface. It contains a series of closed cage molecules, such as C$_{60}$, C$_{70}$, C$_{76}$, C$_{78}$, C$_{80}$, C$_{84}$ and so on. A common form of fullerene molecule C$_{60}$, composed of 60 carbon atoms in a football shape, is known as C$_{60}$ or buckminsterfullerene, which was first discovered in 1985 by Harold Kroto, Richard E. Smalley and Robert F. Curl~\cite{Kroto1985}. Highly symmetrical cage C$_{60}$ is a truncated icosahedron with 60 carbon atoms forming 20 hexagons and 12 pentagons. There are 6 fivefold axes, 10 threefold axes and 60 twofold axes in C$_{60}$, where two ring bonds with a bond length of 6:6 between two hexagons and 6:5 between a hexagon and a pentagon coexist. The van der Waals diameter of C$_{60}$ molecule is about 1.1 nm~\cite{Qiao2007}. Fullerenes can be obtained based on the thermal decomposition of graphite firstly, and then extracted from the thermal decomposition products by solvents and sorbents.

Fullerenes exhibit unique physical and chemical properties. Crystalline fullerenes are semiconductors with a band gap of 1.2$\sim$1.9 eV and good photoconductivity. When they are exposed to the visible light, the electrical resistance decreases, which can be used to make photodetectors and optoelectronic devices. However, C$_{60}$ crystal can be tuned to a metal by doping alkali metal atoms. The C$_{60}$ compounds doped with potassium or Rubidium alkali metal atoms (K$_3$C$_{60}$, Rb$_3$C$_{60}$) are superconductors with the transition temperature of 18-40 K~\cite{Hebard1991}. The ferromagnetic materials can be obtained by adding platinum group to C$_{60}$. Fullerenes possess high electronegativity and act as strong oxidants in chemical process.  Fullerenes have poor solubility, and they are soluble in some organic solvents (such as carbon disulfide, toluene, benzene, tetrachloromethane, decane, hexane, pentane) and insoluble in polar solvents. Fullerenes can absorb, bind or react with hydrogen~\cite{Komatsu2013}, making them useful in hydrogen storage~\cite{Durbin2016}, oxygen production~\cite{Giron2016}, polymerization~\cite{Atovmyan2011}, free radical reaction~\cite{Tzirakis2013} and so on. 

The application of fullerenes is mainly focused on medicine and pharmacology~\cite{LALWANI2013}, and the idea of using the combination of fullerenes' endohedral compounds and radioisotopes to develop anticancer drugs has attracted extensive attention. Fullerenes are also predicted as the basis of the battery, because compared with lithium battery, they have the advantage of high efficiency, light weight and environmental protection. Fullerenes also show nonlinear optical properties due to their solubility in nonpolar solvents, which provide an opportunity for fullerenes to be used as optical closures-limiters of laser radiation intensity. Because the optical absorption spectrum of fullerenes film is between 180 and 680 nm, the application of fullerenes film in solar cells is also possible. 

\subsection{Amorphous carbon}
In contrast to the crystalline diamond and graphite, amorphous carbon contains a certain degree of disorder, in which the carbon atoms does not form the crystalline structure and the bonds between carbon atoms are irregular. $sp^3$, $sp^2$ and even $sp$ hybrid bonds can exist in amorphous carbon, where the dangling $\pi$-bond could be hydrogen-passivated to stabilize it. In addition, amorphous carbon could be turned to other forms of carbon under certain conditions. For examples, microdiamonds can be grown by irradiating the amorphous film with nanosecond lasers at room temperature~\cite{Narayan2015a}; high quality graphene can be obtained by thermal annealing of amorphous carbon~\cite{Zheng2010}; and amorphous carbon powders could be converted to the crystalline graphite by microwave heating~\cite{Kim2016}. 

Among the amorphous carbon, diamond-like carbon (DLC) is a metastable form, where most of the carbon atoms in DLC are connected with $sp^3$ hybrid bond, and a small amount of $sp^2$ bond. DLC is a metastable anisotropic semiconductor. There are also cubic and hexagonal DLC, which inherit the hardness of diamond and are easy to deform. DLC contains both carbon and hydrogen atoms, which can be prepared by physical vapor deposition, CVD and other methods~\cite{Jethanandani1997}. DLC films with different smoothness, $sp^3$ bonding ratio, hydrogen content and deposition amount can be obtained by different methods.

At present, the study of DLC films mainly focuses on the wear resistance, antifriction, antibacterial and corrosion resistance of the coating, as well as improving the electrochemical performance without changing the overall performance of the substrate. They are environment-friendly materials and can be used in mechanical wear-resistant parts, clocks, aerospace and medical materials~\cite{Donnet1999,Mahmud2014,Hauert2013,Tyagi2019}. 

There are four main forms of DLC: hydrogen-free amorphous carbon film (a-C), tetrahedral hydrogen-free amorphous carbon film (ta-C), hydrogenated amorphous carbon film (a-C:H) and tetrahedral hydrogenated amorphous carbon film (ta-C:H). In the ternary diagram of $sp^3$ covalent carbon, $sp^2$ covalent carbon and hydrogen, they belong to different regions~\cite{Jacob1993}. Among them, ta-C has a large number of carbon atoms bonded into tetrahedron by $sp^3$ bond, which is similar to diamond in properties and can be used in electronic devices~\cite{McKenzie1993,McKenzie1996,Neuville2014}. Recently, Ohtake $et$ $al.$ redefined the DLC region on ternary diagram by analyzing the $sp^3$ ratio, hydrogen content, and other properties of 74 kinds of amorphous carbon films~\cite{Ohtake2021}.

Undoped DLC coating has some challenges in thickness, adhesion and stability, which can be solved by preparing DLC containing metal nanoparticles~\cite{Tamulevicius2018}. DLC-based metal nanocomposites can also be applied to novel biosensors that combine the resonant response of waveguide structures~\cite{Tamulevicius2018}. Recently, in view of the well-known bactericidal effect of silver in medicine, Ag-doped DLC coating was synthesized by hot electron vacuum arc deposition, which has good corrosion resistance, electrochemical stability and antibacterial effect against S. aureus~\cite{Mazare2018}.

The $sp^3$ bond in DLC is easy to break at high temperature, which leads to graphitization. Graphitization is usually accompanied by thermal degradation of DLC. Doping silicon, boron, nitrogen, etc. can significantly improve the high-temperature tribological properties of DLC films~\cite{Erdemir_2006}. A recent report showed that in silicon doped DLC, Si atoms often replace $sp^2$ hybrid carbon atoms, which leads to a significant increase in the proportion of $sp^3$ chemical bond, thus inhibiting graphitization~\cite{Zhang2018}. Si doping can improve the nano-wear resistance, hardness and bonding strength. Another report pointed out that the Si-C bond is longer than the C-C bond, which enables the Si-doped DLC film to accommodate high disorder and reduces the frequency of occurrence of highly strained C-C bond~\cite{Hilbert2018}.

\begin{figure*}[!htbp]
  \centering
  \includegraphics[scale=0.48,angle=0]{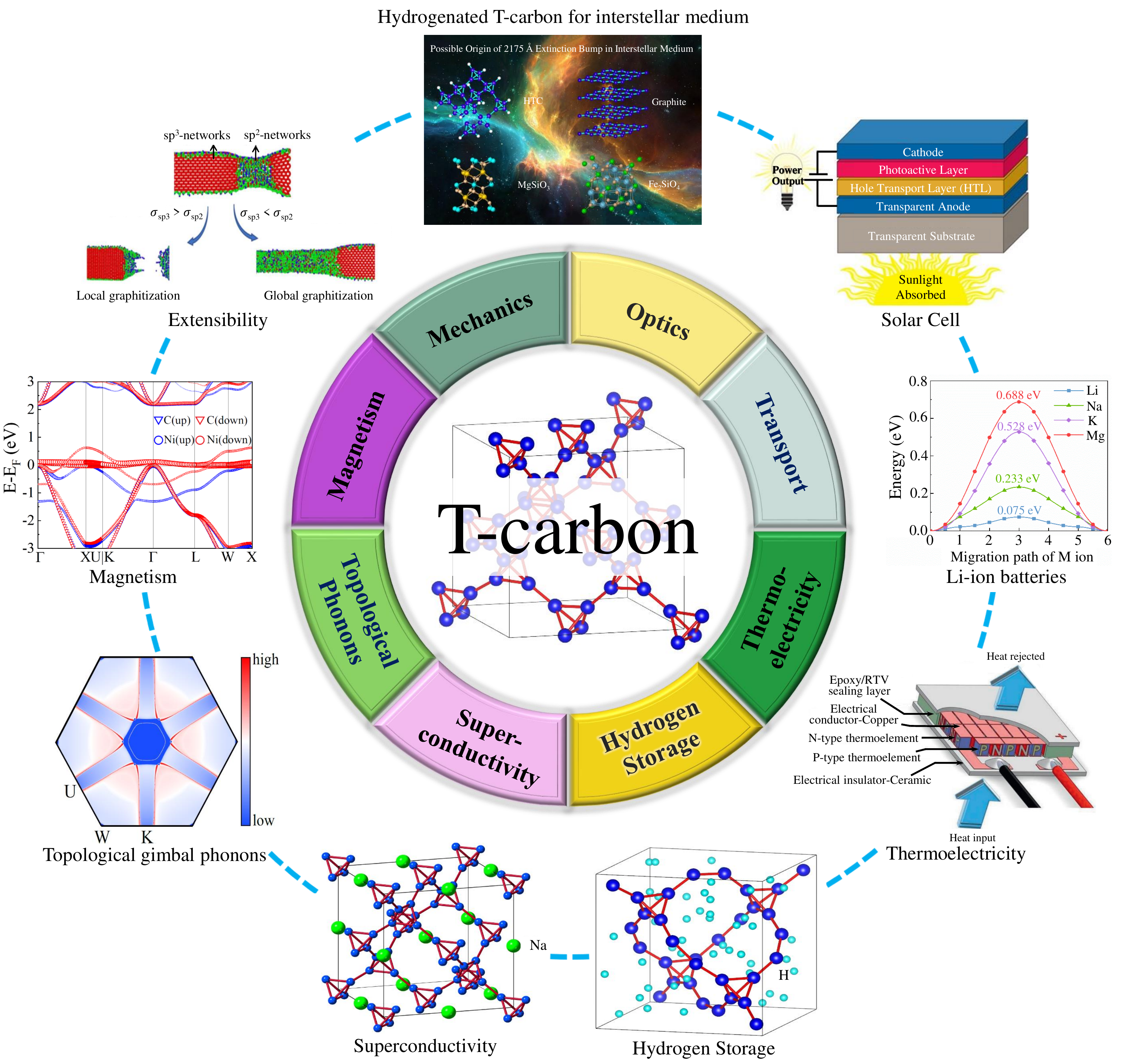}\\
  \caption{Overview of exotic properties and potential applications of T-carbon. Parts of the figure are reprinted (adapted) with permission from ref~\cite{Qin2019}, Copyright (2019) Royal Society of Chemistry; ref~\cite{You2020}, Copyright (2020) American Physical Society; ref~\cite{Faraji2014}, Copyright (2014) Elsevier; ref~\cite{Bai2018}, Copyright (2018) Elsevier; ref~\cite{You2021}, Copyright (2020) Elsevier; https://www.sigmaaldrich.com/technical-documents/articles/technology-spotlights; NASA\&ESA.}\label{fig0}
\end{figure*}

\subsection{Carbon dots}
Carbon Dots (CDs), as a well-known 0D carbon nano allotrope that is mostly smaller than 10 nm, have attracted wide attention in recent years. It consists of nanoscale carbon cores and passivated chemical functional groups shell, which is similar to a soft corona-like shell~\cite{Liang2020}. According to the core structure, CDs can be roughly divided into three main categories: carbon nanodots (CNDs), carbonized polymer dots (CPDs), graphene quantum dots (GQDs). In GQDs, the carbon core possesses one or few graphene sheets with diameters of 2-20 nm~\cite{Georgakilas2015}. CNDs usually have amorphous inner structures, and carbon quantum dots (CQDs) mainly composed of inner graphitic crystalline structures are also regarded as a special case of CNDs~\cite{Langer2021}. CPDs exhibit a polymer/carbon hybrid structure, including rich short polymer chains and functional groups on the surface~\cite{Xia2019}. Although CDs have been found for only 15 years~\cite{Xu2004}, many different approaches have been implemented in the synthesis of CDs~\cite{Liang2020,Semeniuk2019}. For all methods, the advantages of nanoscale carbon particles and sufficient passivation of the surface should be ensured. The carbon cores and surface passivated functional groups make CDs have versatile properties.

The most attractive feature of CDs is its bright and colorful fluorescence emission ranging from ultraviolet (UV) to near-infrared. The main photoluminescence mechanisms include: (1) quantum confinement effect (QCE); (2) surface and defect states; (3) molecular fluorophores (MFs); (4) crosslink-enhanced emission (CEE) effect; (5) thermally activated delayed fluorescence (TADF) and (6) phosphorescence~\cite{Langer2021,Zhu2015}. Any CDs system contains more than one photoluminescence mechanism, which makes the photoluminescence spectrum and quantum yield of CDs very complex. For GQDs with a non-zero tunable band gap, its photoluminescence mainly depends on the QCE and $\pi$-plasmons in the core states. In addition, the effective intersystem crossing (ISC) in the core states allows GQDs to display TADF and phosphorescence. For CNDs, the main contribution for photoluminescence is surface states due to doping effect and MFs. The heteroatoms in CNDs can be controlled by experiments so that the photoluminescence characteristics of functional groups change dramatically. On the other hand, the CEE effect plays a leading role in the photoluminescence of CPDs. It should be noted that molecular photoluminescence is determined only by the fluorescent molecules in CNDs without spatial limitation. However, the potential photoluminescence centers in CPDs are spatially limited~\cite{Langer2021}. Another interesting feature of CDs is its unique photoinduced redox characteristics as both excellent electron donors and acceptors. This can be verified by the experimental high-efficient fluorescence quenching of electron donor and acceptor molecules as quenchers~\cite{Liang2020}.

In addition to the above photoluminescence and redox characteristics, CDs also have the characteristics of chemical inertia, high specific surface area, nontoxicity, low cost and biocompatibility. Using these advantages, CDs are widely used in the fields of photocatalysis, sensors, sterilization, optoelectronics, food and safety. For example, a sensitive and effective wireless bacteria detection and ablation sensing system based on fluorescent CDs (FCD) and CsWO$_3$ nanohybrid has recently been designed. The surface-coatable CsWO$_3$–FCD nanohybrid shows excellent solubility, photothermal effect, and sensing ability~\cite{Robby2021}. Another example is that a new type of CDs was synthesized last year, which integrates FeCo$_2$O$_4$ nanoflowers supported on Ni foam. Such new metal oxide-integrated CDs show excellent performance in electrochemical decomposition of water~\cite{Kundu2020}.

\subsection{T-carbon}

T-carbon was theoretically predicted in 2011~\cite{shengxianlei2011} and first fabricated in experiment in 2017~\cite{Zhang2017}. T-carbon is a 3D carbon allotrope, which was obtained by substituting each carbon atom in diamond with a regular carbon tetrahedron composed of four carbon atoms with $sp^3$ hybridization. T-carbon has a porous structure and is a semiconductor with a direct band gap of about 5 eV. Due to its exotic properties, T-carbon has great applications such as hydrogen storage, photocatalysis, superconductivity, etc.~\cite{Sun2019,Qin2019,You2020} (Fig.~\ref{fig0}).

\begin{figure}[!htbp]
  \centering
  \includegraphics[scale=0.44,angle=0]{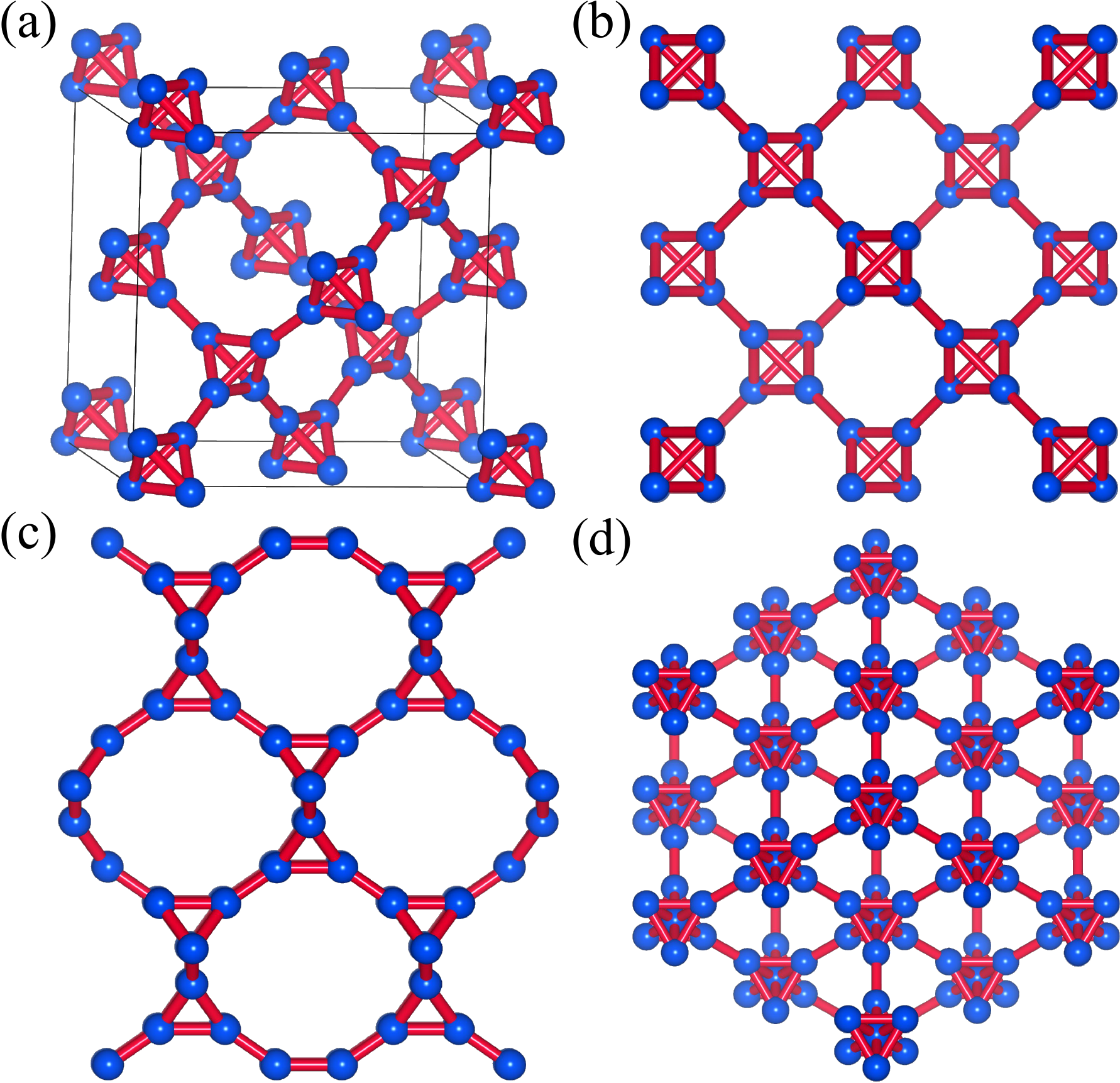}\\
  \caption{The structure of T-carbon. (a) The cubic crystalline structure of T-carbon obtained by replacing each carbon atom in diamond with a carbon tetrahedron. (b)–(d) Views from [100], [110] and [111] directions of T-carbon, respectively. Copyright (2011) American Physical Society. Reprinted (adapted) with permission from ref~\cite{shengxianlei2011}.}\label{fig2-1}
\end{figure} 

\begin{figure}[!htbp]
  \centering
  \includegraphics[scale=0.31,angle=0]{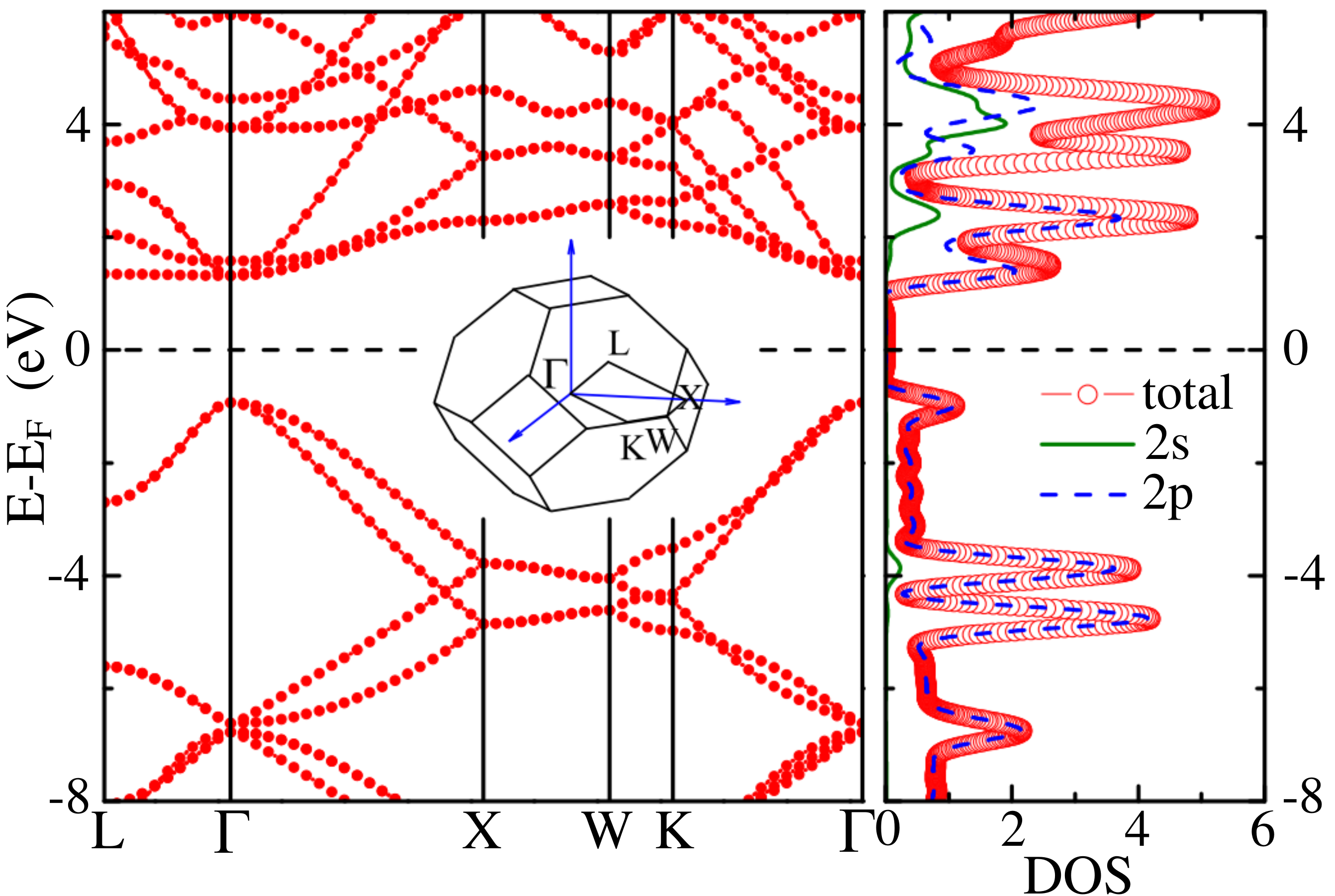}\\
  \caption{Electronic band structures and corresponding DOS for T-carbon. Copyright (2011) American Physical Society. Reprinted (adapted) with permission from ref~\cite{shengxianlei2011}.}\label{fig2-2}
\end{figure}

T-carbon has the same space group of $Fd\bar{3}m$ as diamond, as shown in Fig.~\ref{fig2-1}, and its primitive cell contains two tetrahedral units with eight carbon atoms. The three lattice vectors are $a_1=(1,1,0)l/2$, $a_2=(0,1,1)l/2$ and $a_3=(1,0,1)l/2$ with $l=7.52$ \AA, and the carbon atoms occupy the Wyckoff position $32e(0.0706, 0.0706, 0)$. The comparison of T-carbon with several typical carbon allotropes are listed in Table~\ref{Table2-1}. It is noted that T-carbon has much lower equilibrium density (1.5 g/cm$^3$) and bulk modulus (169 GPa) among 3D carbon materials. The electronic band structure and density of states (DOS) of T-carbon are plotted in Fig.~\ref{fig2-2}. One observes that T-carbon is a semiconductor with a direct band gap by different calculation methods~\cite{shengxianlei2011,Sun2019, Alborznia2019}. The projected DOS shows that 2$p_x$, 2$p_y$ and 2$p_z$ orbitals make more contribution to the valence band than 2$s$ orbitals, indicating the anisotropic $sp^3$ bonds in T-carbon.

In this review, we will first discuss recent experiments on T-carbon. Then, we will present the intriguing properties of T-carbon, including the mechanical, electronic, optical, topological and superconducting  properties, and its potential applications in solar cells, photocatalysis, interstellar medium, etc. Finally, we will outline future prospects and outlook on T-carbon.

\section{Experiments on T-carbon}
Recently, T-carbon has been successfully synthesized in two independent experiments by picosecond laser irradiation~\cite{Zhang2017} and plasma enhanced chemical vapor deposition~\cite{Xu2020}, respectively. 

\subsection{Synthesis of T-carbon nanowires by picosecond laser irradiation}
In 2017, T-carbon nanowires (NWs) have been successfully synthesized by irradiating multi-walled carbon nanotubes (MWCNTs) dissolved in methanol solution with picosecond laser at a specific frequency~\cite{Zhang2017}. The first synthesis experiment of T-carbon lays a basis for further investigation and application of T-carbon.

\begin{figure}[!htbp]
  \centering
  \includegraphics[scale=0.35,angle=0]{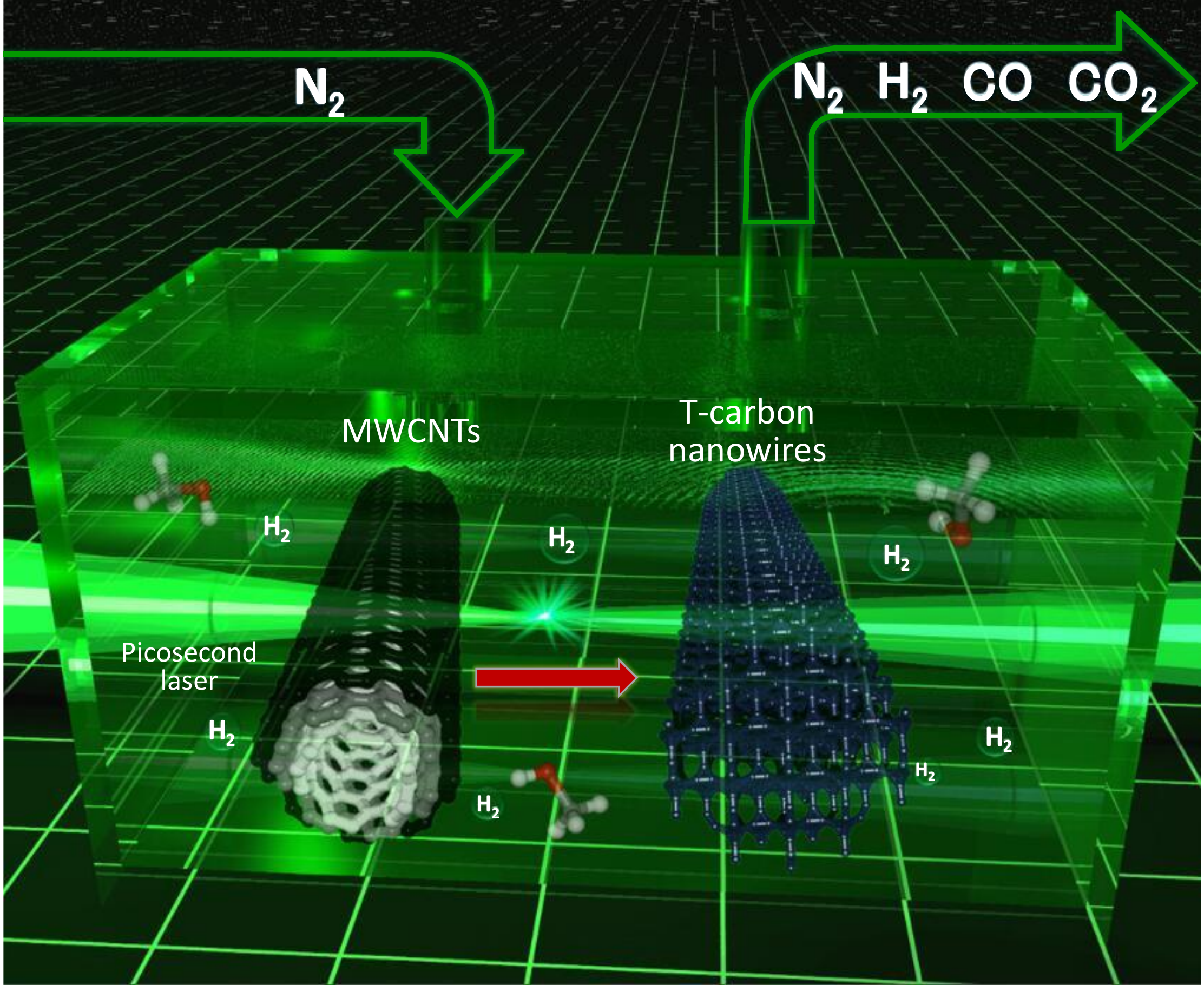}\\
  \caption{The experimental layout of T-carbon nanowires from a pseudo-topotactic conversion of multi-walled carbon nanotubes under picosecond laser irradiation. Reproduction from http://gr.xjtu.edu.cn/web/jinying-zhang publications.}\label{fig3-1}
\end{figure}

\begin{figure}[!htbp]
  \centering
  \includegraphics[scale=0.46,angle=0]{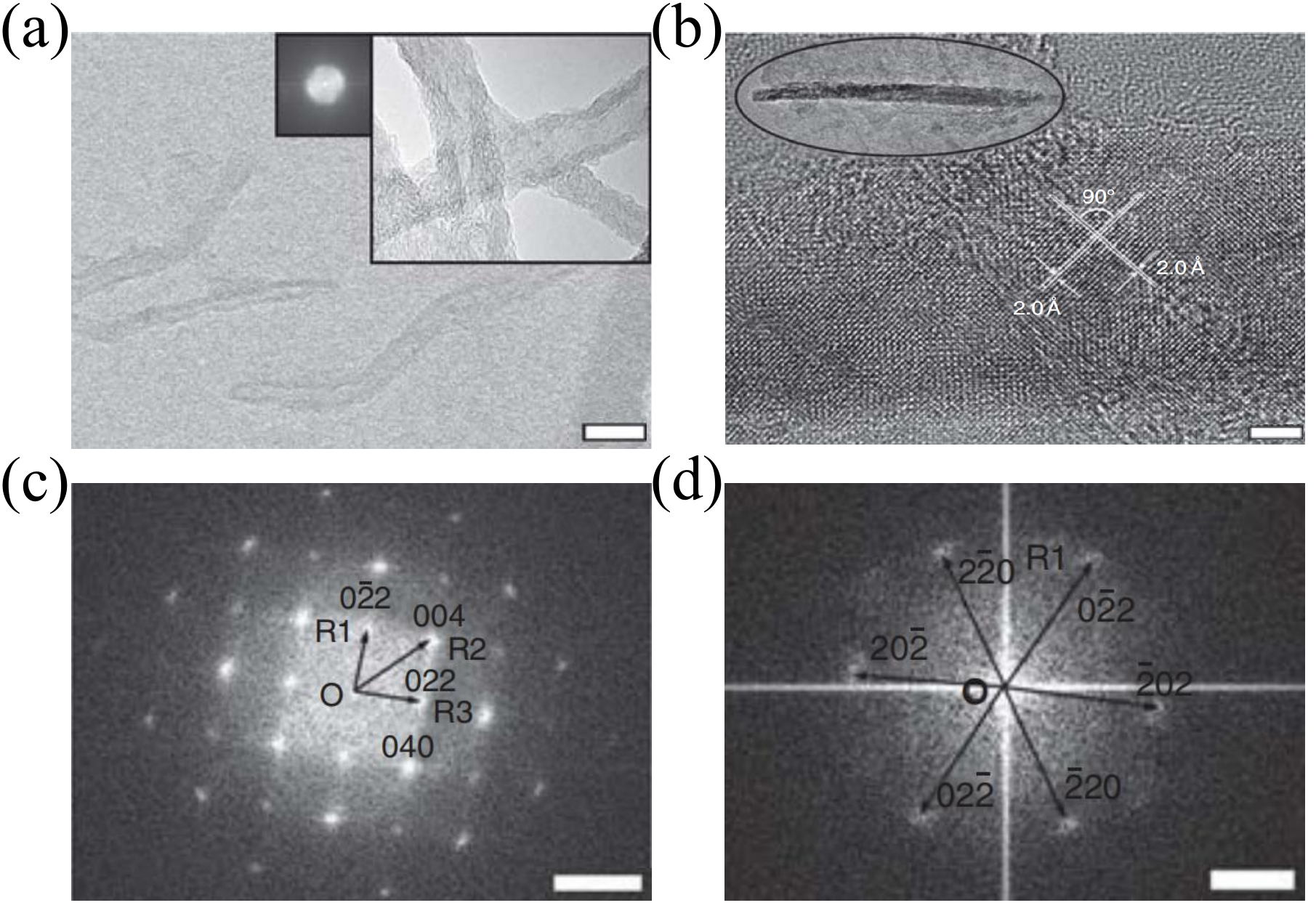}\\
  \caption{HRTEM images of (a) shortened MWCNTs with FFT pattern and the structure of pristine MWCNTs inserted in the upper right corner (scale bar = 20 nm) and (b) carbon nanowire produced from picosecond laser irradiation with the low magnification in the inset (scale bar = 5 nm). FFT patterns of (c) the nanowire corresponding to (b) (scale bar = 5 nm$^{-1}$), and (d) another nanowire (scale bar = 2 nm$^{-1}$ ). Reprinted (adapted) with permission from ref~\cite{Zhang2017}.}\label{fig3-2}
\end{figure}

The MWCNTs used in the experiment were obtained by CVD. 0.5 mg of MWCNTs were dispersed in absolute methanol (40 ml), and then the suspension was irradiated for an hour by the Q-switched laser with the wavelength of 532 nm, pulse duration of 10 ps, repetition rate of 1000 Hz and pulse power of 75 mW. After irradiation, the suspension became transparent and gas was produced during the reaction. The experimental layout of T-carbon nanowires is presented in Fig.~\ref{fig3-1}. Images of the transparent suspension after reaction obtained by the high-resolution transmission electron microscopy (HRTEM) show that there were new NWs structures appearing as presented in Figs.~\ref{fig3-2}(a) and (b). Besides, the suspension also contains unreacted MWCNTs and some amorphous phases. A fast Fourier transform (FFT) pattern calculated from HRTEM images shows that the new NWs had a cubic lattice as shown in Figs.~\ref{fig3-2}(c) and (d). The results of electronic energy loss spectroscopy (EELS) show that the ratio of $sp^3$ components in the suspension is dramatically increased after reaction compared with the suspension before laser irradiation, and only a small amount of carbon atoms with $sp^2$ hybrid bonds (16$\pm$4\%) left, which correspond to amorphous structures. To verify that there were only $sp^3$ hybrid bonds in the NWs, a comparative experiment was conducted, where the methanol solvent without MWCNTs was irradiated under the same conditions, and the EELS results show that the irradiated methanol solution also had a lot of the same amorphous structures. Moreover, the FFT patterns of NWs with different tilting angles were consistent with the diffraction patterns of different crystal planes of T-carbon, and the measured intra-planar and inter-planar angles of NWs were consistent with the calculated results of T-carbon, indicating that the new NWs were T-carbon. The obtained T-carbon NWs have the lattice constant of 7.80 \AA, which is in good agreement with the calculated value of 7.52 \AA. The observed ultraviolet (UV) absorption spectra and photoluminescence spectra of the suspension after irradiation exhibited an obvious absorption peak, which meets well with the calculated UV and photoluminescence spectra of T-carbon (see Fig.~\ref{fig8-3}). These results revealed that the T-carbon nanowires were really generated~\cite{Zhang2017}. 

The average diameter of T-carbon NWs was 11.8$\pm$2.8 nm, which was consistent with the statistical average of MWCNTs (11.7$\pm$2.2 nm), indicating that T-carbon NWs came from the pseudo-topotactic conversion of MWCNTs, which belongs to the first-order phase transition~\cite{Zhang2017}.

\subsection{Preparation of T-carbon by plasma enhanced chemical vapor deposition}
\begin{figure}[!htbp]
  \centering
  \includegraphics[scale=0.32,angle=0]{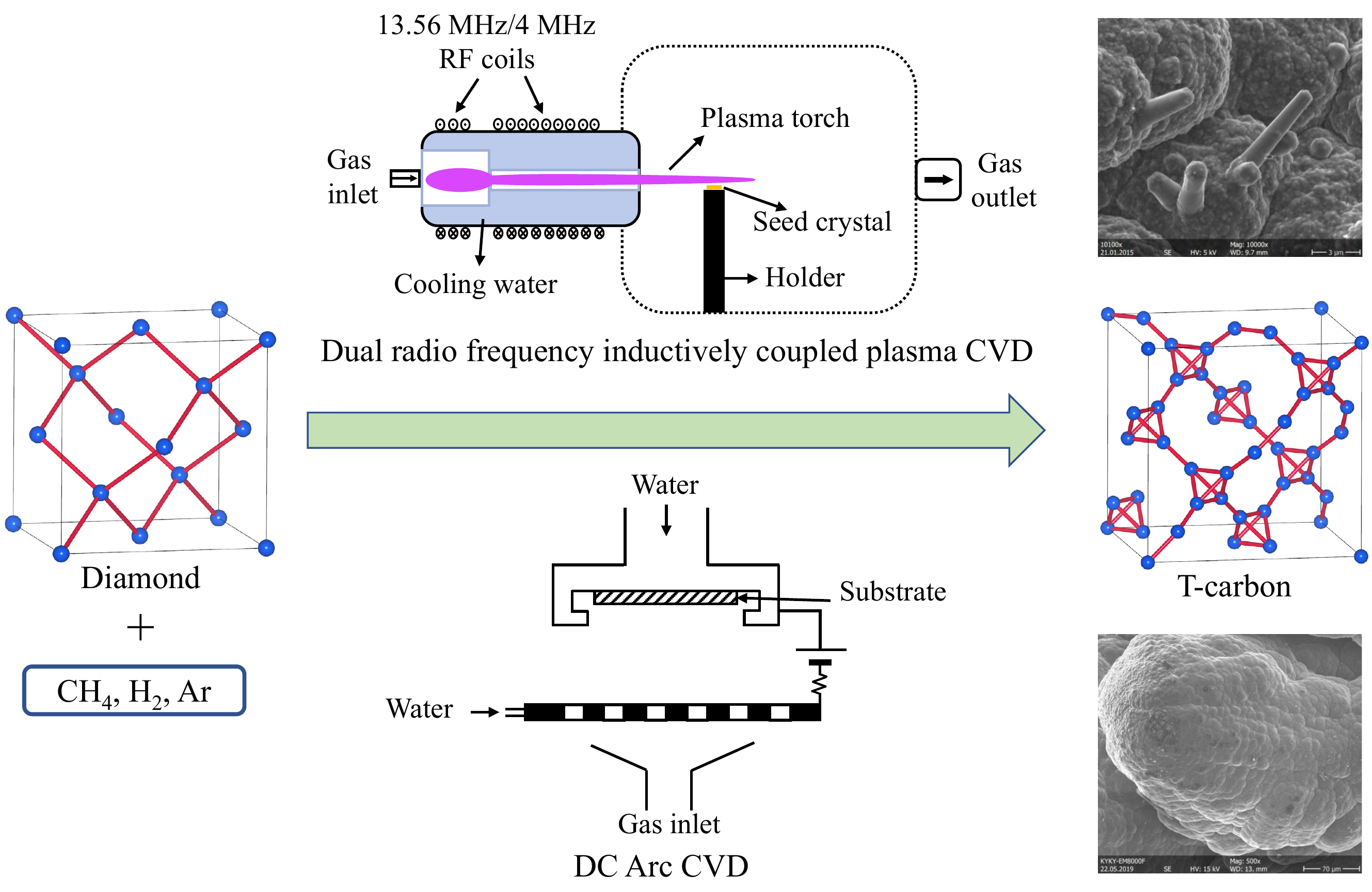}\\
  \caption{The schematic diagrams of T-carbon synthesis by plasma enhanced CVD. Copyright (2020) Elsevier. Reprinted (adapted) with permission from ref~\cite{Xu2020}.}\label{fig3-3}
\end{figure}

Plasma enhanced CVD (PECVD) is a mature technology that is widely used in material growth. Kai Xu $et$ $al$. used PECVD to grow T-carbon samples on the substrate of polycrystalline cubic diamond and single-crystalline cubic diamond under high temperature and high pressure, and obtained T-carbon with two different morphologies as shown in Fig.~\ref{fig3-3}~\cite{Xu2020}. For the substrate of the polycrystalline cubic diamond, the plasma feed gas was the mixture of Ar/H$_2$/CH$_4$ with 3 standard liters per minute (slm) of Ar, 3 slm of H$_2$ and 30 standard cubic centimeters per minute (sccm) of CH$_4$, and the plasma was lighted and sustained by direct current arcjet (DC arc). The pressure in the reaction chamber was kept at 8 kPa, and the substrate was heated to around 1000 ℃ for 4 h. For the substrate of the single-crystalline cubic diamond, the plasma feed gas was the mixture of Ar/H$_2$/CH$_4$ with 4 slm of Ar, 1.2 slm of H$_2$ and 60 sccm of CH$_4$ and the plasma was lighted and sustained by the dual radio frequency inductively coupled plasma (RF ICP), where the pressure was also kept at 8 kPa, and the substrate temperature was maintained at 890 ℃ for 200 h. The obtained product in the former experiment looks like the rods extended from the cauliflower-like diamond and connected with each other, forming kink joints, and the extended length of the sticking carbon rods can reach several microns. The obtained product in the latter experiment is similar to the rod bunch connected together. The thinnest stick-like sample thinned by the focus ion beam method was about 80 nm.

As the sample produced by the above two experiments was small, the XRD spectra were measured based on the newly generated phases together with their surrounding substrate. By comparing the measured XRD pattern of T-carbon with the simulated result by DFT calculations, it was found that both two samples have some peaks which meet well with the calculated result. For example, two peaks of $2\theta=39.41^{\circ}$ and 61.34$^{\circ}$ in the XRD spectrum for the sample grown on the single crystalline diamond substrate correspond to the (222) and (115) planes of T-carbon, respectively. Among all known carbon allotropes only T-carbon has such XRD peaks, indicating that the new rod-like phase was T-carbon. According to Bragg's law, the lattice constant was then deduced to be 7.89 \AA\ and 7.85 \AA, respectively, which are in good agreement with the calculated value of 7.52 \AA~\cite{shengxianlei2011} and the value (7.8 \AA) in the previous experiment~\cite{Zhang2017} for T-carbon. The weight percentage of T-carbon took up 1.46\% in the total mixed sample in terms of the relative ratio of XRD diffraction peaks. The results of selected area diffraction of TEM also show that there was T-carbon in the mixed sample. Moreover, the EELS spectrum showed that there were only $sp^3$ bonds in the sample. Although the ratio of T-carbon in the sample was very small, many peaks of the Raman spectrum corresponding to T-carbon were still obtained by comparing with the calculated results. In addition, the measured and simulated Fourier transform infrared spectroscopy (FT-IR) of the sample also show obvious consistent peaks (see Fig.~\ref{fig8-3}).

In the above experiments, the temperature of the plasma was much higher than that of the substrate, so the hotter carbon atoms moved upward and the colder moved downward. When colder atoms moved downward to deposit onto the substrate, they would be subject to the upward impact from the hotter atoms, which is equivalent to experiencing a negative pressure environment. The negative pressure environment was shown to be more suitable for the synthesis of T-carbon~\cite{You2019}. In addition, carbon atoms are prone to form T-carbon phase on the diamond substrate because T-carbon has a similar lattice structure and chemical bond with diamond. The experimental preparation of T-carbon with PECVD shows that the massive production of T-carbon is feasible.

The realization of T-carbon in experiments not only added a new member to the carbon family, but also offered great opportunities for the further study on T-carbon. In the near future, the massive production and practical applications of T-carbon are highly expected.

\section{Properties of T-carbon}
\subsection{Chemical bonding}
\begin{figure}[!htbp]
  \centering
  \includegraphics[scale=0.4,angle=0]{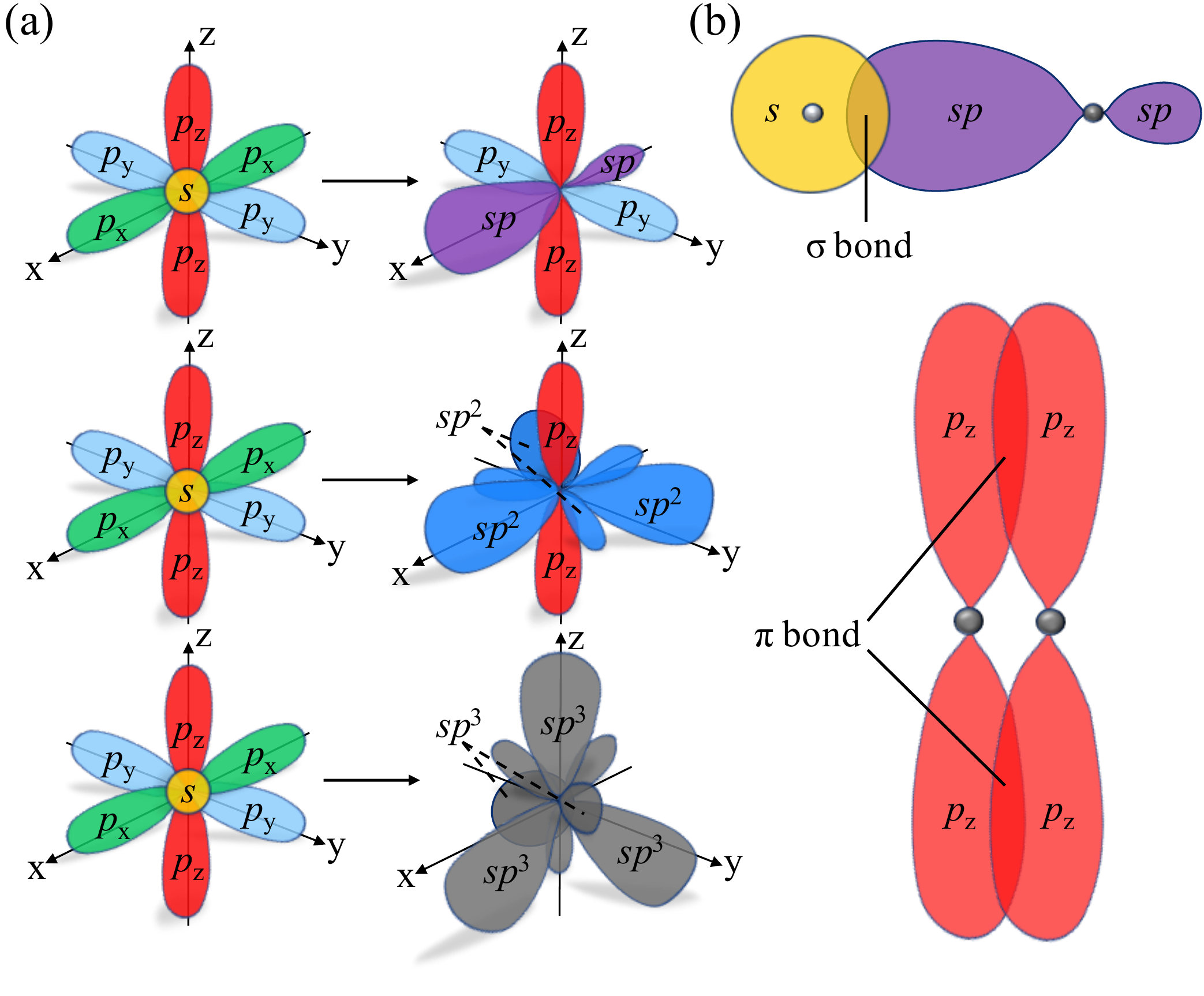}\\
  \caption{Schematic plots of (a) $sp$, $sp^2$ and $sp^3$ hybridization and (b) $\sigma$ and $\pi$ bonds.}\label{fig4-1}
\end{figure}

As we mentioned above, carbon atoms can form $sp$, $sp^2$, and $sp^3$ hybrid chemical bonds, which consist of $\sigma$ and $\pi$ bonds that differ in the overlapping of atomic orbitals as shown in Fig.~\ref{fig4-1}. Covalent bonds are formed by the overlapping of atomic orbitals. $\sigma$ bonds are a result of the head-to-head overlapping of atomic orbitals whereas $\pi$ bonds are formed by the lateral overlapping of two atomic orbitals. The $\sigma$ bonds are generally stronger than the $\pi$ bonds, owing to the significantly lower degree of overlapping of $\pi$ bonds. In general, single bonds ($sp^3$ hybridization) consist of $\sigma$ bonds, whereas a double bond ($sp^2$ hybridization) is made up one $\sigma$ and one $\pi$ bonds, and a triple bond ($sp$ hybridization) is composed of two $\pi$ and one $\sigma$ bonds. The $s$ character, which means the percentage of the $s$ orbital initially involved in the hybridization process, is related to the bond length and strength, because the $s$ orbital is smaller than the $p$ orbitals, and it has stronger interaction between the electrons and nuclei. Therefore, C-C bond in $sp$ hybridization is shortest and strongest, and that in $sp^3$ is the longest and weakest. Different bonding methods or ratios will lead to completely different properties, resulting in the rich diversity of carbon materials.

\begin{figure}[!htbp]
  \centering
  \includegraphics[scale=0.54,angle=0]{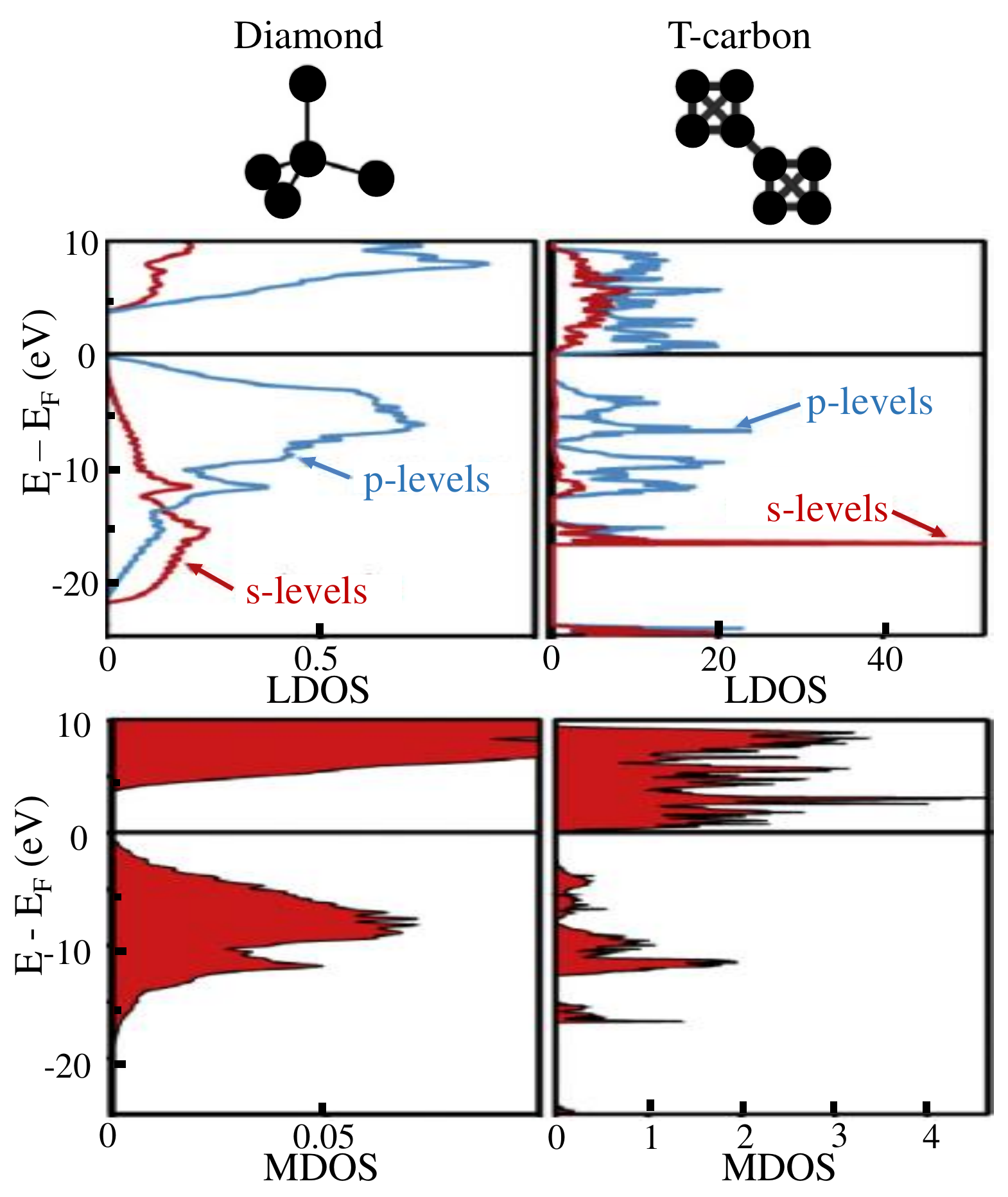}\\
  \caption{The structural motif (top), local DOS (middle) and MDOS for $sp^3$-mixed levels (bottom) of diamond and T-carbon. Copyright (2017) Elsevier. Reprinted (adapted) with permission from ref~\cite{Esser2017}.}\label{fig4-2}
\end{figure}

\begin{figure}[!htbp]
  \centering
  \includegraphics[scale=0.42,angle=0]{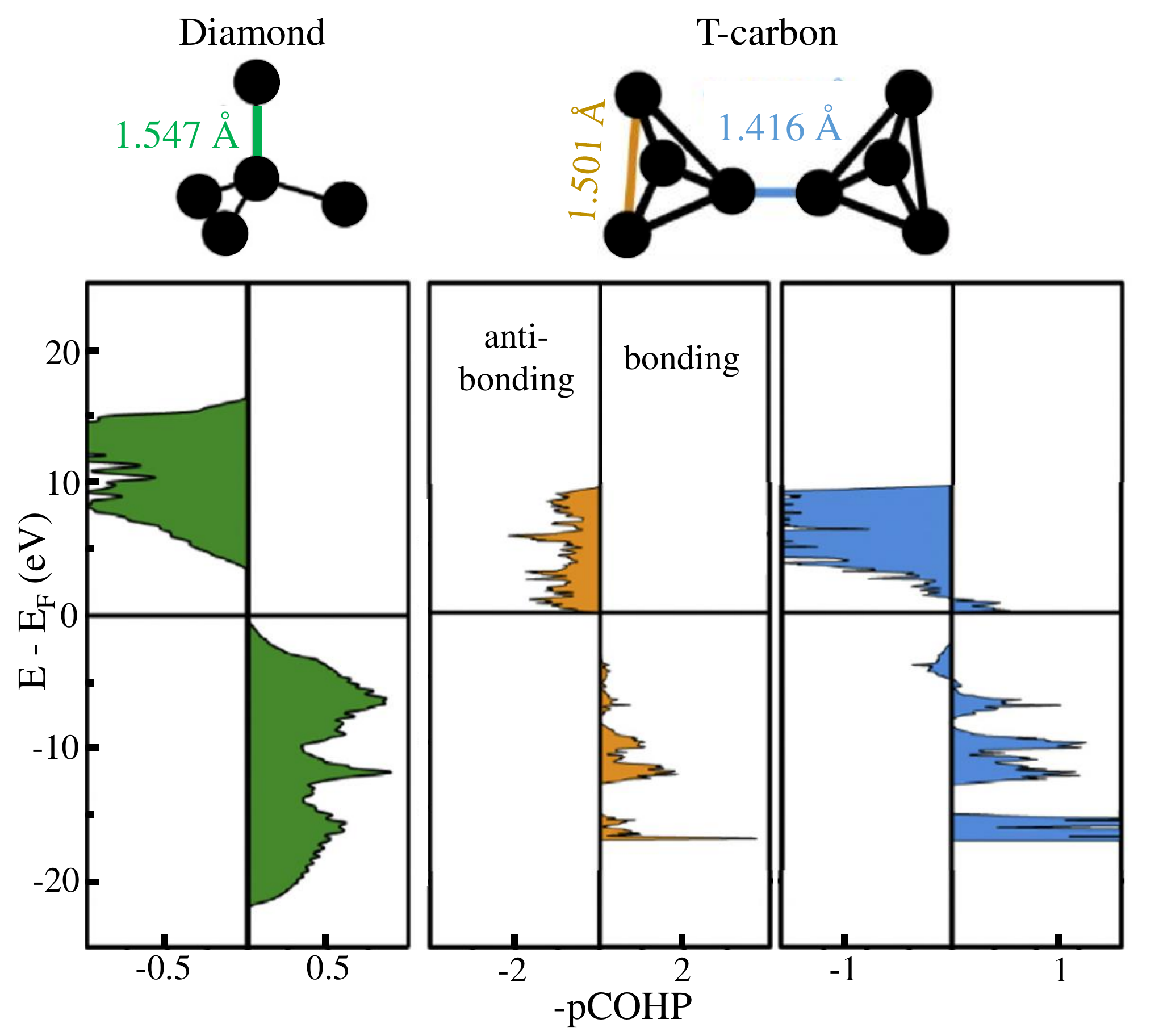}\\
  \caption{Projected COHPs of diamond and T-carbon. From left to right: C-C bond in diamond, C-C bond in the tetrahedron of T-carbon, and C-C bond between the tetrahedrons of T-carbon. Copyright (2017) Elsevier. Reprinted (adapted) with permission from ref~\cite{Esser2017}.}\label{fig4-3}
\end{figure}

T-carbon possesses $T_d$ symmetry around the center of the tetrahedral units consisting of four carbon atoms (C$_4$ units). The calculated total energies of graphite, diamond and T-carbon show that the energy of T-carbon is 125.6 kJ/mol higher than that of graphite, and 113.2 kJ/mol higher than that of diamond as a result of different chemical bonds~\cite{Esser2017}. However, due to the existence of dynamic barriers, T-carbon is still a stable structure and in the metastable state. For diamond, the C-C bond has a bond length of 1.547 \AA\ with the bond angle of 109.5°, which is formed by $sp^3$ hybridization of the $s$ and $p$ orbitals. For T-carbon, the bond length and bond angle in the C$_4$ units are 1.501 \AA\ and 60°, respectively, while between two adjacent C$_4$ units they are 1.416 \AA\ and 144.74°, respectively. To further investigate the hybridization of orbitals in T-carbon, the weighted mixing of local density of states (LDOS) and mixing density of states (MDOS) were performed with the $sp^3$-orbital mixing technique~\cite{Esser2015} as shown in Fig.~\ref{fig4-2}. Compared to diamond, the orbital mixing of $s$- and $p$-level of T-carbon is much lower. The projected Crystal Orbital Hamilton Population (pCOHP) uses the Hamiltonian matrix element between two atoms to effectively weight DOS by the bond energy, and the pCOHP results for diamond and T-carbon are plotted in Fig.~\ref{fig4-3}. Generally, the energy of the bonding state is less than zero (energy lowering), while that of antibonding state is above zero compared to the non-bonding state of zero energy. The shorter bond between two C$_4$ units in T-carbon is 49\% stronger than the longer one within C$_4$ units, and has largely $s$-character in the main bonding peak range from -17 to -15 eV (see Fig.~\ref{fig4-3}), while the higher $p$ orbitals do not contribute much to stabilization in contrast there is a leftward anti-bonding region area -5 to -2 eV. This anti-bonding characteristic is due to the obvious deviation of the bond angle from its parent allotrope, resulting in the systematic strain and the hybrid efficiency reduction of the $s$ and $p$ orbitals. Overall, T-carbon is different from traditional carbon allotropes in the way of bonding. On the one hand, the unique bonding brings a variety of good properties for T-carbon, on the other hand, it makes people have a deeper understanding of the law of carbon bonding, which plays a great instructive role in the search for new carbon allotropes.

\subsection{Mechanical properties}
The unique lattice structure and bond nature of T-carbon imply that it has excellent mechanical properties, which have been studied in several works.

\begin{figure}[!htbp]
  \centering
  \includegraphics[scale=0.28,angle=0]{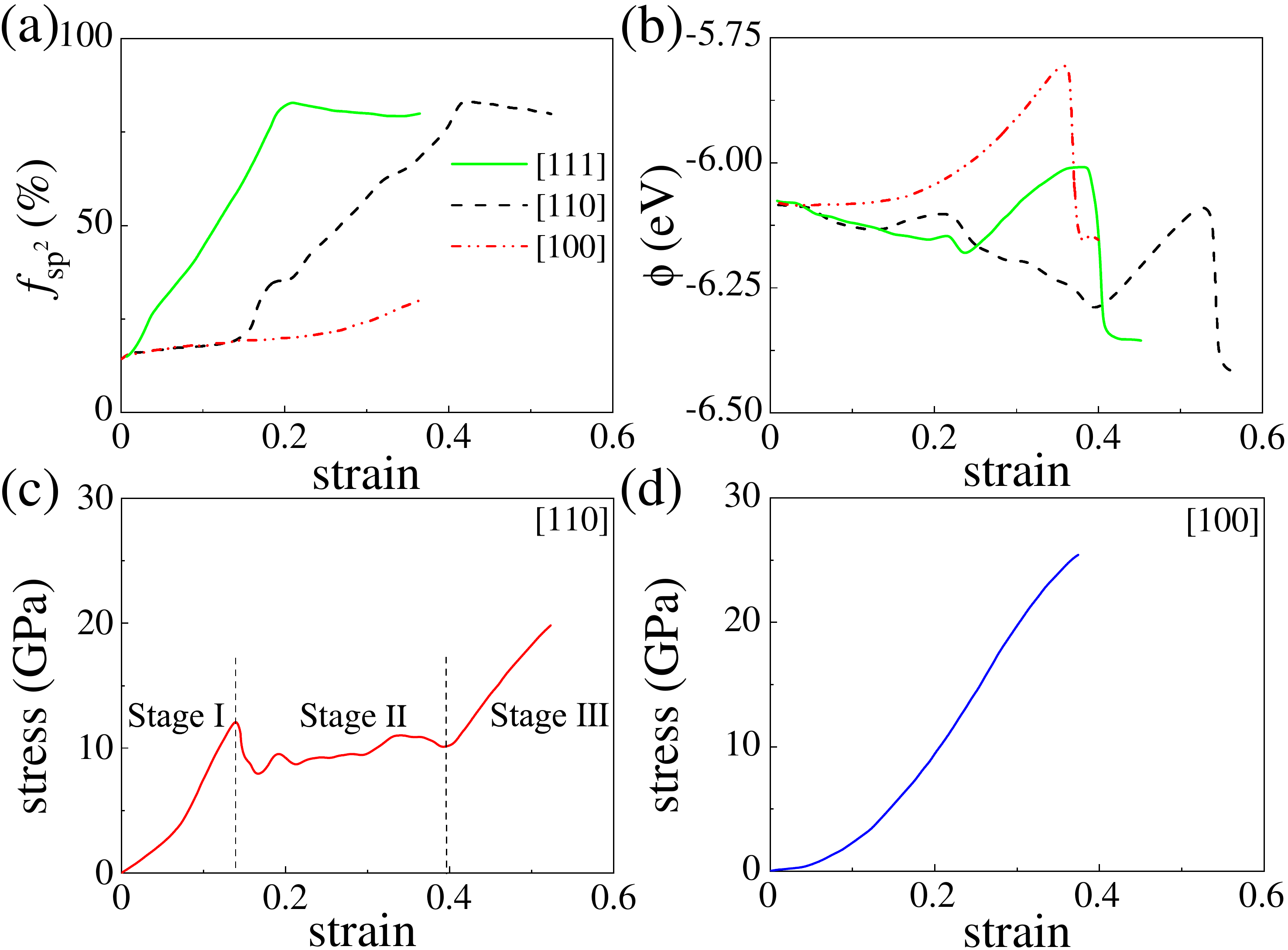}\\
  \caption{(a) The average atomic potential as a function of strain in [100], [110] and [111] directions, and (b) corresponding evolution of the ratio ($f_{sp^2}$) of $sp^2$ hybrid bond for T-carbon NWs. Stress-strain curves for T-carbon NWs along (c) [110] and (d) [100] directions. Copyright (2018) Elsevier. Reprinted (adapted) with permission from ref~\cite{Bai2018}.}\label{fig5-1}
\end{figure}

Bai $et$ $al.$ studied the structures and mechanical properties of T-carbon NWs under tensile strains in three different directions [100], [110] and [111] at the temperature of 300 K by molecular dynamics simulation~\cite{Bai2018}. The results show that T-carbon exhibits excellent ductility with high failure strain and mechanical anisotropy in Fig.~\ref{fig5-1}. It is noted that the ratio of $sp^2$ hybrid bond to all chemical bonds ($f_{sp^2}$) grows rapidly after a certain strain and can reach up to 80$\%$ with the increase of strain in [111] and [110] directions as shown in Fig.~\ref{fig5-1}(a), indicating the appearance of a global graphitization, while $f_{sp^2}$ slightly increases until the failure of T-carbon NWs with the increase of strain in [100] direction. The variation of average atomic potential energy ($\phi$) under tensile strains further reflects the evolution of global graphitization [Fig.~\ref{fig5-1}(b)]. The smaller $\phi$ is, the more stable the system is. With the increase of strain in [110] and [111] directions, the $\phi$ decreases first due to the global graphitization, indicating that the carbon networks become stable after the global graphitization, and then dramatically increases, corresponding to the deformation of the $sp^2$ network. Figure~\ref{fig5-1}(c) shows that the slope of stress-strain ($\sigma-\varepsilon$) in the [110] direction is large before and after graphitization (stage I and stage III), while during the graphitization, the stress increases slowly and fluctuates (stage II). The fluctuation of stress during the graphitization enables T-carbon NWs to withstand a larger strain. However, stress rapidly increases with the increase of strain in the [100] direction, indicating that no global graphitization occurs in this case.

\begin{figure}[!htbp]
  \centering
  \includegraphics[scale=0.7,angle=0]{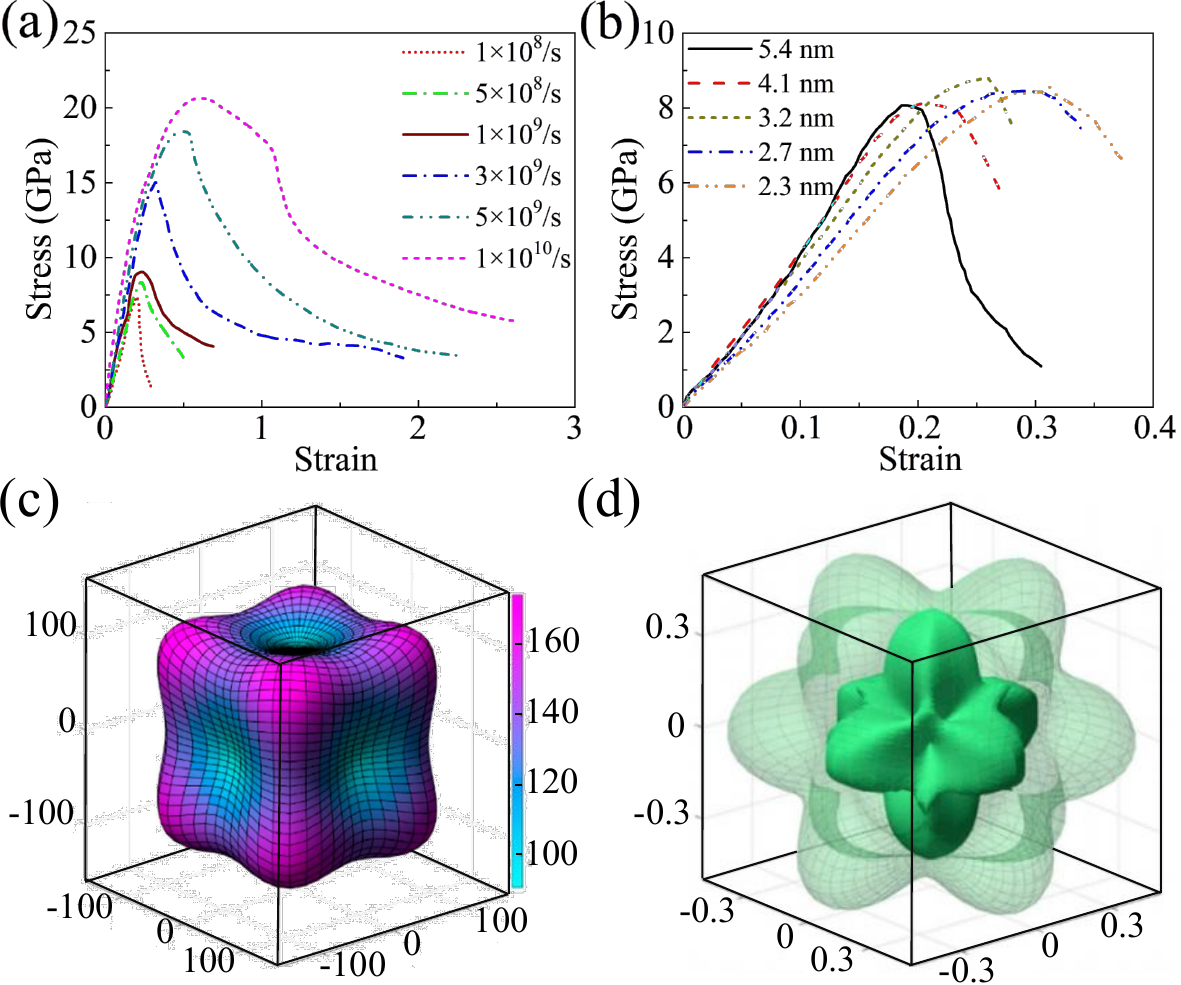}\\
  \caption{(a) The tensile stress-strain curves for nanocrystalline T-carbon with an average grain size of 5.4 nm under different strain rates. (b) The stress-strain curves of nanocrystalline T-carbon with different grain sizes at a strain rate of 1$\times$10$^8/s$. Three-dimensional contour plots of (c) Young's modulus and (d) Poisson's ratio for T-carbon. Copyright (2019) Elsevier. Reprinted (adapted) with permission from ref~\cite{Wang2019}. Copyright (2019) Elsevier. Reprinted (a) and (b) with permission from ref~\cite{Wang2019}. Reprinted (c) and (d) with permission from ref~\cite{Cheng2020}.}\label{fig5-2}
\end{figure}

The process of global graphitization can be also viewed from the microstructural evolution of T-carbon. With the increase of strain in [110] and [111] directions, $sp^2$ clusters appear at the boundary and then propagate to the entire NWs, leading to the the evolution of NWs from $sp^3$ to $sp^2$ network. The change in the number of chemical bonds during the graphitization can be implemented to analyze the fracture of chemical bonds. It was found that when applying strain to the above three directions, the number of chemical bonds between the C$_4$ units was remained, while the total number of chemical bonds in NWs is slightly increased, which shows that the internal chemical bonds within the C$_4$ units were broken during the global graphitization. This graphitization process in T-carbon is different from previously reported carbon materials, because the $sp^2$ clusters in other carbon materials like diamond~\cite{Guo2011}, diamond-like carbon~\cite{Bai2016} tend to be localized and do not spread to the entire network, resulting in brittle fracture. Moreover, for diamond and diamond-like carbon, their failure stress is larger than that of T-carbon NWs, because the $sp^2$ dominated network cannot withstand much high stress and with a few strains the $sp^2$ clusters of T-carbon NWs will be cracked and fractured. The unique global graphitization process for T-carbon makes it the excellent ductility, whereas the failure stress of T-carbon is much smaller than that of diamond and diamond-like carbon, because T-carbon has a lower density and does not have the dense structure to withstand great stress.

The above study was based on the single-crystal T-carbon NWs. The nanocrystalline T-carbon samples composed of a large number of single-crystal grains with random directions, sizes and shapes were also investigated~\cite{Wang2019}, and the relation between stress and strain under different strain rates is shown in Fig.~\ref{fig5-2}(a), from which the Young's modulus, failure stress and failure strain can be obtained. The authors found that with the increase of strain rate, the failure stress, failure strain and Young's modulus of nanocrystalline T-carbon are increased. Compared with nanocrystalline diamond, the failure strain of nanocrystalline T-carbon was much higher, while the failure stress and Young's modulus were lower than that of diamond. These results are similar to that of single-crystalline T-carbon NWs we mentioned above~\cite{Bai2018}. Compared to nanocrystalline materials including the carbon materials and metal materials, the strain rate sensitivity (defined as $\partial \sigma/\partial\dot{\varepsilon}$) of T-carbon NWs are much higher.

The stress-strain relation for different grain sizes at a fixed strain rate was also investigated as shown in Fig.~\ref{fig5-2}(b). It was shown that with the increase of grain size, the failure strain is decreased, while the failure stress and Young's modulus have no obvious changes. 

The atoms inside the grains mainly contain $sp^3$ bonds, while the boundary of grains consists of the atoms with $s$, $sp$ and $sp^2$ bonding states, corresponding to the carbon atoms connected to one, two and three adjacent atoms within the 0.2 nm radius, respectively. With the increase of tensile strain, the number of $s$, $sp$, $sp^2$ bonds increases, corresponding to the increase of the boundary thickness. Because of more disordered bonds at the grain boundaries and the relatively concentrated stress at their junction interface, the cracks always initiate and propagate from the grain boundaries, resulting in the brittle fracture of the nanocrystalline T-carbon.

The mechanical properties including Young's modulus, shear modulus and Poisson's ratio were also investigated in Ref.~\cite{Cheng2020}. 3D diagrams of Young's modulus and Poisson's ratio for T-carbon are shown in Figs.~\ref{fig5-2}(c) and (d). The Young's modulus and Poisson's ratio show the elastic anisotropy of T-carbon, which is higher than that of diamond, but lower than that of Y-carbon and TY-carbon, because of their different arrangements of atoms or types of chemical bonds. The results of the ratio of the maximum and minimum of Young's modulus reveal that the elastic anisotropy of the above carbon allotropes in each plane is arranged in the following order: (111) $<$ (001) = (010) = (100) $<$ (011) = (110) = (101) plane, because they have similar cubic symmetry and their respective three lattice constants are the same.

T-carbon was calculated to exhibit a Vickers hardness of 61GPa by Sheng $et$ $al.$~\cite{shengxianlei2011}. However, one other study~\cite{Chen2011} reported that the hardness of T-carbon is less than 10 GPa, and the reasons are listed in the following:
(1) The empirical formula for calculating the hardness of T-carbon used by Sheng $et$ $al.$ proposed by Gao $et$ $al.$~\cite{Gao2003} as well as the commonly used Simunek and Vackar’s (SV) method~\cite{Simunek2006} is suitable for the materials with uniform bond distribution. As we mentioned above that the $sp^3$ bond in T-carbon is anisotropic, the Vickers hardness obtained by the above two methods may be not accurate.
(2) Considering the high density of C-C covalent bonds in carbon tetrahedrons of T-carbon, the C$_4$ unit can be regarded as a whole, called as an artificial super-atom. For an artificial super-atom, the four $sp^3$ chemical bonds are uniform. By applying this concept to the above two methods, the hardness of T-carbon was estimated to be 8.2 and 7.7 GPa, respectively;
(3) The formula of ideal elastic hardness given by Sneddon~\cite{Sneddon1965} was used to estimate the approximate range of hardness, which is defined as $H_{id}=\dfrac{E_{cot\phi}}{2(1-\nu^2)}$, where $E$ is Young's modulus, $\nu$ is Poisson's ratio, and cot$\phi \approx$ 0.5 for pyramid indentation and the real hardness would be (0.01-0.2)$H_{id}$ at high strains. By substituting the Young's modulus value $E$ = 185 GPa provided by Sheng $et$ $al.$ into Sneddon's formula, the hardness value of T-carbon was estimated in the range of 0.5 to 10 GPa;
(4) J. S. Tse showed that the shear stress of 3 GPa can be used as the upper limit of hardness of T-carbon~\cite{Tse2010}; 
(5) When the Pugh modulus ratio is greater than 0.571, the material is generally considered to be brittle and hard~\cite{Pugh1954}. The Pugh modulus ratio of T-carbon is only 0.414, indicating that it has strong ductility and low hardness; 
(6) The hardness of T-carbon was calculated to be 5.6 GPa by Chen $et$ $al.$ with their own model~\cite{Chen2011a}. This model was also used to estimate the hardness of super-hard materials, such as diamond, and the obtained results were consistent with those obtained by empirical formula and SV model. As the calculations on hardness of a material depend strongly on the designed model, T-carbon provides a great opportunity to test various models on hardness because of its exotic chemical bondings. Therefore, how to get a reasonable hardness of T-carbon is an intriguing issue that needs more explorations. 

\subsection{Thermoelectric properties}
The thermoelectric effect is a phenomenon that can directly convert a temperature difference to an electric voltage or vice versa. A thermoelectric device creates a voltage when a different temperature is distributed on each side, whereas when a voltage is applied to it, a temperature difference will be created on the two ends. The high thermoelectric figure of merit (ZT) means the low-cost conversion, which plays an essential role in the reuse of waste heat. 

\begin{figure}[!htbp]
  \centering
  \includegraphics[scale=0.17,angle=0]{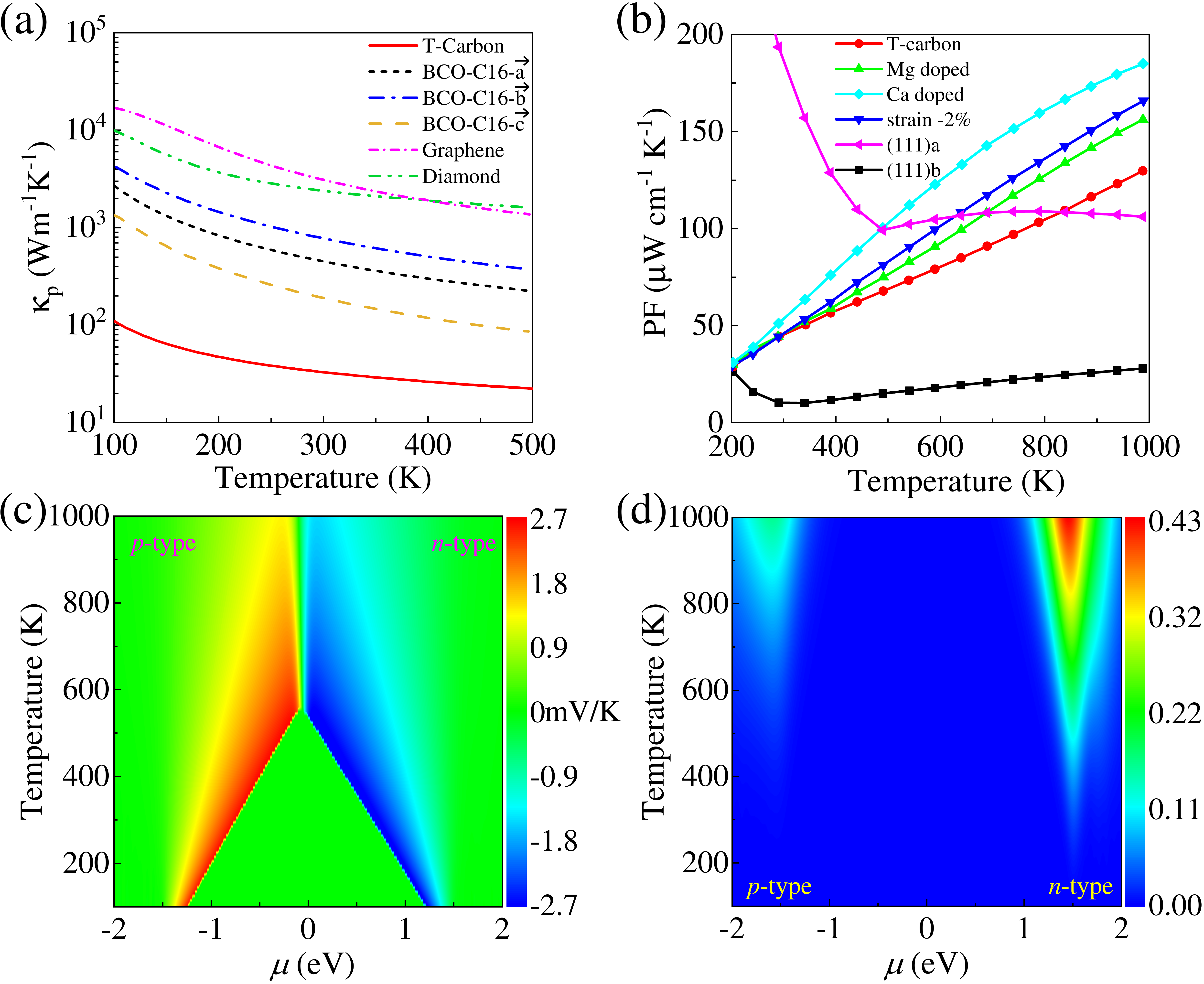}\\
  \caption{ (a) Temperature dependent lattice thermal conductivity for several carbon allotropes: T-carbon, BCO-C$_{16}$ ($\vec{a}$, $\vec{b}$, $\vec{c}$), graphene and diamond. (b) The power factor (PF) of calcium (Ca) or magnesium (Mg) doped T-carbon, T-carbon with compressive strain, and T-carbon cut into two-dimensional structures along the (111) direction. (c) Seebeck coefficient (thermopower) and (d) the figure of merit $ZT$ in contour plots in the plane of chemical potential ($\mu$) and temperature. Copyright (2017) American Physical Society. Reprinted (a) with permission from ref~\cite{Yue2017}. Copyright (2019) Royal Society of Chemistry. Reprinted (b), (c) and (d) with permission from ref~\cite{Qin2019}.}\label{fig6-1}
\end{figure}

Yue $et$ $al.$ studied the heat transport properties of T-carbon~\cite{Yue2017} and the lattice thermal conductivity $\kappa_p$ as a function of temperature was plotted in Fig.~\ref{fig6-1}(a). At room temperature (300 K), the lattice thermal conductivity of T-carbon is 33.06 W/mK, which is much smaller than that of graphite, diamond, and BCO-C$_{16}$. The low thermal conductivity of T-carbon comes from the strong anharmonicity of phonons in T-carbon.

Thermoelectric efficiency (ZT) defined as $ZT=S^2\sigma T/\kappa$, where $S$, $\sigma$, $\kappa$, T represent Seebeck coefficient, electrical conductivity, total thermal conductivity, and absolute temperature, respectively, is inversely proportional to $\kappa$. The low lattice thermal conductivity of T-carbon shows that it is likely to be a good thermoelectric material. Figure~\ref{fig6-1}(c) shows that the Seebeck coefficient of T-carbon can be as high as 2000 $\mu$V/K~\cite{Qin2019}, which is higher than that of many superior thermoelectric materials, such as SnSe (550 $\mu$eV/K)~\cite{Zhao2016}. The maximum ZT value of T-carbon is around 0.43 as shown in Fig.~\ref{fig6-1}(d), which is much smaller than that of SnSe with the maximum ZT of 2.6, because the former possesses lower electrical conductivity and higher thermal conductivity.

The conductivity of T-carbon can be improved by applying strain~\cite{Qin2014}, doping elements~\cite{Gharsallah2016}, or cutting it into low-dimensional structures~\cite{Qin2016a}. Figure~\ref{fig6-1}(b) shows the power factor defined as PF= $S^2\sigma$ as a function of temperature for T-carbon doped with Ca and Mg elements, applied with compressive strain, or cut into two-dimensional (2D) structures along the (111) direction. On one hand, doping elements can effectively improve the PF of T-carbon; on the other hand, after doping the thermal conductivity of T-carbon will decrease, leading to the significant enhancement of ZT of T-carbon. The PF of T-carbon can also be improved by applying a few compressive strain. In addition, by cutting T-carbon into a 2D structure along the (111) direction, PF can be greatly improved at low temperatures, which means that T-carbon can be transformed from high-temperature to low-temperature thermoelectric materials. In general, low thermal conductivity and high Seebeck coefficient make T-carbon a potential thermoelectric material. In addition, its thermoelectric properties can be further improved by improving the conductivity through various ways. Thermoelectric devices based on T-carbon may enable solid-state cooling to replace vapor-compression cycle technologies and convert energy from waste heat sources under ambient conditions.

\subsection{Excitonic effect}
Electron-phonon (e-ph) interactions play an important role in carriers lifetime and transport properties of materials~\cite{Bardeen1955,Minnigerode1983}. The absorption and recombination of excitons directly influence the light absorption and luminescence of semiconductors. The excitonic effect and the relative energy position of the bright and dark excitons have an important effect on the photo-physical properties for semiconductors~\cite{Kilina2009}. 

\begin{figure}[!htbp]
  \centering
  \includegraphics[scale=0.43,angle=0]{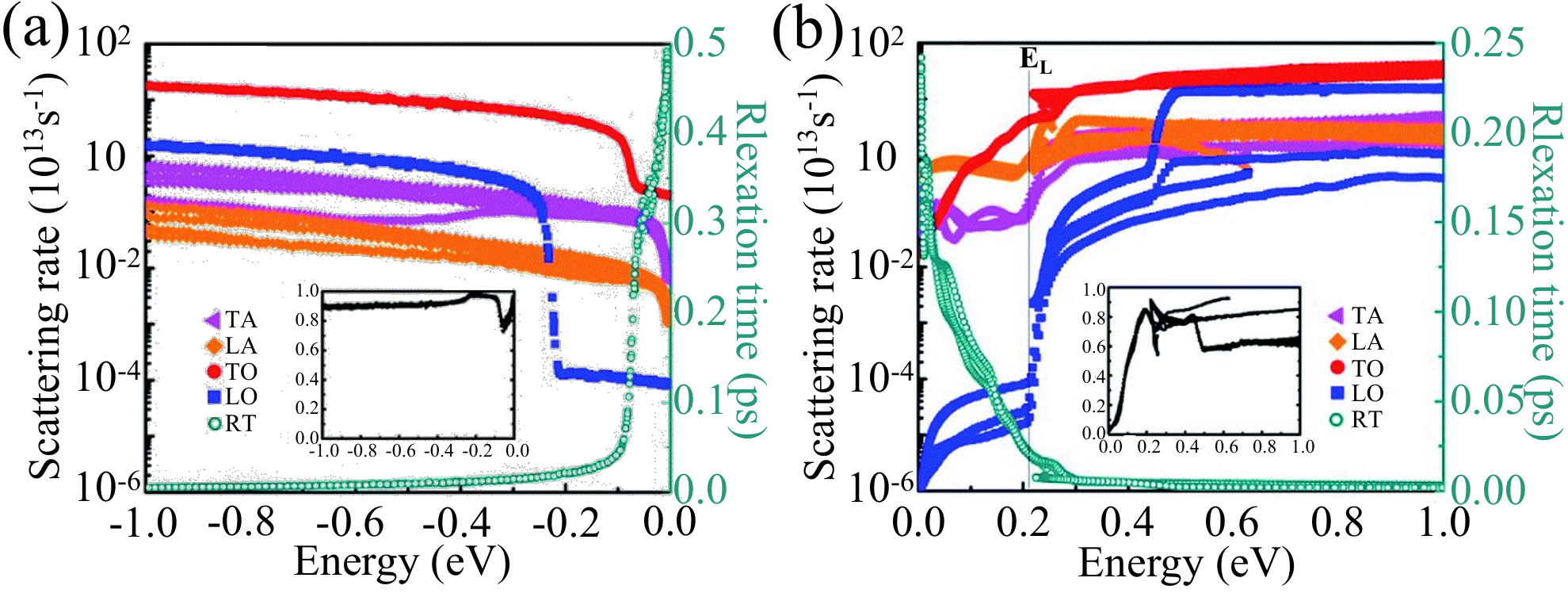}\\
  \caption{Mode-resolved scattering rates and total relaxation time of (a) holes and (b) electrons, where the relative contribution from the TO branch is inserted. Reprinted (adapted) with permission from ref~\cite{Bu2020}.}\label{fig7-1}
\end{figure}

The scattering rate and relaxation time (RT) of different phonon modes, including lateral acoustics (TA), longitudinal acoustics (LA), lateral optics (TO), and longitudinal optics (LO) phonon modes, are shown in Fig.~\ref{fig7-1}(a)~\cite{Bu2020}. In general, the contribution to scattering from optical phonons (mainly TO mode) is much larger than that from acoustic phonons in a wide energy range. It can be found that the acoustic phonons dominate the scattering at the CBM, while the optical phonons dominate the scattering at the VBM. However, the electron RT ($\sim$ 0.5 ps) from optical phonons scattering at the VBM is higher than that ($\sim$ 0.25 ps) from acoustic phonons scattering at the CBM, indicating that TO mode phonons make the most contributions to electroacoustic scattering. Figure~\ref{fig7-1}(b) shows an energy band valley of about 0.22 eV at point L (E$_L$) above the CBM. When the energy is higher than the valley, the inter-valley scattering occurs between $\Gamma$ valley at CBM and L valley, and the scattering rate is greatly enhanced. The mean free path (MFP) of hot carriers is directly related to the RT, because MFP is the product of group velocity and RT. Since the group velocity of holes is greater than that of electrons, and the RT of holes is about twice than that of electron, the free path of hot holes (up to 80 nm) is much larger than that of hot electrons (only 15 nm).

\begin{figure}[!htbp]
  \centering
  \includegraphics[scale=0.37,angle=0]{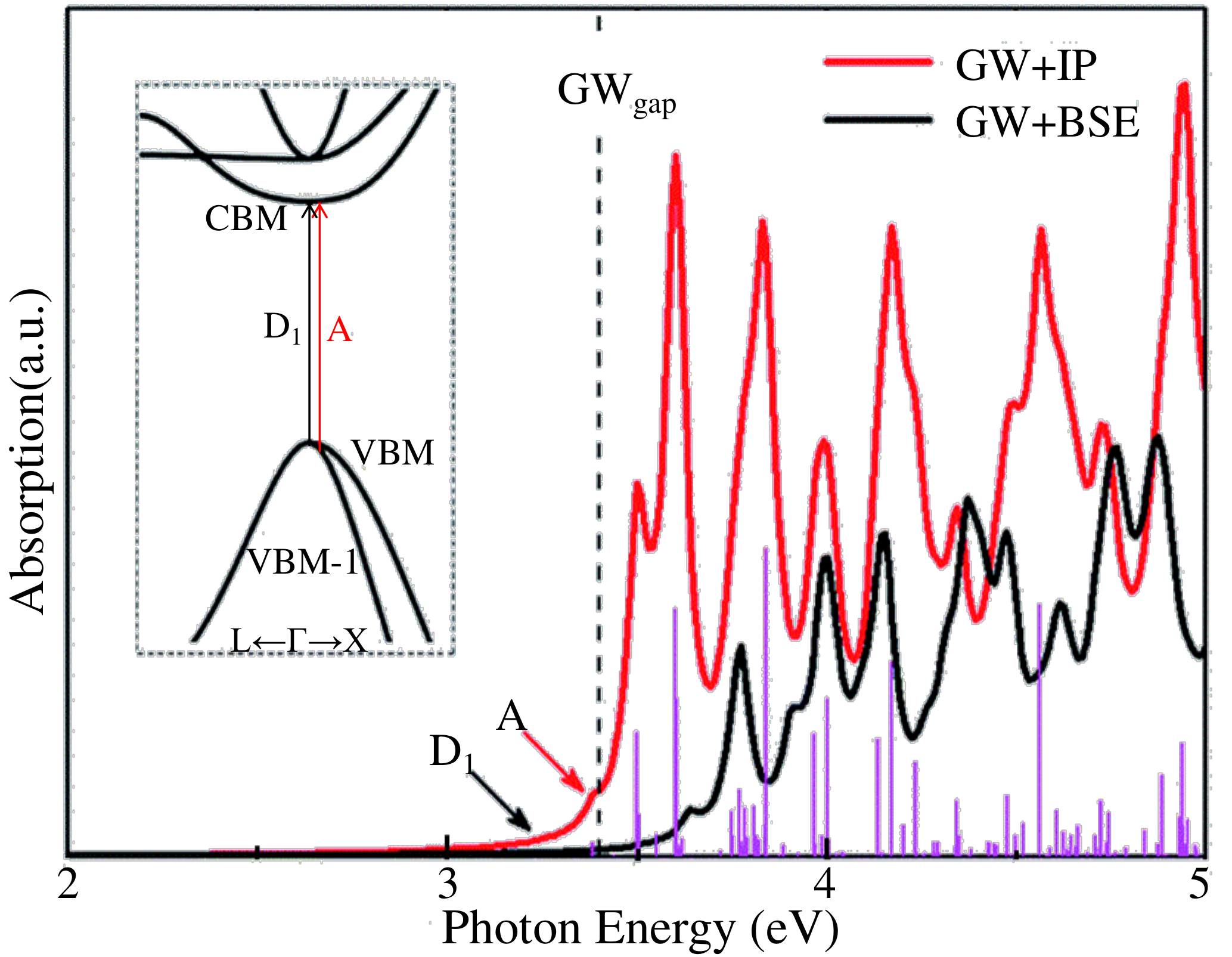}\\
  \caption{The optical absorption of T-carbon with inclusion of excitonic effect based on GW plus independent-particle (IP)          approximation and Bethe-Salpeter equation (BSE), respectively. The first dark and the first bright exciton are denoted as D$_1$ and A, respectively, and their corresponding absorption process was inserted. Reprinted (adapted) with permission from ref~\cite{Bu2020}.}\label{fig7-2}
\end{figure}

The electron-phonon scattering mainly affects the lifetime of hot carriers, and the electron-hole interactions mainly influence the lifetime of excitons. The optical absorption including the excitonic effect is shown in Fig.~\ref{fig7-2}. The position marked by letter A corresponds to the lowest photon energy of the bright exciton ($\sim$ 3.39 eV), indicating a transfer from the first valence band to the first conduction band near $\Gamma$ point. There are four more dark excitons with the first one below the bright exciton locating at 3.23 eV denoted by D$_1$, which originates from the transfer between VBM/VBM-1 and CBM. The radiative lifetimes of the first bright exciton and the first dark exciton were calculated to be 2.8 ns and 3.4 ns, respectively. The longer exciton lifetime of dark excitons will trap more excitons population, leading to the low photoluminescence quantum yield of T-carbon. It is an important research topic and highly expected to improve the photoluminescence quantum yield of T-carbon.

\subsection{Optical properties}
By studying optical absorption characteristics of materials, various information such as the electronic band structure, impurity defects, etc. can be directly obtained. The optical properties of T-carbon were extensively studied both in theories and experiments~\cite{Zhang2017,Alborznia2019,Xu2020,Cheng2020}. 

\begin{figure}[!htbp]
  \centering
  \includegraphics[scale=0.26,angle=0]{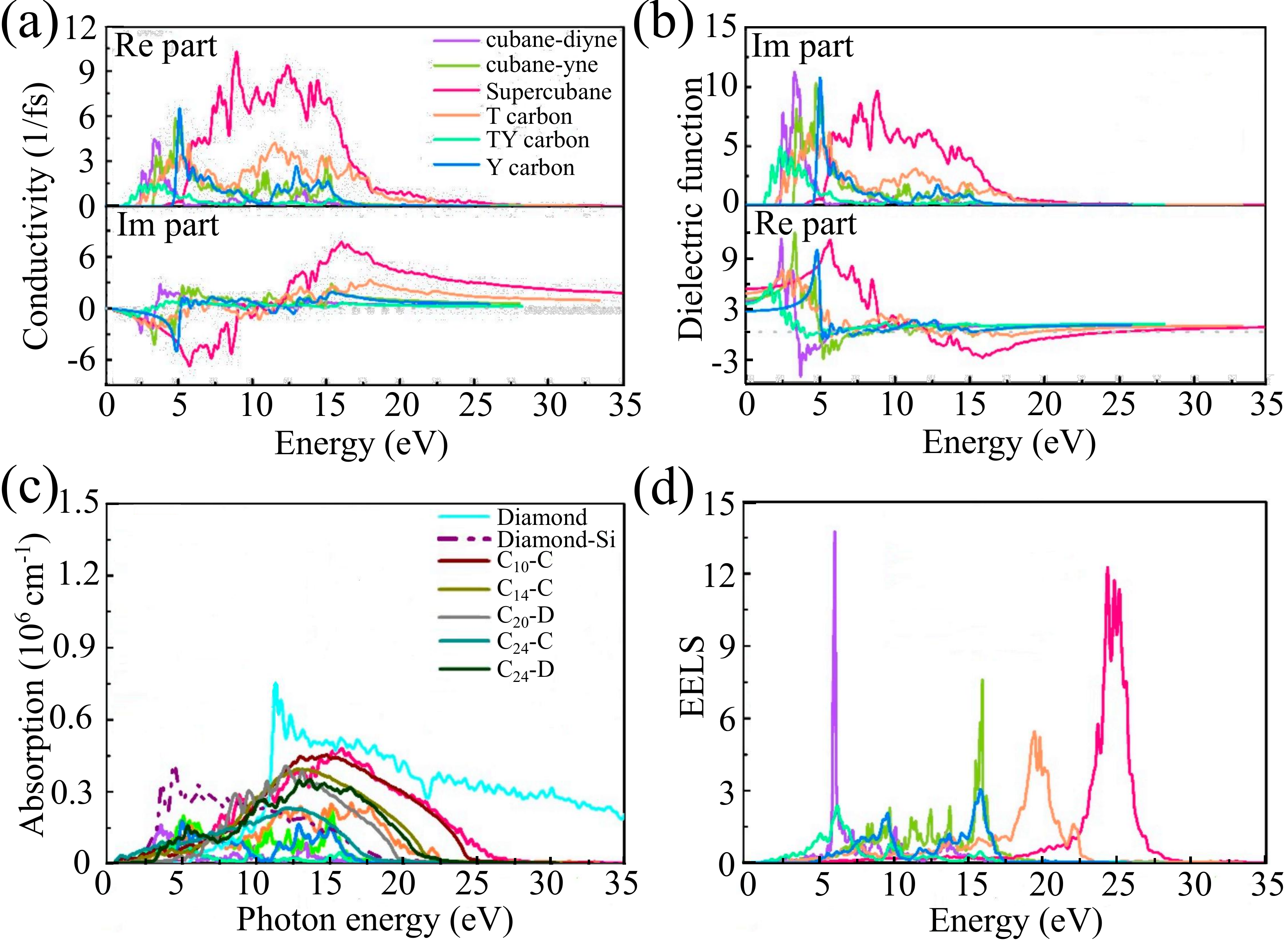}\\
  \caption{Calculated optical properties for several carbon allotropes with similar structures: (a) Conductivity, (b) dielectric function, (c) absorption spectra, and (d) EELS. Reprinted (adapted) with permission from ref~\cite{Cheng2020}.}\label{fig8-1}
\end{figure}

The optical properties of several carbon allotropes are shown in Fig.~\ref{fig8-1}. For T-carbon, the real part of the conductivity becomes positive when the photon frequency is higher than 1.780 1/fs, and T-carbon has the widest photon energy range among these carbon materials as shown in Fig.~\ref{fig8-1}(a). Fig.~\ref{fig8-1}(b) shows the static dielectric constant of T-carbon is about 4.590. The results of absorption coefficient in Fig.~\ref{fig8-1}(c) show that only T-carbon, Cubane-yane and TY carbon can absorb visible light, and T-carbon can absorb the light with photo energy above 1.8 eV. The absorption coefficients of T-carbon in the visible light range are comparable to that of GaAs with the highest absorption coefficient currently used, exhibiting the potential applications of T-carbon in optoelectronic materials. The electron energy loss spectrum (EELS) describes the energy loss of electrons in the process of rapidly passing through the material, which becomes larger near 20 eV as shown in Fig.~\ref{fig8-1}(d).

\begin{figure}[!htbp]
  \centering
  \includegraphics[scale=0.30,angle=0]{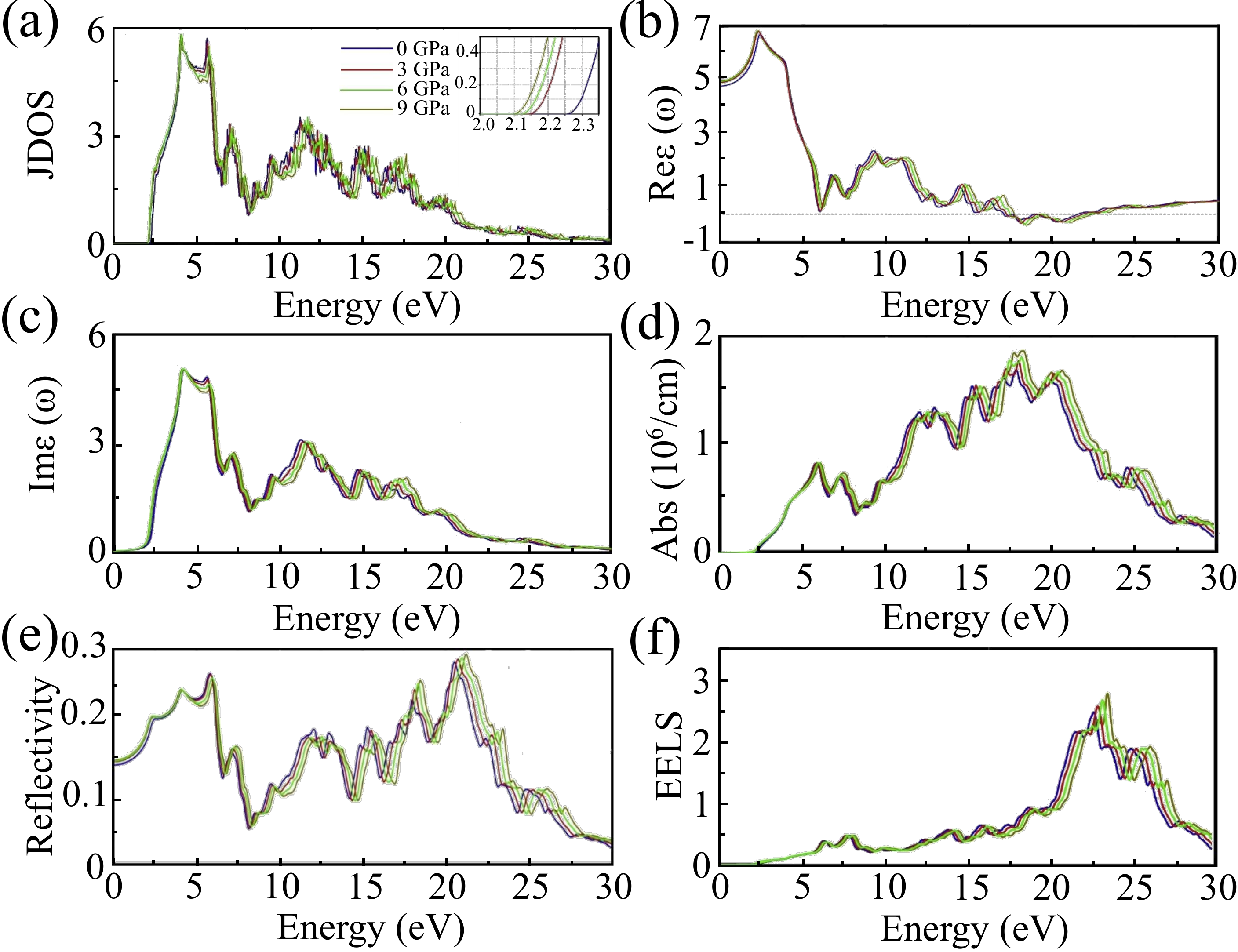}\\
  \caption{(a) Joint density of states (JDOS) with partial enlarged views at specific energies in the inset, (b) real part and (c) imaginaty part of the dielectric function, (d) absorption and (e) reflectivity spectra, and (f) EELS of T-carbon in $z$ polarization direction under different pressure. Copyright (2019) Elsevier. Reprinted (adapted) with permission from ref~\cite{Alborznia2019}.}\label{fig8-2}
\end{figure}

The optical properties of T-carbon under different hydrostatic pressures have also been studied~\cite{Alborznia2019}. When the T-carbon is subjected to the hydrostatic pressure of 0, 3, 6, and 9 GPa, the change of the joint density of states (JDOS), the complex dielectric function, light absorption spectra, reflection spectra, and ELSS in the incident light energy at the range of 0-30 eV in $z$ polarization direction are presented in Fig.~\ref{fig8-2}. One may observe that with the increase of static pressure, the spectra are slightly shifted to the region with higher intensity of the incident light (blue shift). In addition, the spectra in the $x$-axis and $z$-axis polarization direction are consistent due to cubic crystal structure of T-carbon with high symmetries. From the JDOS in the inset of Fig.~\ref{fig8-2}(a) the band gap of T-carbon can be obtained, and as the pressure increases, the band gap of T-carbon gradually decreases as shown in the inset of Fig.~\ref{fig8-2}(a). 

\begin{figure}[!htbp]
  \centering
  \includegraphics[scale=0.19,angle=0]{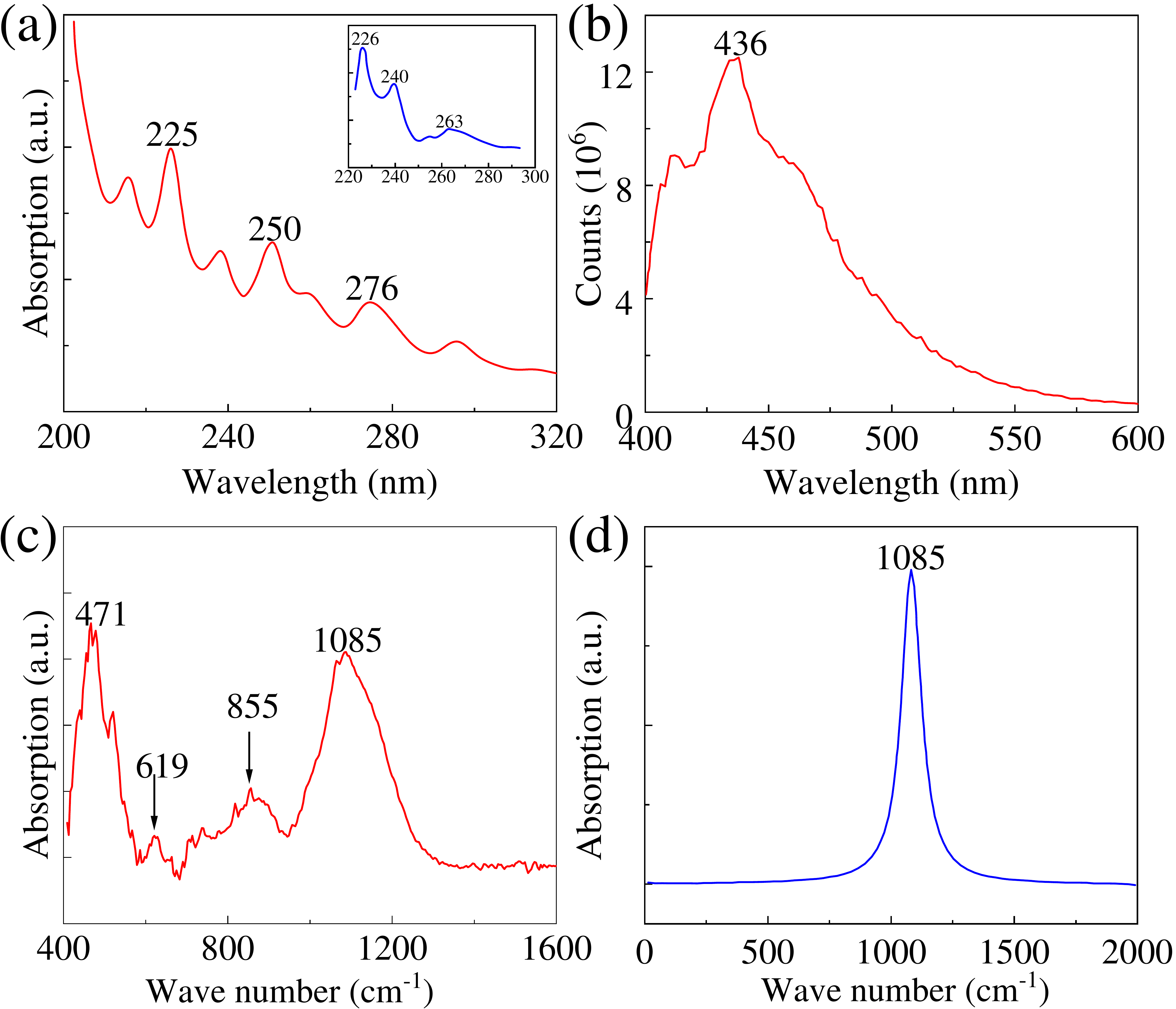}\\
  \caption{(a) UV–visiable absorption of T-carbon NWs with the simulated absorption spectrum inserted. (b) The photoluminescence spectrum. (a) and (b) Reprinted (adapted) with permission from ref~\cite{Zhang2017}. (c) FT-IR spectra of the sample grown on the substrate of single crystalline diamond. (d) The simulated FT-IR spectra of T-carbon. Copyright (2020) Elsevier. (c) and (d) Reprinted (adapted) with permission from ref~\cite{Xu2020}.}\label{fig8-3}
\end{figure}

Figure~\ref{fig8-3}(a) shows the experimental data of UV absorption spectrum for T-carbon NWs. There are three obvious absorption peaks at 225, 250, and 276 nm, which are comparable with the calculated values of 226, 240, and 263 nm for T-carbon as shown in the inset of Fig.~\ref{fig8-3}(a). Figure~\ref{fig8-3}(b) shows the photoluminescence spectrum of T-carbon NWs. The quantum yield defined as the ratio of the number of emitted photons to the absorbed photons at the wavelength of 436 nm is 5.4$\pm$0.2$\%$. Figure~\ref{fig8-3}(c) shows the experimental data of FT-IR spectra for the sample synthesized by PECVD on a single crystal diamond substrate, which shows a consistent peak with the calculated result at 1085 cm$^{-1}$ as shown in Fig.~\ref{fig8-3}(d). 

\subsection{Topological phonons}
So far, the topological band structure is mainly discussed in electronic systems~\cite{Bansil2016,Chiu2016,You2020a,Wan2011,You2019b,Yan2017,You2019c,You2021b}. Recently, topological phonons have also attracted much attention~\cite{Zhang2010,Li2012,Liu2017,Jin2018,Liu2019,Wang2020}, because of their potential implications in electron-phonon coupling, dynamic instability~\cite{Prodan2009}, and phonon diodes~\cite{Liu2017}. The spin-orbit coupling (SOC) of carbon atom is very small, and carbon materials are well known as the candidates of topological semimetal (metal)~\cite{Chen2020}, which may have a family of ideal candidates of topological phonons. However, it is not easy to find a realistic carbon material featuring topological phonons, because it requires the topological surface states of phonons that should be well separated from the bulk phonon spectrum. 

\begin{figure}[!htbp]
  \centering
  \includegraphics[scale=0.45,angle=0]{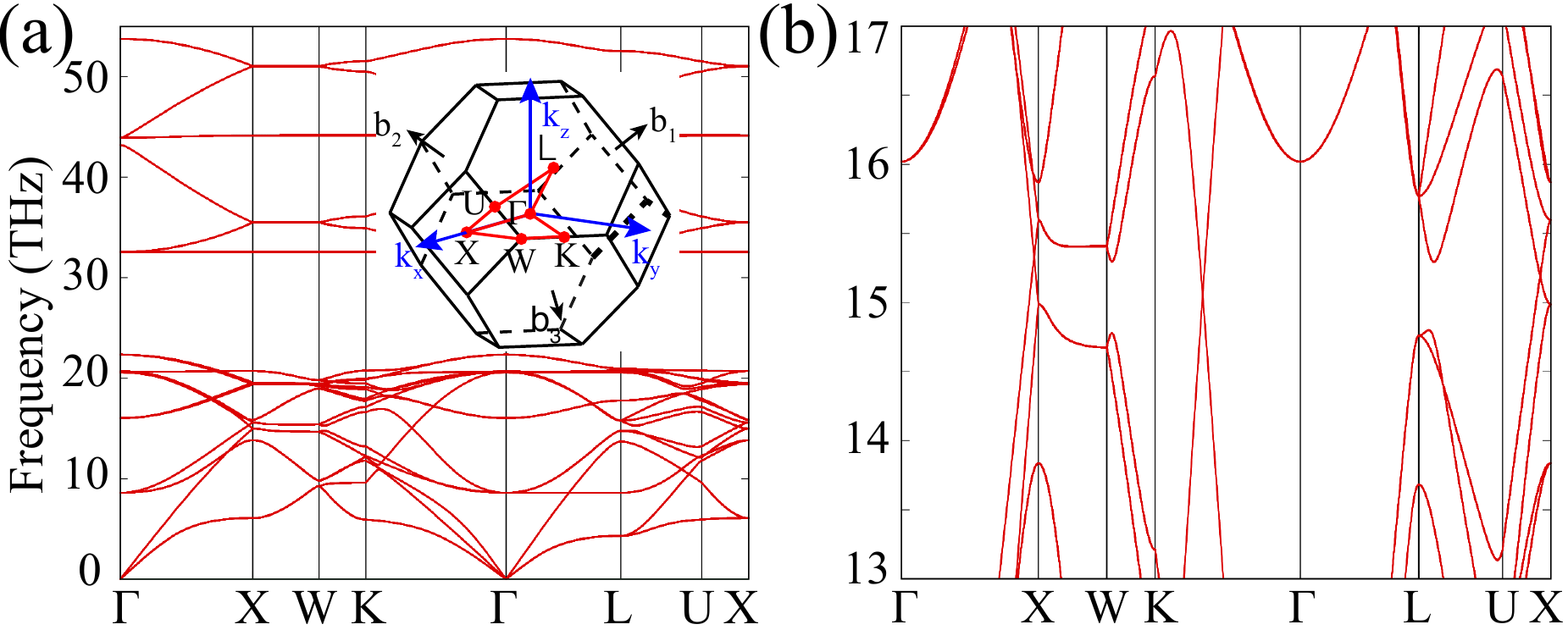}\\
  \caption{(a) The phonon spectra of T-carbon along high-symmetry paths, where the BZ of the primitive cell of T-carbon with high symmetry points and paths is inserted. (b) The enlarged views of phonon spectra of T-carbon in the range of 13 to 17 THz. Copyright (2021) American Physical Society. Reprinted (adapted) with permission from ref~\cite{You2021}.}\label{fig9-1}
\end{figure}

\begin{figure}[!htbp]
  \centering
  \includegraphics[scale=0.3,angle=0]{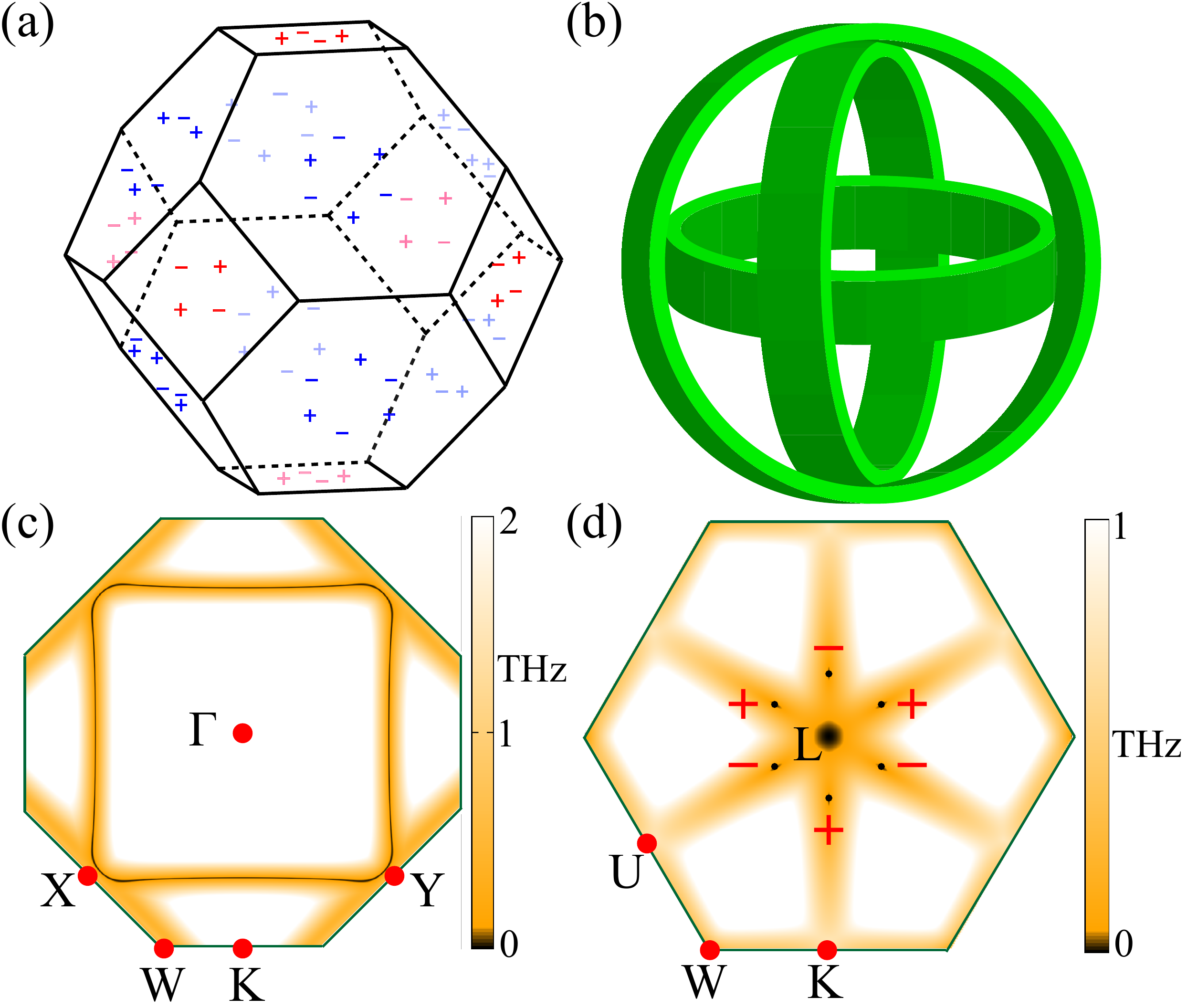}\\
  \caption{(a) The distribution of Weyl points at the boundaries of BZ in T-carbon, i.e, in $k_{x,y,z}=2\pi/a$ and $(\pm1,\pm1,\pm1)$ planes, where the red- and blue-colored signs represent the Weyl point at the square and hexagonal surfaces, respectively, and the "+" and "-" indicate the chirality of Weyl points. (b) Schematic depiction of the gimbal. (c) Shape of the hourglass Weyl loop (black-colored loop) in $k_z=0$ plane obtained from the DFT calculations. The color map corresponds to the frequency difference between the two crossing phonon bands. (d) Distribution of type-II Weyl phonons at about 14.5 THz in (111) plane obtained from the DFT calculations. Copyright (2021) American Physical Society. Reprinted (adapted) with permission from ref~\cite{You2021}.}\label{fig9-2}
\end{figure}

T-carbon as a realistic carbon material has a high-symmetry crystal structure with the same space group as diamond, which exhibits exotic topological phonons~\cite{You2021}. The phonon spectrum of T-carbon and its enlarge view in the range of 13 to 17 THz were shown in Fig.~\ref{fig9-1}. It is clear that there are two flatbands above a large phonon band gap of about 10 THz between 22 to 32 THz, which are related to the specific geometry of T-carbon~\cite{You2019a}. Since high frequency phonons make little contribution to the electron-phonon coupling and thermal transport~\cite{You2020}, we only study the relative low-frequency phonons below 20 THz. From Fig.~\ref{fig9-1}(b), we find that two phonon bands from optical branch cross linearly at about 15 THz on $\Gamma-X$, $K-\Gamma$ and $U-X$ paths, and at about 14.5 THz on $L-U$ path. After a careful study within $k_z=0$ plane, we find that the crossing points on $\Gamma-X$ and $\Gamma-K$ paths are not isolated, but form a nodal loop in $k_z=0$ plane. Besides these band crossing points, there is double degeneracy appearing on $X-W$ path, which is protected by the combination of glide mirrors $\tilde{M_x}$ and $\tilde{M_z}$ because of their anticommutating relation on $X-W$ path. Considering the crystal symmetry of T-carbon, we can obtain three intersecting nodal loops (named as nodal gimbal phonons) around $\Gamma$ point and 6 pairs of type-I Weyl phonons around the center of squares at the boundary of BZ at about 15 THz, and 12 pairs of type-II Weyl phonons around the center of hexagons at the boundary of BZ at about 14.5 THz as shown in Fig.~\ref{fig9-2}. By $k\cdot p$ perturbative analysis, the above crossing points (loops) were demonstrated to be protected by the symmetries of T-carbon~\cite{You2021}.

\begin{figure}[!htbp]
  \centering
  \includegraphics[scale=0.42,angle=0]{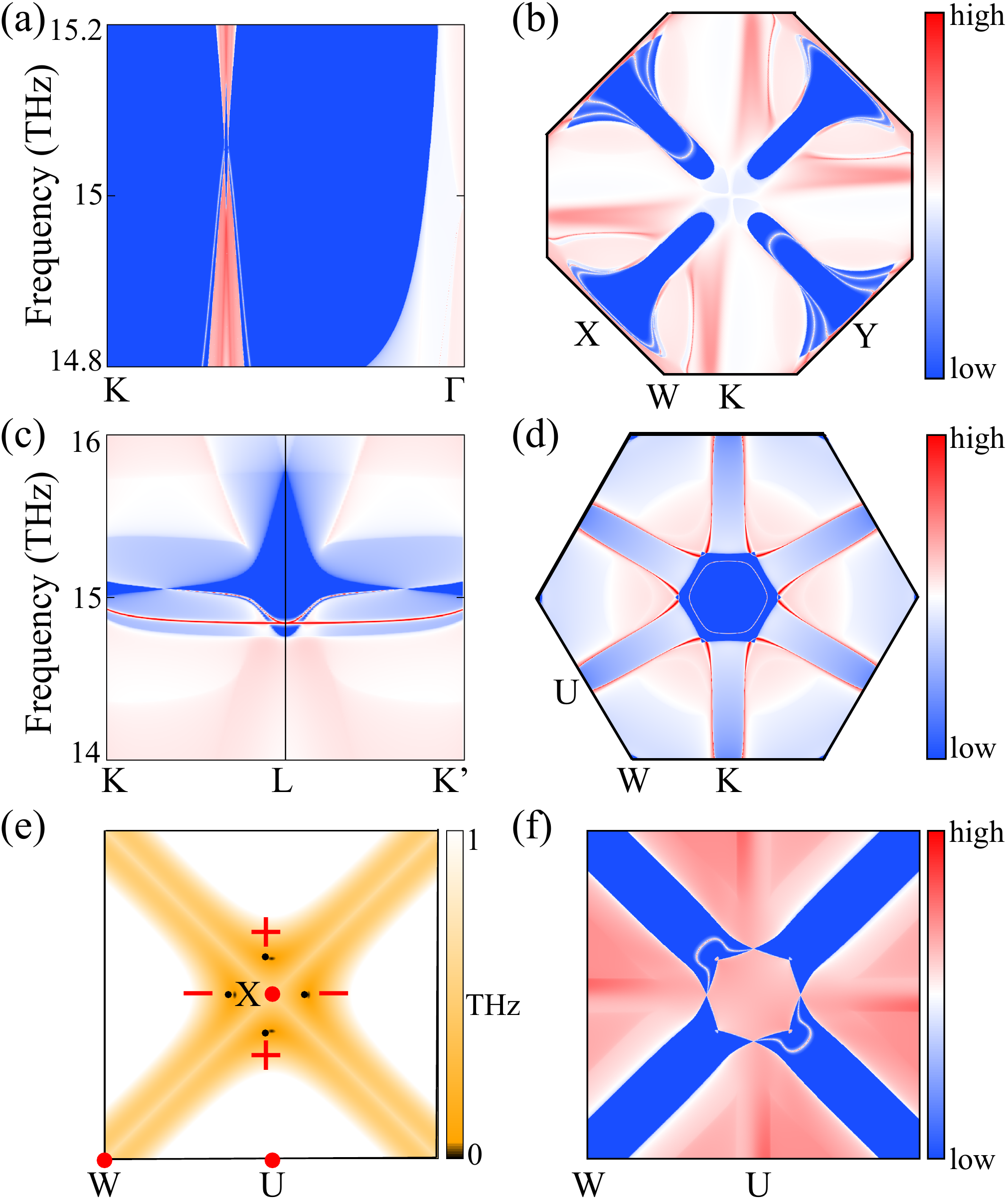}\\
  \caption{(a) Local density of states (LDOS) and (b) a constant energy slice at 15 THz projected on the semi-infinite (001) surface of T-carbon. (c) LDOS and (d) a constant energy slice at 15 THz projected on the semi-infinite (111) surface. (e) Distribution of Weyl phonons in $k_x=2\pi/a$ plane of T-carbon obtained from the DFT calculations, where the Weyl phonons with opposite chirality are marked as "+" and "-", respectively, and the redefined Cartesian coordinate system $(q_x, q_y, q_z)$ in $k_x=2\pi/a$ plane is also indicated. (f) The phonon surface arcs projected on the semi-infinite (100) surface of T-carbon at a constant frequency slice 15.2 THz. Copyright (2021) American Physical Society. Reprinted (adapted) with permission from ref~\cite{You2021}.}\label{fig9-3}
\end{figure}

The surface of a nodal line semimetal features the drumhead like states. From Figs.~\ref{fig9-3}(a) and \ref{fig9-3}(c), where we show the phonon surface states of T-carbon on (001) and (111) surfaces, respectively, one may observe the drumhead surface bands that emanate from the bulk nodal points, which connects the two nodal lines through the surface BZ boundary. In Figs.~\ref{fig9-3}(b) and \ref{fig9-3}(d), we plot the constant frequency slice at 15 THz, which cuts through the drumhead, forming a few arcs or circles, because the drumhead is not completely flat in frequency. The surface of a Weyl semimetal features the Fermi arcs connected two Weyl points with opposite chirality. From Figs.~\ref{fig9-3}(e) and \ref{fig9-3}(f), it is clear to see two Fermi arcs connected with two opposite Weyl points at the ends, which is the fingerprint of nontrivial Weyl phonons.

\section{Potential applications of T-carbon}

\begin{figure}[!htbp]
  \centering
  \includegraphics[scale=0.3,angle=0]{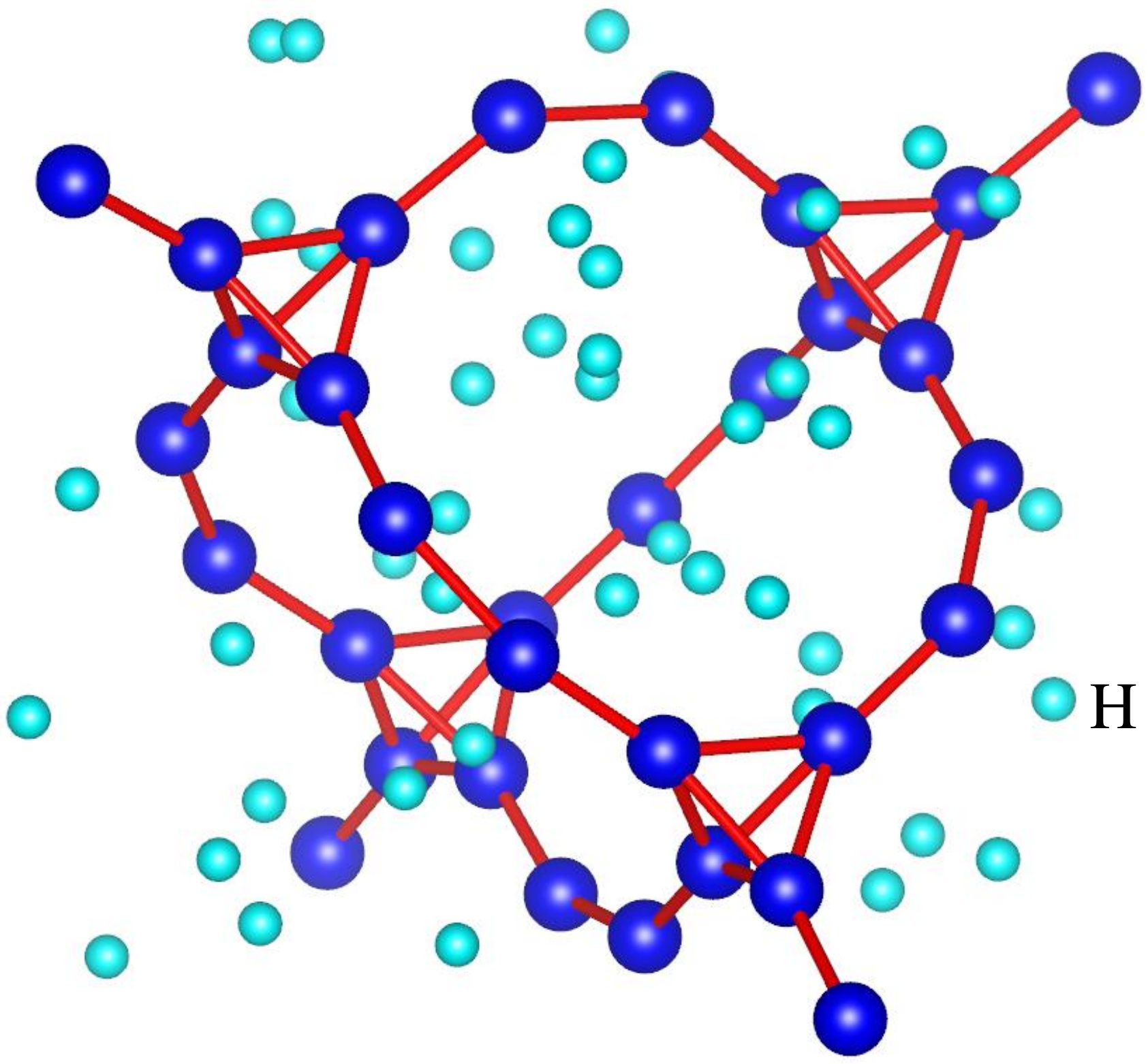}\\
  \caption{Hydrogen storage in T-carbon.}\label{fig10-1}
\end{figure}

Although T-carbon itself possesses many intriguing properties, as we discussed above, the wide energy gap of T-carbon might make it less practical application. Thus, it is important to tune the band gap of T-carbon. The most commonly used method in tuning the band gaps is to introduce electronic or hole states by doping that may optimize the performance of materials. Owing to the large interspaces between carbon atoms along certain directions, many kinds of atoms can be doped into T-carbon to form stable compounds, which makes T-carbon have important applications in hydrogen storage~\cite{shengxianlei2011}, ion battery~\cite{Qin2019}, solar cells~\cite{Sun2019}, photocatalysis~\cite{Ren2019,Tian2020}, magnetism and superconductivity~\cite{You2020}. These advances will be discussed in subsequent sections. Moreover, B and Al doped T-carbon are predicted to be a kind of stable structure with impurity bands and zero band gap, which promotes the conductivity of T-carbon. Meanwhile, the absorption, reflection and loss peaks of Al doped T-carbon in UV light are decreased, indicating its potential application in optoelectronic and microelectronic devices~\cite{Lu2021}. In addition, by substituting one C atom with one N atom in each C$_4$ unit in T-carbon, a new type of stable carbonitride C$_3$N can be obtained, which is a multi-layered structure, and its single-layer film has great potential application in water nanofiltration, providing a new solution for wastewater treatment and seawater desalination~\cite{Zhou2020}.

\subsection{Hydrogen storage}
Since the 1970s, hydrogen has been regarded as one of the most promising alternatives to fossil fuels, because the only by-product of its combustion is water, which has great advantages in reducing greenhouse gases and harmful pollutants. On the other hand, hydrogen is light in weight and can provide higher energy density, making that hydrogen powered engine is more efficient than internal combustion engines. The main reason that hinders the popularization of hydrogen energy is that it is challenging to obtain and store a large amount of hydrogen safely, densely, and quickly. For this reason, discovering the next generation of new materials for hydrogen storage is essential. As carbon materials have the characteristics of high specific surface area and light weight, searching for new porous carbon materials for hydrogen storage has always been a hot topic~\cite{Tour2010,Borchardt2017,Xu2017,Srinivasu2012}. 

Due to the fluffy structure, T-carbon has potential application in hydrogen storage~\cite{shengxianlei2011}. T-carbon has a low density, about two-third that of graphite and half that of diamond. According to the maximum number of hydrogen molecules adsorbed, the hydrogen storage value can be estimated. The hydrogen adsorption energy for T-carbon can be defined as $E_{adsorption}=(E_{T-carbon}+nE_{H_2}-E_{compound})/n$, where $E_{adsorption}$, $E_{H_2}$ and $E_{compound}$ are the total energy of T-carbon, one hydrogen molecule and T-carbon absorbing $n$ hydrogen molecules, respectively. For T-carbon absorbing 8 hydrogen molecules, the adsorption energy is about 0.173 eV. By increasing the number of absorbed hydrogen molecules until the number of hydrogen molecules reaches 16, the absorption energy becomes negative, indicating that the highest absorption capacity of hydrogen molecules is reached. Due to the strong C-C bond in T-carbon and the weak repulsion between the adsorbed hydrogen molecules H$_2$, T-carbon is still stable after absorbing 16 hydrogen molecules in a unit cell. According to the maximum number of adsorbed hydrogen molecules as shown in Fig.~\ref{fig10-1}, the hydrogen storage capacity of T-carbon is calculated to be about 7.7 wt$\%$~\cite{shengxianlei2011}, which is quite competitive compared with some high-capacity hydrogen storage materials~\cite{Qin2019}. 

\subsection{Li-ion batteries}
\begin{figure}[!htbp]
  \centering
  \includegraphics[scale=0.28,angle=0]{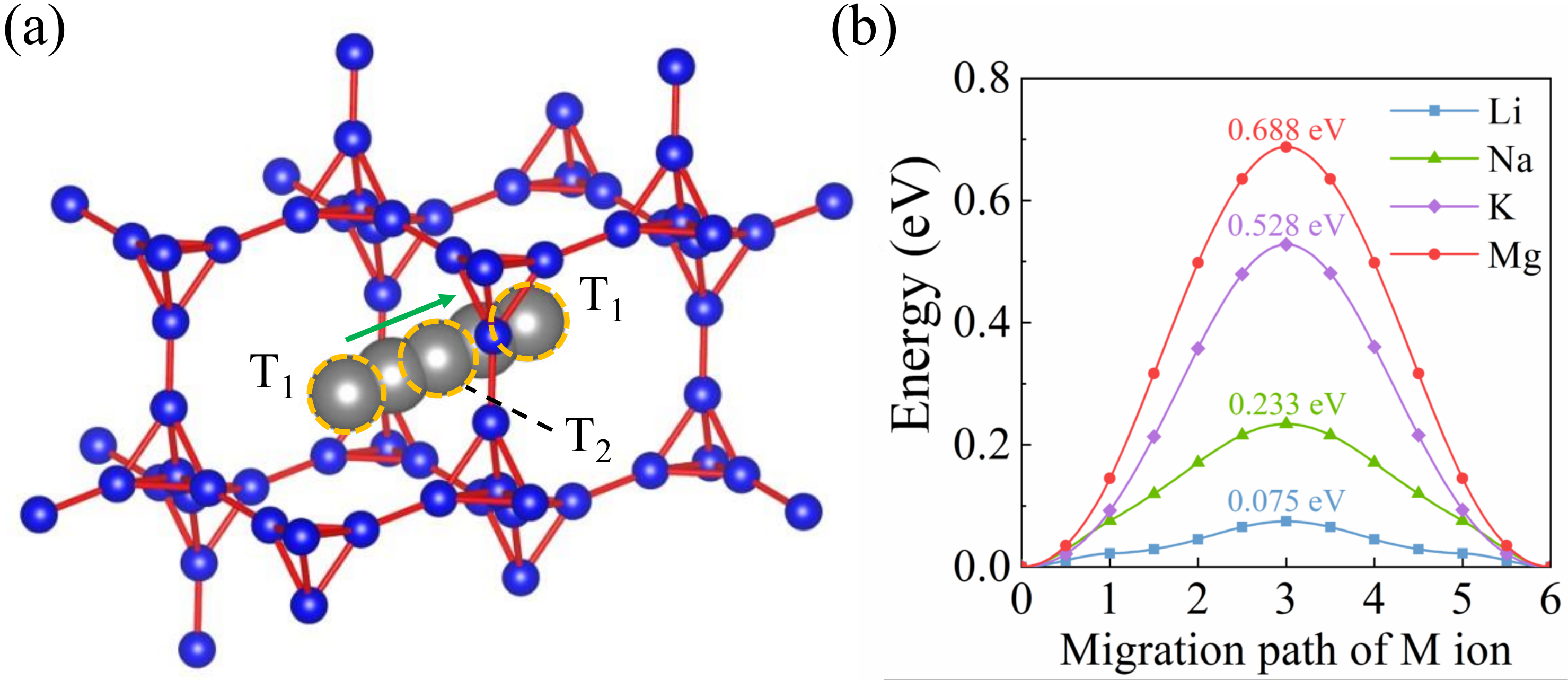}\\
  \caption{The potential application of T-carbon in ion batteries: (a) the migration of M (Li, Na, K, Mg) atoms in T-carbon, (b) the energy change of M atoms diffusing along the minimum migration path as indicated in (a). Copyright (2019) Royal Society of Chemistry. Reprinted (adapted) with permission from ref~\cite{Qin2019}.}\label{fig12-1}
\end{figure}

Rechargeable energy storage devices, such as lithium-ion batteries and portable power sources, play an important role in electronic devices, biomedicine, aerospace and electric vehicles, etc~\cite{Nitta2015}. Carbon-based materials are widely used in lithium-ion batteries, among which graphite is the most commonly used negative electrode material, because graphite has the high specific surface area and large interlayer space to accommodate lithium atoms, resulting in a high energy storage capacity of 372 mA$\cdot$h$\cdot$g$^{-1}$~\cite{Uthaisar2010}. 
 
As a new member of the carbon family, T-carbon has good application potential in lithium-ion batteries and other rechargeable energy storage devices due to its fluffy structure~\cite{Qin2019}. The specific volume of metal atoms is defined as $C=nF/M_{C_x}$~\cite{Nitta2015}, where $n$, $F$ and $M_{C_x}$ represent the number of electrons per metal atom involved in the electrochemical process (for Li, Na and K, $n$ = 1; for Mg, $n$ = 2), the Faraday constant with the value of 26.801 A$\cdot$h$\cdot$g$^{-1}$, and the mass of $C_x$ ($x$ is the number of carbon atoms, and for T-carbon, $x$ is 8), respectively. The specific volumes of Li (Na or K) and Mg are 588 and 1176 mA$\cdot$h$\cdot$g$^{-1}$, both are much higher than that of graphite.

The most stable adsorption position for metal ions M (Li, Na, K, Mg) is the center T$_1$ of the cage in T-carbon. The migration of M atoms in T-carbon is shown in Fig.~\ref{fig12-1}(a). The migration barriers of M atoms in T-carbon are 0.075, 0.233, 0.528 and 0.688 eV, respectively, as shown in Fig.~\ref{fig12-1}(b). The Li-ion migration barrier in T-carbon is much smaller than that in graphite with the battier of 0.327 eV~\cite{Uthaisar2010}. According to Arrhenius law, the diffusion constant of lithium ion in T-carbon, which is defined as $D\sim exp(-E/k_BT)$, where $E$ is the energy barrier, was calculated to be 1.7$\times$10$^4$ times that in graphene~\cite{Hao2018}. Therefore, T-carbon is a good lithium ion diffusion material and would have great application potential in ultra-fast charging and discharging of rechargeable energy storage devices. 
    
Recently, theoretical calculations show that T-carbon has a maximum lithium capacity of 1116 mA$\cdot$h$\cdot$g$^{-1}$ and tends to form an energy favorable mixture of C$_{64}$Li, C$_{64}$Li$_{16}$ and C$_{64}$Li$_{32}$~\cite{Cheng2021}. Since too high and too low lithium concentrations are not conducive to the adsorption and release of lithium by T-carbon anode respectively, only half of the maximum lithium capacity ($\sim$ 487 mA$\cdot$h$\cdot$g$^{-1}$) in T-carbon can be used for anode discharge, which may provide guidance for the future research of lithium, sodium or potassium ion battery electrodes in new energy resources.
    
\subsection{Solar cells}
\begin{table*}
	\renewcommand\arraystretch{1.25}
	\caption{The effective masses of electron ($m_e/m_0$) and hole ($m_h/m_0$), electron ($\nu_e$) and hole ($\nu_h$) mobility, band gaps ($E_g$) and energy level of CBM ($E_{\rm CBM}$) for T-carbon and several common used solar cell materials.}
	{\centering
		\begin{tabular}{lp{2cm}<{\centering}p{2cm}<{\centering}p{2cm}<{\centering}p{2cm}<{\centering}p{2cm}<{\centering}p{2cm}<{\centering}p{2cm}<{\centering}}
			\hline
			\hline
	&$m_e/m_0$  &$\nu_e$   (${\rm cm^2\cdot s^{-1}\cdot V^{-1}}$) &$m_h/m_0$  &$\nu_h$   (${\rm cm^2\cdot s^{-1}\cdot V^{-1}}$)  &$E_g$ (eV)   &$E_{\rm CBM}$ (eV) & Reference\\
	\hline
T-carbon         &0.22             &2360      &0.66             &270        &2.27  &-4.4     & \cite{Sun2019} \\
MAPbI$_3$    &0.14-0.26     &5-10       &0.2-0.44       &1-5        &1.5     &-3.9     & \cite{Umari2014,Oga2014,Motta2015} \\
P                    &0.17-1.12     &1000       &0.15-6.35     &10000  &1.51     &-         &  \cite{Qiao2014}\\
TlO$_2$        &0.096-0.424   &3300    &0.144-4.092   &4300    &1.56    &-          &  \cite{Ma2017}\\
Si                   &1.08             &1500       &0.59               &450       &1.12    &-          & \cite{MurphyArmando2008}\\
TiO$_2$        &5-10             &18   &0.6-1.0   &0.16   &3.2     &-4.1      &  \cite{Enright1996,Tiwana2011}\\
MoS$_2$      &0.46-0.48      &60-72     &0.57-0.60   &152-200 &1.6            &-            &  \cite{Cai2014}\\ 
SnO$_2$      &0.26,0.21               &166,229        &1.27,1.6                &15.7,11             &3.66         &-4.42     &\cite{Jiang2018,Hu2019}\\ 
ZnO              &0.29             &205          &0.78           &50              &3.4          &-4.4         &  \cite{Coleman2006,Jiang2018,Enright1996}\\ 
			\hline
			\hline
	\end{tabular}}
	\label{Table8-1}
\end{table*}

As the most widely available renewable energy in nature, solar energy is of great significance in the development of next-generation green energy. In recent years, organic-inorganic hybrid perovskite solar cells, as a new member of the solar cell family, have attracted great attention for the next generation of solar cell technology due to their excellent properties~\cite{Jeon2015,Zhou2014,Yang2015}. Since 2010, the energy conversion efficiency of perovskite solar cells has increased from 3.8$\%$ to 22.7$\%$~\cite{Kojima2009,NREL2018}. At present, the most commonly used electron transport material in perovskite batteries is TiO$_2$, which is applied to collect and transport electrons from the perovskite absorber layer~\cite{Moehl2014,Ravi2018}. Although electrons can enter TiO$_2$ from the perovskite absorption layer quickly, on one hand, the electron recombination rate is high because of the low electron mobility in TiO$_2$, and on the other hand, the required sintering temperature is high for the preparation of TiO$_2$ conductive crystals, limiting the application of TiO$_2$ in perovskite solar cells~\cite{Liu2013}. Therefore, it is necessary to design new and efficient electron transport materials, which requires a high electron mobility of the materials to meet the rapid electron transport and avoid interface electronic recombination, and the materials to have energy levels matched with perovskite in order to facilitate electron injection and hole blocking~\cite{You2015}. 

Due to the cubic structure, T-carbon exhibits isotropic transport properties. The effective mass $m^{\*}/m_0$ of the electron and hole of T-carbon was calculated to be 0.22 and 0.66, respectively, which are comparable with that (0.2-1.3) of perovskite MAPbI$_3$ and MASnI$_3$~\cite{Umari2014}. The electron mobility of T-carbon was calculated to be 2360 cm$^2\cdot$s$^{-1}\cdot$V$^{-1}$, which is about 9 times of the hole mobility (270 cm$^2\cdot$s$^{-1}\cdot$V$^{-1}$), which indicates that the electron transport capacity of T-carbon is better than the hole transport capacity~\cite{Sun2019}. The electron mobility of T-carbon is comparable with or even higher than that of some commonly used materials, such as black phosphorus (1000 cm$^2\cdot$s$^{-1}\cdot$V$^{-1}$)~\cite{Qiao2014}, TlO$_2$ (3300 cm$^2\cdot$s$^{-1}\cdot$V$^{-1}$)~\cite{Ma2017}, Si (1500 cm$^2\cdot$s$^{-1}\cdot$V$^{-1}$)~\cite{MurphyArmando2008} and TiO$_2$ (0.0524 cm$^2\cdot$s$^{-1}\cdot$V$^{-1}$)~\cite{Tiwana2011}, and the hole mobility of T-carbon is much larger than that of MoS$_2$ (200 cm$^2\cdot$s$^{-1}\cdot$V$^{-1}$)~\cite{Cai2014}, perovskite MAPbI$_3$ (1-5 cm$^2\cdot$s$^{-1}\cdot$V$^{-1}$)~\cite{Oga2014,Motta2015} and TiO$_2$ (0.16 cm$^2\cdot$s$^{-1}\cdot$V$^{-1}$)~\cite{Tiwana2011} as listed in Table~\ref{Table8-1}. The high electron mobility of T-carbon can further improve the short-circuit current and efficiency~\cite{Yang2016}. Therefore, T-carbon may be a viable alternative to traditional electron transport material TiO$_2$.

In perovskite solar cells, the bottom of CBM for the electron transport materials should be lower than the perovskite to achieve rapid electron injection~\cite{Zhou2015,Rao2015,Yang2018,Li2016}. T-carbon has a direct energy gap of about 2.273 eV with the energy level of the bottom of CBM located at 4.4 eV~\cite{Sun2019}, which is close to the energy levels of SnO$_2$ and ZnO that have faster electron injection efficiency than TiO$_2$~\cite{You2015,Xiong2018,Yang2016a,Choi2018}, indicating that T-carbon may have a higher electron injection rate than TiO$_2$. 

\subsection{Photocatalysis}
To solve the problem of environmental pollution caused by the overuse of fossil fuels, producing hydrogen from water using sunlight is an essential and natural choice. Therefore, searching for stable and effective photo-catalysts is really meaningful. The typical photo-catalysis materials include TiO$_2$, CdS, C$_3$N$_4$ and so on~\cite{Zhou2016,Gopannagari2017,Ida2014,Cao2014,Acar2016}. 

\begin{figure}[!htbp]
  \centering
  \includegraphics[scale=0.43,angle=0]{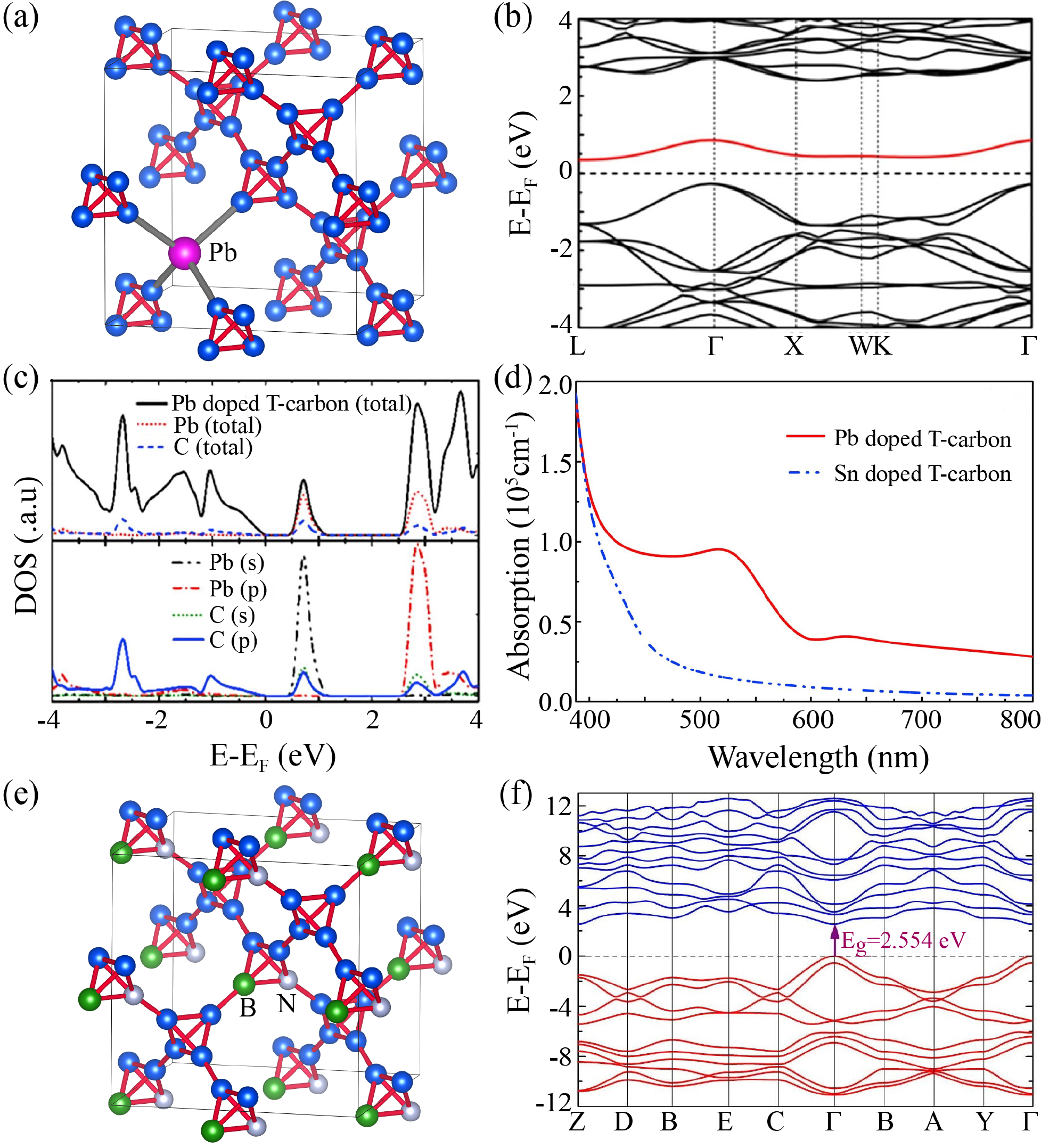}\\
  \caption{(a) The crystal structure of Pb doped T-carbon with one C$_4$ unit substituted by one Pb atom in primitive cell of T-carbon. (b) The band structure of Pb doped T-carbon with HSE06 method, where the red line is the impurity state of Pb atom. (c) The total and partial DOS of Pb doped T-carbon. (d) Absorption spectra of Pb (red solid line) and Sn (blue dash line) doped T-carbon, respectively. (e) The crystal structure of B-N codoped T-carbon and (f) its correspdong band structure. Copyright (2019) Elsevier. (a), (b), (c) and (d) Reprinted (adapted) with permission from ref~\cite{Ren2019}.}\label{fig14-1}
\end{figure}

Appropriate band gap is an important criterion for photocatalytic materials. To realize it, doping is usually an effective method to manipulate the electronic band structures. In 2018, Ren $et$ $al.$ adopted various doping methods to tune the band gap of T-carbon, and predicted a stable structure obtained by replacing a C$_4$ unit in T-carbon with a single atom in IVA group~\cite{Ren2019}. Moreover, in 2020, Tian $et$ $al.$ proposed another stable structure by $p-n$ codoping with one boron and one nitrogen atoms substituting two carbon atoms in each C$_4$ unit of T-carbon to form an intra-tetrahedron B-N bond~\cite{Tian2020}. The corresponding two structures are shown in Figs. 24(a) and (e). According to the calculation of phonon spectra and cohesive energies, the two doping structures are demonstrated to be kinetically stable and slightly thermodynamically less stable than pristine T-carbon. Unlike T-carbon itself, which has a direct band gap, the single-atom M (M=Si, Ge, Sn, Pb) in IVA group doped T-carbon has an indirect band gap, and the impurity states were successfully introduced into the energy gap of T-carbon as shown in Fig.~\ref{fig14-1}(b). Among these doped structures, Si and Ge doped T-carbon have a larger energy gap than T-carbon, while Sn and Pb doped T-carbon have a smaller energy gap. The change of energy gap is mainly due to the shift of the bottom of CBM relative to the Fermi level, which is attributed to the doped atomic states, indicating the strong localization of electronic state near the impurity. For B-N codoped T-carbon, it remains a direct band gap semiconductor with a smaller energy gap than T-carbon. It is noted that the energy gap of Pb doped T-carbon and B-N codoped T-carbon is 1.62 and 2.55 eV, respectively. In addition, Pb doped T-carbon performs the good optical absorption properties in the visible light region with a broad peak at 500-550 nm and a weak peak at 600-650 nm, corresponding to green and red light radiation, respectively. Because of the suitable band gap, localized electronic states and good optical absorption, Pb doped T-carbon is expected to be a high performance photocatalyst material~\cite{Deng2015}. 

\subsection{Magnetism in Ni-doped T-carbon}
\begin{figure}[!htbp]
  \centering
  \includegraphics[scale=0.44,angle=0]{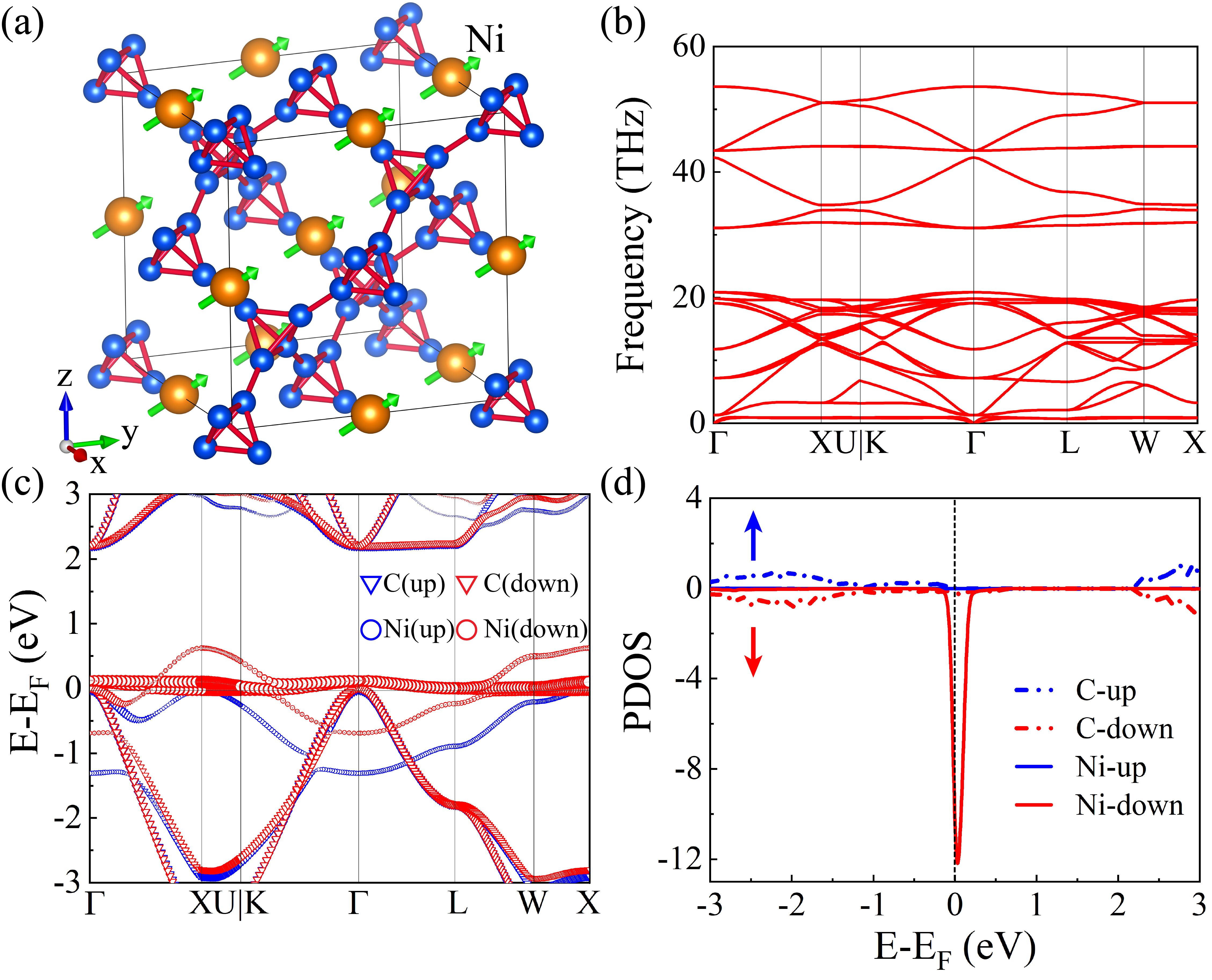}\\
  \caption{(a) The cubic crystalline structure of Ni-doped T-carbon NiC$_8$ with Ni atoms (orange balls) occupying the Wyckoff position $4a$. The green arrows indicate the magnetization direction of the Ni atom. (b) The phonon spectra, (c) electron band structure, and (d) partial density of states (PDOS) of the Ni-doped T-carbon.}\label{fig15-1}
\end{figure}

\begin{table}[htbp]
  \caption{The lattice constants $l$ (\AA), magnetic moments $\langle S \rangle$ ($\mu_B$), energy differences per magnetic atom between ferromagnetic and antiferromagnetic configurations $\Delta E$ (meV/Ni), and Curie temperatures $T_C$ (K) for Ni-doped T-carbon and face centered cubic (fcc) Ni, respectively.}
  \label{tab:15-1}
  \begin{tabular}{ccccc}
		\hline
		   &$l$ (\AA) &$\langle S \rangle$ ($\mu_B$) &$\Delta E$ (meV/Ni)  &$T_C$ (K)  \\
		\hline  \hline
		Ni-doped T-carbon    &7.52   &1.2 &-5.5     &33      \\
		Fcc Ni                        &3.52  &0.6 &-106   &627~\cite{Yu2008}     \\
		\hline
	\end{tabular}
\end{table}

The observation of magnetism in carbon materials has attracted much attention, such as a charge transfer complex of C$_{60}$ and tetrakis ethylene~\cite{ALLEMAND1991}, carbon nanofoams, carbon nanodiamonds~\cite{Rode2004,Talapatra2005}, untreated graphite with intrinsic point defects~\cite{Cervenka2009,Esquinazi2002}, highly oriented pyrolytic graphite irradiated with high-energy protons~\cite{Esquinazi2003} and so on. As we discussed above, T-carbon has many potential applications, and the exploration of possible magnetic properties based on T-carbon would greatly enrich the practical application of T-carbon. 

A natural idea to introduce magnetism into T-carbon is to dope magnetic atoms, where transition metals are good choice. When a transition metal atom is intercalated into the primitive cell of T-carbon to form C$_8$M (M = transition metal atom), we find that the most energetically preferred structure with the transition metal atoms forms a fcc structure, as shown in Fig.~\ref{fig15-1}(a). After checking phonon spectra of all doped structures, only Ni-doped T-carbon is dynamically stable with no imaginary frequency modes in the whole Brillouin zone. The optimized lattice constant of Ni-doped T-carbon is about 7.52 \AA. Moreover, Ni-doped T-carbon is demonstrated to possess a ferromagnetic ground state with the magnetization along [111] direction [see Fig.~\ref{fig15-1}(a)]. The band structure  and PDOS of Ni-doped T-carbon is depicted in Figs.~\ref{fig15-1}(c) and (d). Unlike T-carbon itself, Ni-doped T-carbon behaves as a half-metal with only one spin species (spin down) passing through the Fermi level, and the magnetism is introduced due to the different occupations of spin up and down electrons. The DOS near Fermi level is mainly dominated by the $e_g$ orbitals of Ni atoms. In Ni-doped T-carbon, T-carbon maintains the direct band gap semiconductor, but the band gap slightly decreases. To estimate the Curie temperature of Ni-doped T-carbon, we calculate the energy differences ($\Delta E$) between ferromagnetic and antiferromagnetic configurations of Ni-doped T-carbon with fcc Ni using the same method [GGA+U (U=4 eV)], and $\Delta E$ of the former is about 20 times smaller than that of the latter as listed in Table~\ref{tab:15-1}. It is known that the Curie temperature of fcc Ni is 627 K~\cite{Yu2008}, and the Curie temperature of Ni-doped T-carbon is estimated to be about 33 K. The magnetism in T-carbon and the half-metal property make Ni-doped T-carbon might have potential applications in spintronics, such as spin valve, spin transistor, magneto-optical devices and so on.

\subsection{Superconductivity in Na-doped T-carbon}
The observation of superconductivity has long been a popular topic in condensed matter physics and materials science, which has received wide interest in recent years~\cite{Xi2015,You2021a}. Except for the pure carbon superconductors, such as single-walled carbon nanotubes and twisted bilayer graphene with low superconducting transition temperatures~\cite{Tang2001,Cao2018,Cao2018a}, several carbon intercalation compounds were reported to be superconductors, including graphite intercalation compounds~\cite{Lauginie1980}, alkali-metal
doped C$_{60}$~\cite{Rosseinsky1991,Holczer1991,Murphy1992,Tanigaki1991,Tanigaki1993,Palstra1995} and diamond intercalation compounds~\cite{Ekimov2004}.

\begin{figure}[!htbp]
  \centering
  \includegraphics[scale=0.38,angle=0]{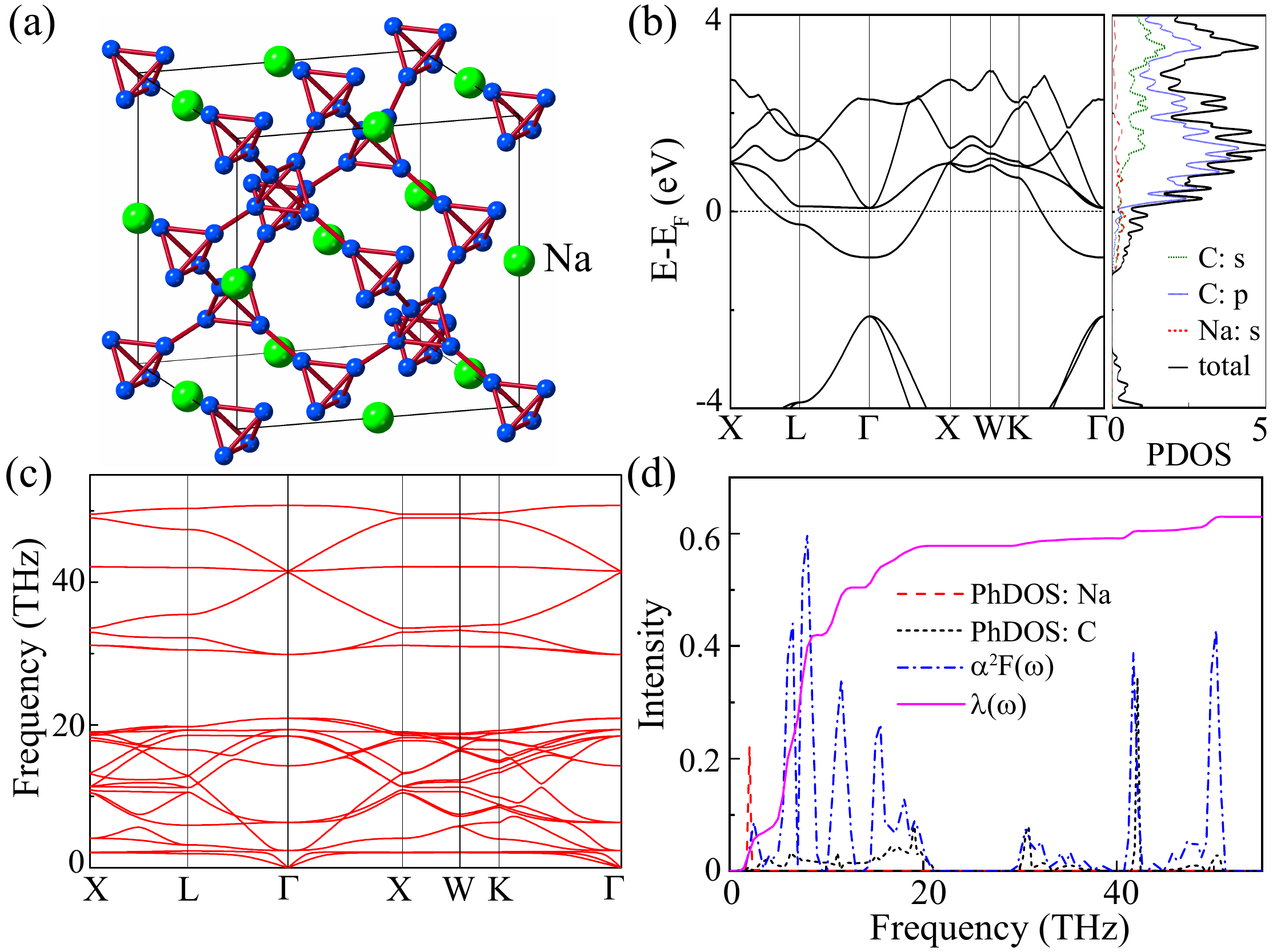}\\
  \caption{(a) The cubic crystalline structure of Na-doped T-carbon NaC$_8$ with Na atoms (green balls) occupying the Wyckoff position $4a$. (b) The electron band structure and partial density of states (PDOS) of the Na-doped T-carbon with inclusion of the SOC. (c) The phonon spectra and (d) phonon density of states (PhDOS) (divided by 40), Eliashberg spectral function $\alpha^2F(\omega)$, and cumulative frequency-dependent of EPC $\lambda(\omega)$ of Na-doped T-carbon at ambient pressure. Copyright (2020) American Physical Society. Reprinted (adapted) with permission from ref~\cite{You2020}.}\label{fig16-1}
\end{figure}

Due to the large interspaces between carbon atoms, many different atoms can be intercalated into T-carbon. The crystal structure of Na-doped T-carbon was shown in Fig.~\ref{fig16-1}(a), which was a compensate metal with a dispersive band contributed by the $s$ and $p$ orbitals of C atoms and $s$ orbital of a Na atom passing through the Fermi level [Fig.~\ref{fig16-1}(b)]~\cite{You2020}. There is no imaginary frequency mode in Fig.~\ref{fig16-1}(c), indicating the dynamical stability of Na-doped T-carbon. From the PhDOS in Fig.~\ref{fig16-1}(d), one can observe a peak from Na atoms at the low frequency because the relative larger atomic mass of Na atom than that of C atom. As the Na-doped T-carbon is a metal with an electronic Fermi surface, superconductivity may be induced by phonon-mediated electron pairing. The electron-phonon coupling strength $\lambda(\omega)$ was calculated to be 0.63. Utilizing the calculated Eliashberg spectral function $\alpha^2F(\omega)$ and $\lambda(\omega)$ together with a typical value of the effective screened Coulomb repulsion constant $\mu^{\*}$ = 0.1, the superconducting transition temperature Tc of Na-doped T-carbon was calculated to be about 11 K at ambient pressure with the McMillan-Allen-Dynes (MAD) approach~\cite{McMillan1968,Allen1975}.

\begin{table}[htbp]
  \caption{The pressure dependent superconductive parameters of N(E$_F$) (in unit of states/spin/eV/cell), volume (\AA$^3$), $\omega_{log}$ (in K), $\lambda$, and Tc (in K) for Na-doped T-carbon.}
  \label{tab:16-1}
  \begin{tabular}{cccccccc}
		\hline
		P(GPa)   &N(E$_F$) &V(\AA$^3$)  &$\omega_{log}$(K)   &$\lambda$ &Tc(K) \\
		\hline  \hline
		0    &0.85  &108.85   &413.2     &0.63     &10.9          \\
		3    &0.83  &106.81   &389.8     &0.67     &12.1        \\
        5    &0.82  &105.54   &369.8     &0.70     &12.9       \\
		7    &0.82  &104.35   &348.0     &0.74     &13.8       \\
        10   &0.80  &102.67   &297.2     &0.86     &16.0          \\
        12   &0.79  &101.61   &254.4     &0.98     &17.2             \\
        14   &0.79  &100.60   &181.0     &1.36     &18.7           \\
		\hline
	\end{tabular}
\end{table}

\begin{figure}[!htbp]
  \centering
  \includegraphics[scale=0.4,angle=0]{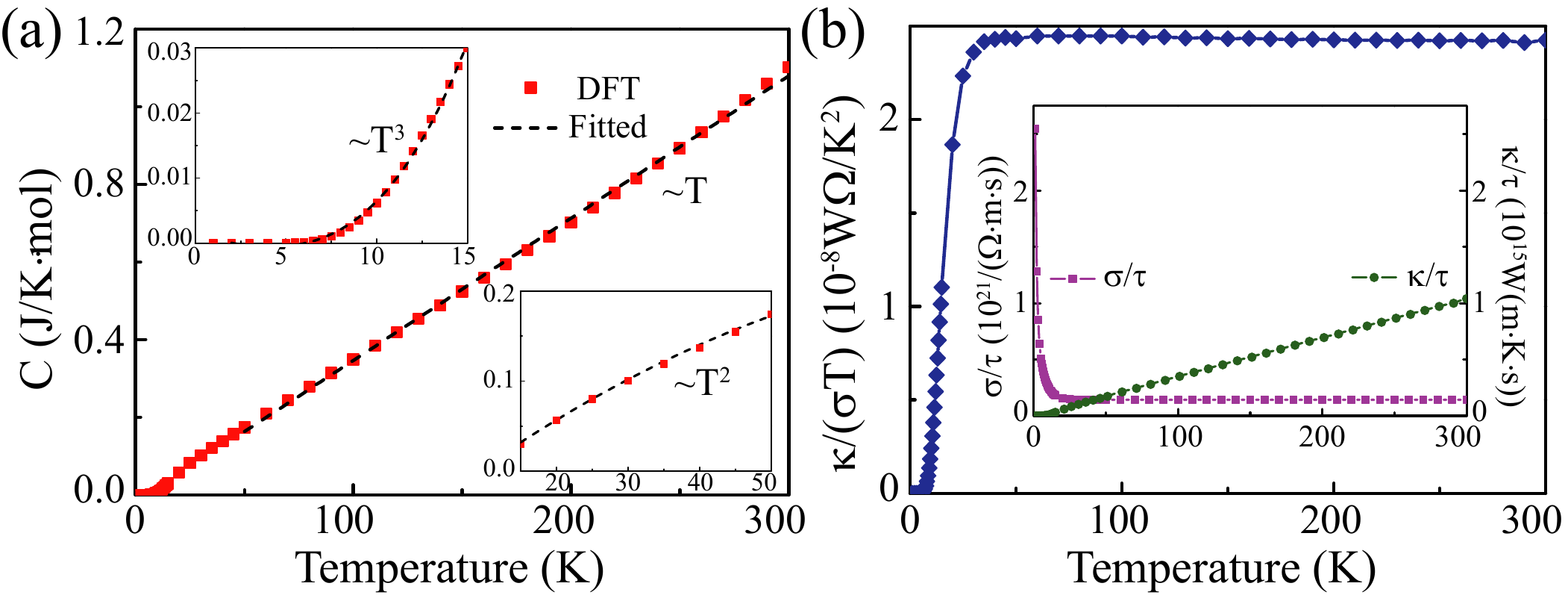}\\
  \caption{(a) Temperature dependence of electronic specific heat C in the normal state of Na-doped T-carbon superconductor. The upper and lower insets are the enlarged parts of the specific heat at temperature T $<$ 15 K and 15 K $<$ T $<$ 50 K, respectively. The dash lines are fitting curves at different temperature regions. (b) Temperature dependent Lorenz number $L$ [$L=\kappa$/($\sigma$T)]. The inset exhibits the temperature dependent electrical ($\sigma$) and thermal ($\kappa$) conductivities over the relaxation time ($\tau$). Copyright (2020) American Physical Society. Reprinted (adapted) with permission from ref~\cite{You2020}.}\label{fig16-2}
\end{figure}

The temperature dependent specific heat, electrical conductivity and thermal conductivity are presented in Fig.~\ref{fig16-2}(a). It was noted that, at low temperature below 15 K, specific heat $C(T) \sim T^3$, at $15<T<50$, $C(T) \sim T^2$, and at high temperature above 50 K, $C(T) \sim T$. In addition, the temperature dependent Lorenz number defined by $L =\kappa/(\sigma T)$, as given in Fig.~\ref{fig16-2}(b), was not a constant at temperature below 50 K, which dramatically violates the Wiedemann-Franz law, and it shows a constant at high temperature, indicating a Fermi liquid behavior. Therefore, the normal state of Na-doped T-carbon superconductor exhibits a non-Fermi liquid behavior at temperature below 50 K, and such a non-Fermi liquid normal state may be an anomalous metal.

The effect of pressure on superconducting transition temperature Tc for Na-doped T-carbon superconductor was also studied and the results are listed in Table~\ref{tab:16-1}. As can be seen, with the increase of pressure, Tc increases owing to the increase of both electron-phonon coupling strength $\lambda(\omega)$ and spectral function $\alpha^2F(\omega)$. The increase of Tc with pressure for Na-doped T-carbon superconductor results from an enhancement of the electron-phonon coupling $\lambda$ due to the shift of the phonon spectrum to lower frequencies, whereas for doped fullerenes A$_3$C$_{60}$ (A = K, Rb, Cs), the Tc is rapidly decreased under pressure because of the sharp decrease of electronic density of states under pressure caused by the rapid increase of the width of the conduction band when C$_{60}$ molecules are compressed together~\cite{Kadish1996,Diederichs1996,Diederichs1997}.

\subsection{Hydrogenated T-carbon molecules in interstellar medium}
The 2175 \AA\ UV extinction bump in interstellar medium (ISM) of the Milky Way was firstly discovered in 1965 by Stecher~\cite{Stecher1965}. Since then, several satellite observatories have confirmed the existence of this bump feature in tens stars of the Milky Way~\cite{Fitzpatrick1986,Savage1996,Fitzpatrick2007,Zafar2018}. Many subsequent studies on numerous galaxies show that such a 2175 \AA\ extinction bump feature in interstellar medium (ISM) is almost ubiquitous and pronounced in the Milky Way~\cite{Draine1989,Zafar2018}. Carbonaceous materials are commonly regarded as the origin of the 2175 \AA\ extinction bump in ISM~\cite{Draine1989}. However, the exact physical origin of the bump has been debated for a long time, which is still hitherto unclear after intensive explorations of more than a half century. Although a few potential carbonaceous candidates including graphite, nongraphic carbon, diamond and polycyclic aromatic hydrocarbons were proposed to explain the physical origin, a consensus is yet far to achieve. Therefore, at present a great puzzle still remains in the interpretation of this mysterious bump in ISM. It is noteworthy that, for T-carbon with anisotropic $sp^3$ hybrid bonds and low density, a prominent peak appears at the wavelength of 2250 \AA\ (not far from 2175 \AA) in its UV-visible optical absorption spectra~\cite{Zhang2017}. Because T-carbon is more readily formed under negative pressure~\cite{You2019}, it is easier to form T-carbon or its fragments in interstellar space with negative pressure circumstance. In addition, neutral hydrogen is full of the interstellar space, thus we believe that T-carbon would most likely exist in interstellar space as the hydrogenated T-carbon (HTC) molecules or clusters~\cite{Ma2020}.

\begin{figure}[!htbp]
  \centering
  \includegraphics[scale=0.44,angle=0]{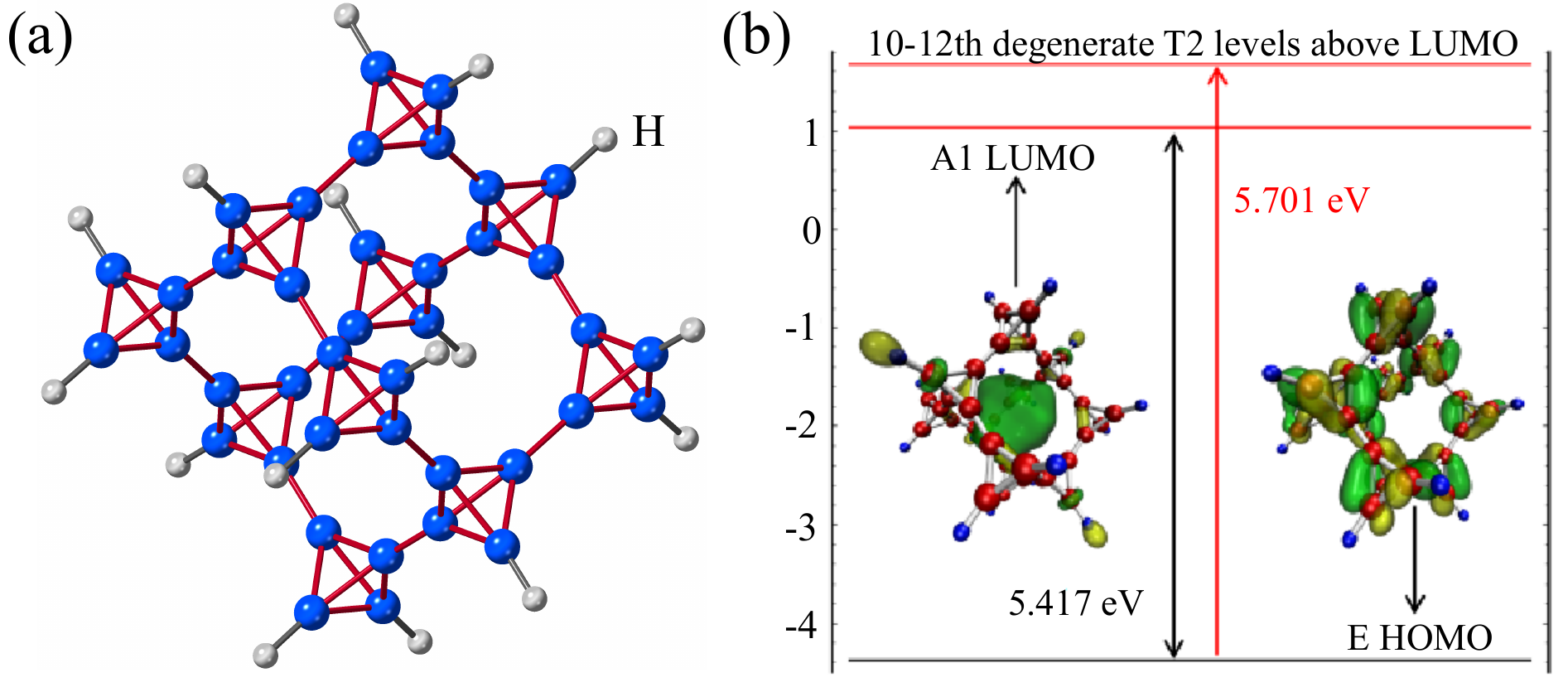}\\
  \caption{(a) The structure of hydrogenated T-carbon (HTC) molecule (C$_{40}$H$_{16}$). (b) Calculated molecular orbital energy levels of HTC molecule together with the molecular orbital maps of the lowest unoccupied molecular frontier orbitals (LUMOs) and the highest occupied molecular frontier orbitals (HOMOs). The green and yellow regions represent the positive and negative phases of the orbital wave function. The calculations show that the 2175 \AA\ absorption peak of HTC molecule is intrinsically generated from the excitation from the HOMO (with the orbital symmetry $E$) to the 10th level (degenerate in energy with the 11th and 12th levels with the orbital symmetry T2) above the LUMO (with the orbital symmetry A1). Reprinted (adapted) with permission from ref~\cite{Ma2020}.}\label{fig17-1}
\end{figure}
\begin{figure}[!htbp]
  \centering
  \includegraphics[scale=0.41,angle=0]{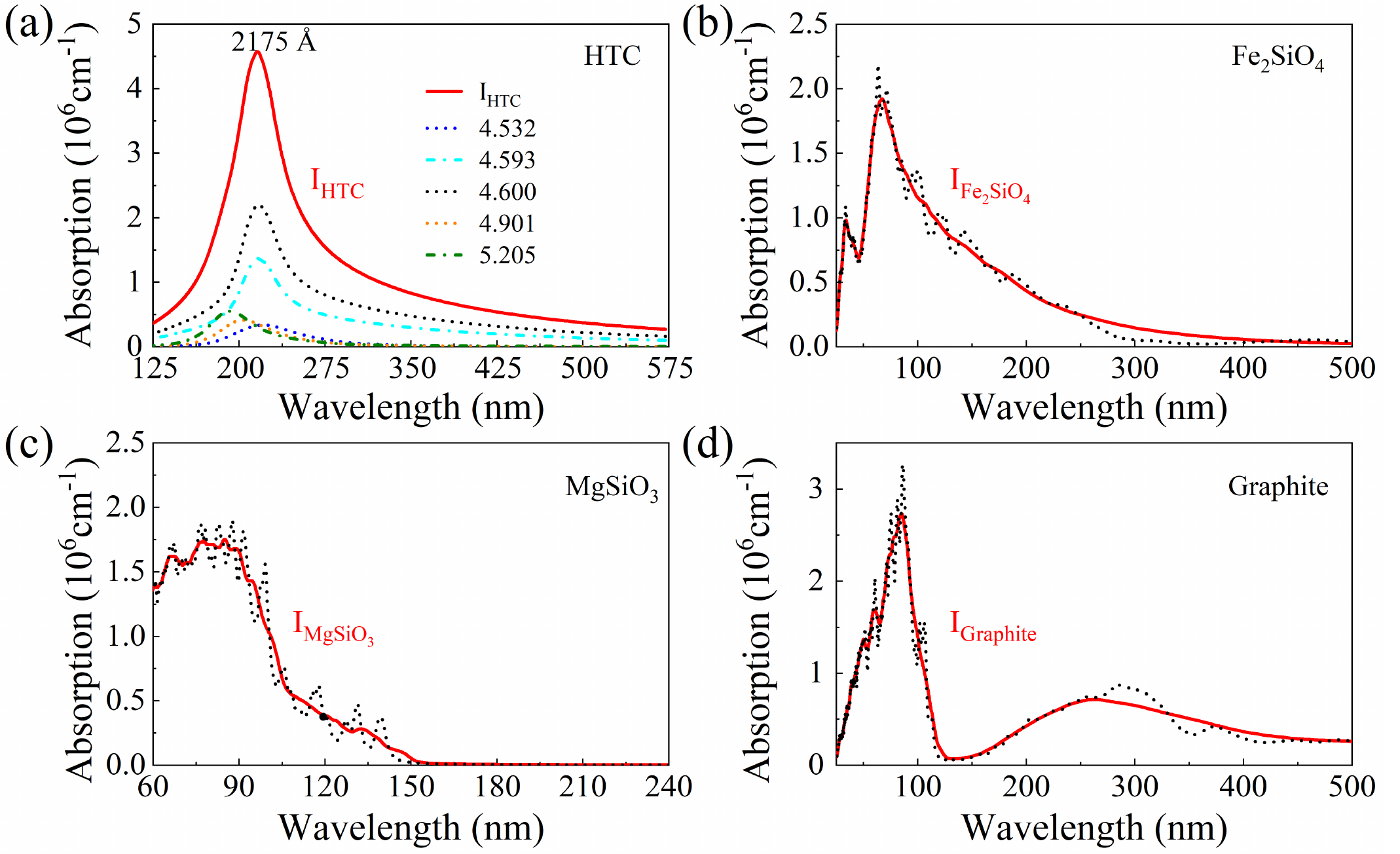}\\
  \caption{Calculated UV absorption spectra of (a) HTC molecule and HTC mixtures, (b) Fe$_2$SiO$_4$, (c) MgSiO$_3$ and (d) graphite. The red curve represents the overall superposed broadened absorption spectrum of HTC mixtures, where the peak is positioned at 2175 \AA. The straight solid lines are for HTC molecule and dashed curves are for HTC mixtures at different absorbed wavelengths labelled in the legend of (a). Red solid lines in (b), (c) and (d) represent the smoothed spectral curves by Savitzky–Golay polynomial regression. Reprinted (adapted) with permission from ref~\cite{Ma2020}.}\label{fig17-2}
\end{figure}

\begin{figure}[!htbp]
  \centering
  \includegraphics[scale=0.47,angle=0]{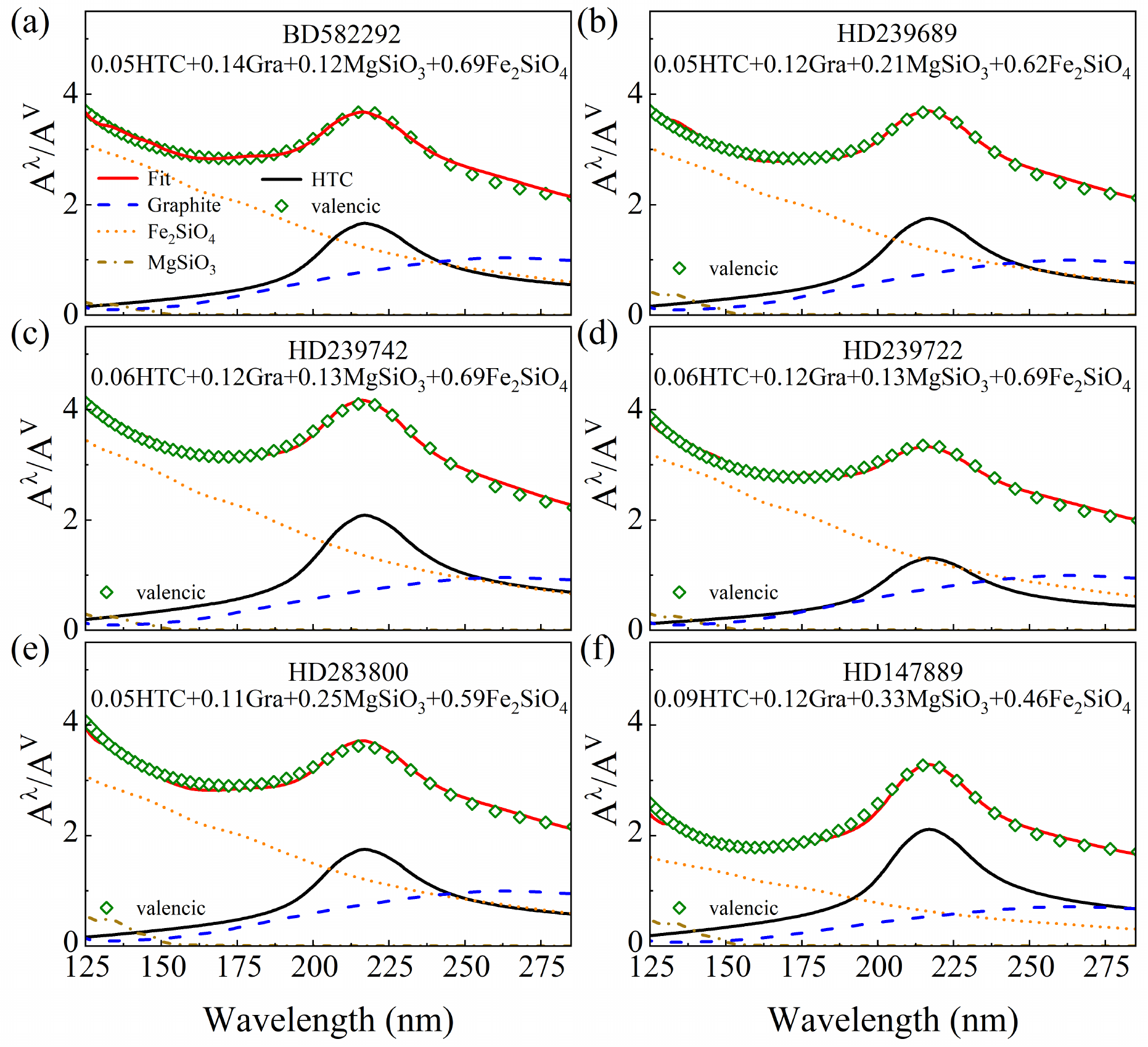}\\
  \caption{Fitting results for the normalized extinction curves of six stars including (a) BD582292, (b) HD239689, (c) HD239742, (d) HD239722, (e) HD283800, and (f) HD147889. Diamonds denote the observed data adapted from the references as indicated. The red solid lines are fitting curves of the observed data according to the equation $I_{Fit}=\alpha_1I_{HTC}+\alpha_2I_{Graphite}+\alpha_3I_{MgSiO_3}+\alpha_4I_{Fe_2SiO_4}$, where $\alpha_i$ ($i$=1,2,3,4) are linear fitting parameters, $I_{HTC}$, $I_{Graphite}$, $I_{MgSiO_3}$, and $I_{Fe_2SiO_4}$ stand for the absorption intensities of HTC mixtures, graphite, MgSiO$_3$, and Fe$_2$SiO$_4$, respectively. The other lines are marked in the legend of (a). Reprinted (adapted) with permission from ref~\cite{Ma2020}.}\label{fig17-3}
\end{figure}

The most possible HTC molecule consists of 40 carbon and 16 hydrogen atoms with $sp^3$ hybrid bonds as depicted in Fig.~\ref{fig17-1}(a). A sharp absorption peak at the wavelength of 2175 \AA\ is found in the UV-visible absorption spectra as shown in Fig.~\ref{fig17-2}(a), which is stable and does not change with the calculation method, indicating the possible physical origin of the 2175 \AA\ UV extinction bump. The absorption peak for HTC molecule is in accord with the excitation from the highest occupied molecular orbitals (HOMO) to the 10th level above the lowest unoccupied molecular orbitals (LUMO). In ISM there may exist HTC mixtures, which might lead to the broadened absorption spectra with the central peak kept at 2175 \AA\ through a mixture model based on clusters of HTC molecules. Since silicate~\cite{Gilman1969}, Mg/Fe-containing olivine~\cite{Campins1989,Jaeger1998} and graphite grains may exist in the interstellar dust, fayalite (Fe$_2$SiO$_4$), enstatite (MgSiO$_3$), and graphite grains should be taken into account when calculating the UV absorption spectra. By fitting the UV extinction curves of optional six stars~\cite{Jenniskens1993,Valencic2003,Whittet2004} with linearly combining the calculated UV absorption spectra of HTC mixtures, graphite, MgSiO$_3$ and Fe$_2$SiO$_4$~\cite{Gilman1969,Fitzpatrick1990,Li2001}, the UV extinction curves in these six stars are perfectly fitted as shown in Fig.~\ref{fig17-3}. To distinguish the contributing proportion of the four ingredients in fitting the UV extinction feature bump in ISM, the column density ratio is calculated. The column density ratio of carbonaceous dust (HTC mixtures and graphite) and silicate dust (MgSiO$_3$ and Fe$_2$SiO$_4$) in fittings is about 0.24. And the fitting results indicate that the central peak at 2175 \AA\ is mainly attributed to HTC clusters as shown in Fig.~\ref{fig17-3}. Thus, HTC molecules and clusters are demonstrated to play an important role in explaining the UV extinction curves of interstellar dust in the Milky. 

In summary, T-carbon has a wide range of potential applications due to its unique structure and properties. The fluffy structure and low density make T-carbon have good potential in various energy storage devices. As the battery anode material, T-carbon shows high lithium-ion diffusion constant and lithium capacity. As a hydrogen storage material, it displays high storage capacity. High electron mobility and appropriate energy levels matched well with perovskite also make T-carbon expected to be a great electron transport material in solar cells. Furthermore, the band structure of T-carbon can be effectively tuned by different doping schemes. Pb-doped T-carbon and B-N co-doped T-carbon hold appropriate band gap, localized electronic states and good optical absorption, which have potential applications in photocatalyst materials. Ni-doped T-carbon with magnetic and half-metal properties can be used in spintronics. Na-doped T-carbon shows a high superconducting transition temperature of 11 K at ambient pressure, which can be enhanced to 19 K at 14 GPa. In addition, it is worth mentioning that the hydrogenated T-carbon molecules may be related to the physical origin of the UV extinction feature in interstellar media.

Besides those mentioned potential applications in this section, other appealing properties of T-carbon are also accompanied with possible applications. For instance, the high failure strain of T-carbon achieved by global graphitization can be used in industrial and mechanical equipment with high ductility requirements. As a potential thermoelectric material, thermoelectric devices based on T-carbon are also very promising. New devices based on the unique topological phonons of T-carbon can be used to explore more novel physical effects of phonon topological states in energy information science.

\section{Derivative structures from T-carbon}

Inspired by T-carbon, many new carbon materials were predicted, which inherit some excellent properties from T-carbon and even perform some better properties than T-carbon. In the following, as examples, we will introduce several derivatives of T-carbon, including TY-carbon~\cite{Jo2012}, T-II carbon~\cite{Li2014}, bct-C$_{16}$~\cite{Ding2020}, $sp^2$-diamond~\cite{He2013}, L-carbon~\cite{Yang2013}, carboneyane~\cite{You2019}, cyclicgraphyne and cycligraphdiyne~\cite{You2019a}.

\begin{figure}[!htbp]
  \centering
  \includegraphics[scale=0.53,angle=0]{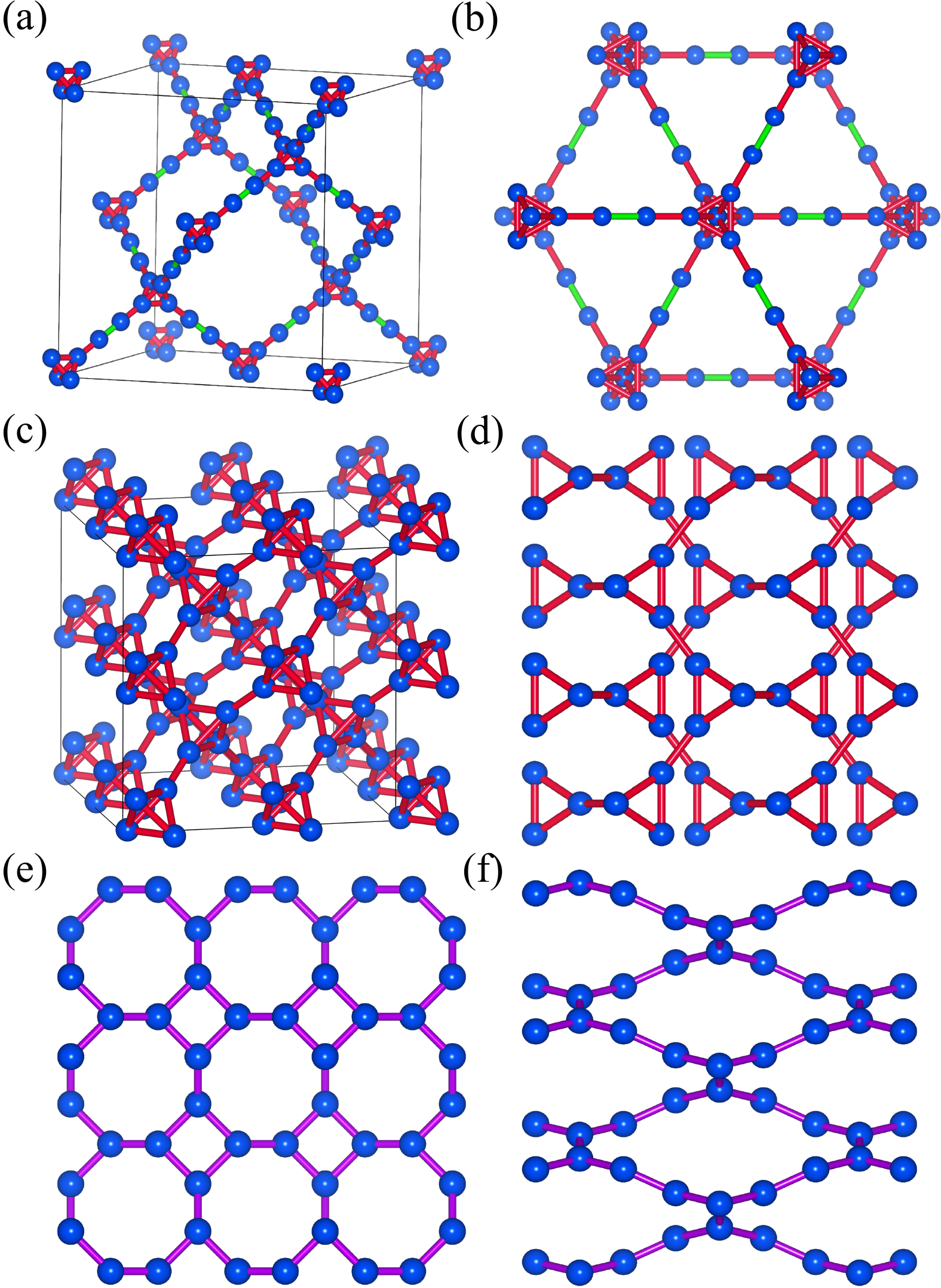}\\
  \caption{Several derivative structures of T-carbon: (a) TY-carbon obtained by introducing $sp$ hybrid bonds into T-carbon and (b) its [111] view; (c) T-II carbon derived by stacking two T-carbon together and (d) its [110] view; (e) top and (f) side views of bct-C$_{16}$, which is obtained by a structural phase transition from T-carbon. Copyright (2012) by the American Physical Society. (a) and (b) Reprinted (adapted) with permission from ref~\cite{Jo2012}. Copyright (2013) Royal Society of Chemistry. (c) and (d) Reprinted (adapted) with permission from ref~\cite{Li2014}. (e) and (f) Reprinted (adapted) with permission from ref~\cite{Ding2020}.}\label{fig2-3}
\end{figure}

\subsection{TY-carbon}
After the proposal of T-carbon, a new stable carbon allotrope named as tetrayne-carbon (TY-carbon) was predicted by introducing $sp$ hybrid bonds into T-carbon~\cite{Jo2012}. The main characteristics of TY-carbon come from the introduced C-C triple bonds. By introducing $sp$ hybrid bonds to diamond, another novel carbon allotrope coined as yne-carbon (Y-carbon) was obtained. The structure of TY-carbon is shown in Figs.~\ref{fig2-3}(a) and (b), and its main properties are summarized in the following: 
(1) TY-carbon was obtained by inserting two carbon atoms connected by an alkyne ($sp$) bond between two neighboring C$_4$ units in T-carbon, thus it has the same space group as T-carbon;
(2) The phonon spectra of TY-carbon and T-carbon are very similar except for the influence of $sp$ bonds;
(3) The XRD spectra of TY-carbon are very similar to that of T-carbon;
(4) TY-carbon has a direct band gap of 2.234 eV, and a phase transition from insulator to metal can occur with increasing the bond length because of the dimerization of carbon atoms as a result of Peierls instability~\cite{Rosenfeld1956};
(5) TY-carbon has a much lower density of 0.523$g/cm^3$ than that of T-carbon, implying its potential application in  hydrogen storage;
(6) By injecting inert gas such as Xe into the cavity of TY-carbon, it is expected to dramatically increase its stiffness~\cite{Itzhaki2005}.

\subsection{T-II carbon}
Li $et$ $al.$ designed a new carbon allotrope named as T-II carbon by moving carbon atoms in T-carbon half of the lattice vector along the crystallographic $a$-axis and then stacking it with T-carbon as shown in Figs.~\ref{fig2-3}(c) and (d)~\cite{Li2014}. T-II carbon has the space group of $Pn\bar{3}m$ with each primitive cell containing two C$_4$ units. The specific characteristics of T-II carbon include:
(1) T-II carbon exhibits a larger stiffness constant $C_{11}$ (560 GPa) than that of T-carbon (198 GPa) due to the denser stacking and more compact structure of the former, indicating the high incompressibility along the principal axis of cubic T-II carbon. (2) T-II carbon is metastable, and it is more stable than T-carbon under the pressure above 6.8 GPa. Thus it is expected to fabricate T-II carbon under high-pressure condition. (3) T-II carbon is a semiconductor with an indirect energy gap of about 0.88 eV;
(4) The calculated hardness of T-II carbon (27 GPa) is higher than that of T-carbon (5.6 GPa); (5) The calculations on the ideal strength and electron localization function show the bonds of T-II carbon can withstand larger shear strain than those of T-carbon.

\subsection{Bct-C$_{16}$}
By the molecular dynamics simulation, it was found that T-carbon underwent a structural phase transition at about 500 K, where the carbon tetrahedron in T-carbon will become a non-planar square, forming a new carbon material with body-centered tetragonal lattice coined as bct-C$_{16}$ as shown in Figs.~\ref{fig2-3}(e) and (f)~\cite{Ding2020}. In addition to high-temperature heating, the phase transition may be induced by doping elements (Cu, Fe) to break the carbon tetrahedrons in T-carbon. There are two kinds of non-planar rings in bct-C$_{16}$ including the ring composed of four carbon atoms with the bond length of 1.483 \AA\ and the ring consisting of eight carbon atoms with the bond length of 1.361 and 1.483 \AA. Bct-C$_{16}$ with quasi $sp^2$ binding feature was revealed to be a topological nodal ring semimetal protected by the time reversal and inversion symmetries. Because of weak and negligible spin-orbit coupling for carbon atoms, the nodal ring in bct-C$_{16}$ is stable. 

Bct-C$_{16}$ can be regarded as the three-dimensional extended structure of octagraphene~\cite{Sheng2012}, which is a versatile structurally favorable periodic $sp^2$-bonded carbon atomic planar sheet with $C_{4v}$ symmetry. The crystal structure of octagraphene is the same as the top view of bct-C$_{16}$ in Fig.~\ref{fig2-3}(e). Octagraphene has attracted widespread attention due to its excellent thermoelectric, optical and mechanical properties, and the potential applications in high-temperature superconductivity and hydrogen storage, etc~\cite{Kochaev2021,Podlivaev2013,Fthenakis2015,Rekha2017,Kang2019,Li2020}.

\begin{figure}[!htbp]
  \centering
  \includegraphics[scale=0.43,angle=0]{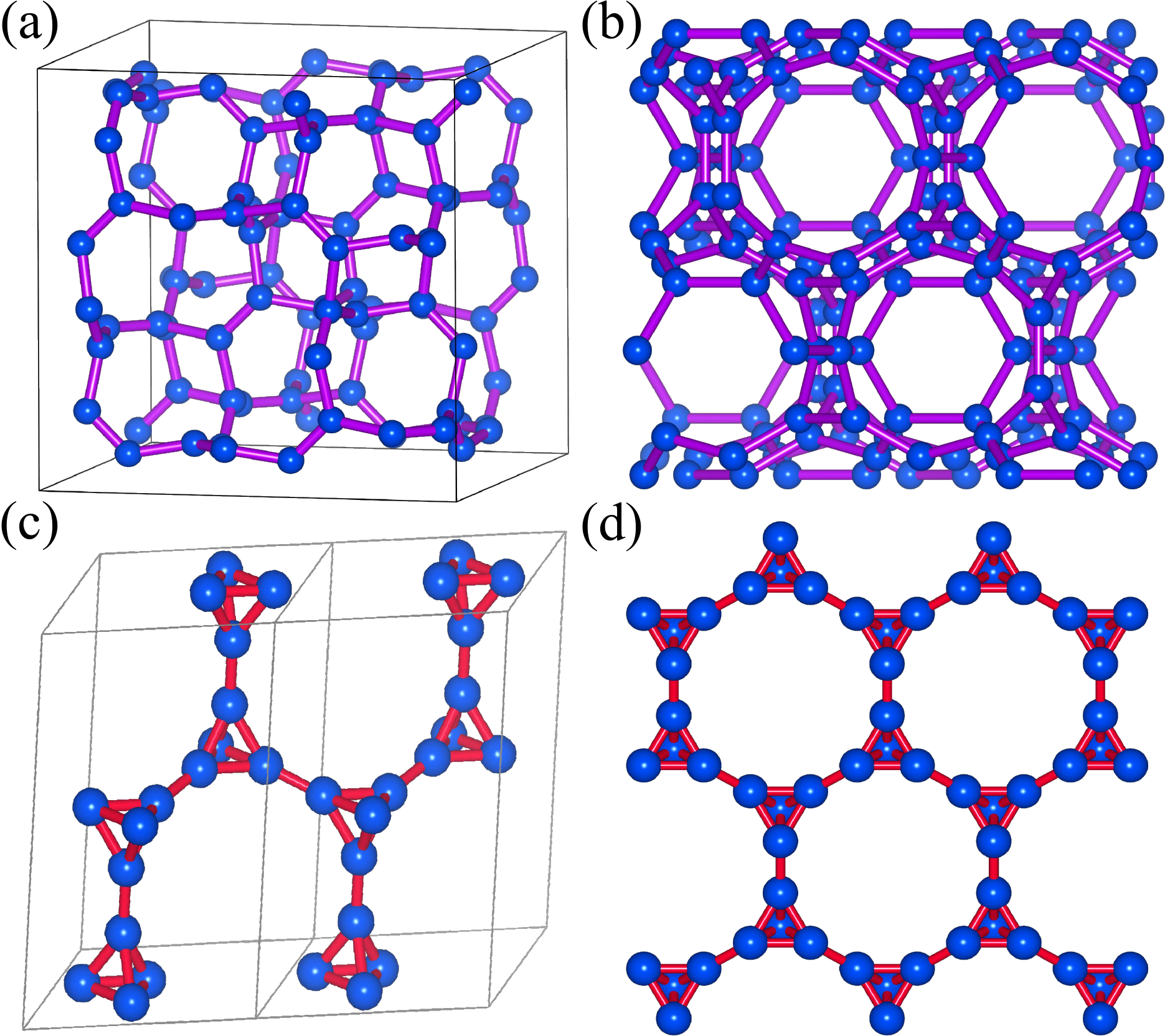}\\
  \caption{Two crystal structures inspired by T-carbon: (a) $sp^2$ diamond obtained by substituting each C–C bond in diamond with a distorted ring composed of six carbon atoms and (b) its [111] view; (c) L-carbon obtained by replacing each carbon atom in h-diamond with one carbon tetrahedron and (d) its [001] view. Copyright (2013) Royal Society of Chemistry. (a) and (b) Reprinted (adapted) with permission from ref~\cite{He2013}. (c) and (d) Reprinted (adapted) with permission from ref~\cite{Yang2013}}\label{fig2-4}
\end{figure}

\subsection{sp$^2$-diamond}
Inspired by the proposal of T-carbon, He $et$ $al.$ proposed another carbon allotrope coined as $sp^2$-diamond by substituting the C-C bond in diamond with a ring composed of six carbon atoms in a specific direction and position as shown in Figs.~\ref{fig2-4}(a) and (b)~\cite{He2013}. $sp^2$-diamond also has the Fd$\bar{3}$m space group with carbon atoms occupying the Wyckoff position (0.451, 0.451, 0.726). The interesting properties of sp$^2$-diamond are listed here:
(1) $sp^2$-diamond has a low density (2.116 $g/cm^3$), about 1.5 times that of T-carbon, but lower than that of most three-dimensional carbon materials. The very low density and porous structure indicate its potential for hydrogen storage, catalysts, and molecular sieves. 
(2) Comparing the phonon spectrum, PDOS and temperature dependent Helmholtz free energies of $sp^2$-diamond to that of T-carbon, $sp^2$-diamond is dynamically stable;
(3) Unlike the most previously reported 3D carbon materials with $sp^2$ network, which are metals, $sp^2$-diamond is a semiconductor with a direct band gap of 1.66 eV because in $sp^2$-diamond the 2$s$, 2$p_x$, 2$p_y$, and 2$p_z$ orbitals are all involved in the hybridization.

\subsection{L-carbon, C–c–C$_4$, H–c–C$_4$, H–2c–C$_4$}
Diamond has two different structures: c-diamond and h-diamond, which have two and four carbon atoms in the primitive cell, respectively~\cite{Kasper1968}. T-carbon was obtained by replacing each carbon atom in c-diamond with one carbon tetrahedron. Applying the same substitution rule to h-diamond, a new structure similar to T-carbon coined as L-carbon was obtained~\cite{Yang2013}. Based on the C$_4$ units substitution method, if half of the carbon atoms in the two kinds of diamond are replaced by C$_4$ units, several novel carbon materials can be obtained, including C-c-C$_4$ with half of carbon atoms in c-diamond substituted by C$_4$ units, H-c-C$_4$ with the non-adjacent carbon atoms in h-diamond replaced by C$_4$ unit and H–2c–C$_4$ with two adjacent carbon atoms in the primitive cell of h-diamond replaced by C$_4$ units. The crystal structure of L-carbon was shown in Figs.~\ref{fig2-4}(c) and (d). The density, cohesive energy, band gap, density of vibrational modes and bulk modulus of L-carbon are similar to those of T-carbon. Most of the characteristics of C-c-C$_4$ are similar to H-c-C$_4$ because of their close structure of C$_4$ units. In order to further understand the micromechanism, the root-mean-square (RMS) of bond angles and bond lengths were calculated for these materials, and the results show that the larger the proportion of C$_4$ units is, the greater the change of RMS of bond angles is, i.e. the greater the strain of the material is. Because the same proportion of C$_4$ units in L-carbon and T-carbon, as well as C-c-C$_4$ and H-c-C$_4$, the strains and several properties of the two are similar.

The electronic band structures of these four materials show that they are all semiconductors, but different from T-carbon with the direct band gap, and they exhibit indirect band gaps in the range of 3.2-4.7 eV. The Raman shifts of four carbon allotropes in the high-frequency region are related to the local vibration modes of C$_4$ units, while in the medium frequencies are associated with the collective vibration modes. Because of the different structures and components of C$_4$ units, these carbon allotropes possess different Raman shifts.

Similar to T-carbon, these carbon materials also have low density due to the substitution of C$_4$ units, especially L-carbon (1.50 $g/cm^3$). It is worth noting that L-carbon has a one-dimensional channel with a radius of 2.65 \AA, which can be used to screen very small molecules in gas mixtures.

Inspired by the structure of T-carbon, three stable counterparts cF-B$_8$~\cite{Getmanskii2017a}, cF-Al$_8$~\cite{Getmanskii2017} and T-Si~\cite{Fu2021} were proposed by replacing the carbon atoms in T-carbon with B, Al and Si atoms, respectively. The metal structure cF-B$_8$ has a low density of 0.9 g/cm$^3$, which is close to that of water. cF-Al$_8$ is a semimetal with high plasticity and a very low density of 0.61 g/cm$^3$. T-Si is a metal silicon allotrope with a density of 0.869 g/cm$^3$. The calculated results of adsorption and migration of Li and  Li$^+$ in T-Si indicate its potential application in the field of lithium ion batteries.

\subsection{Carboneyane}
\begin{figure}[!htbp]
  \centering
  \includegraphics[scale=0.58,angle=0]{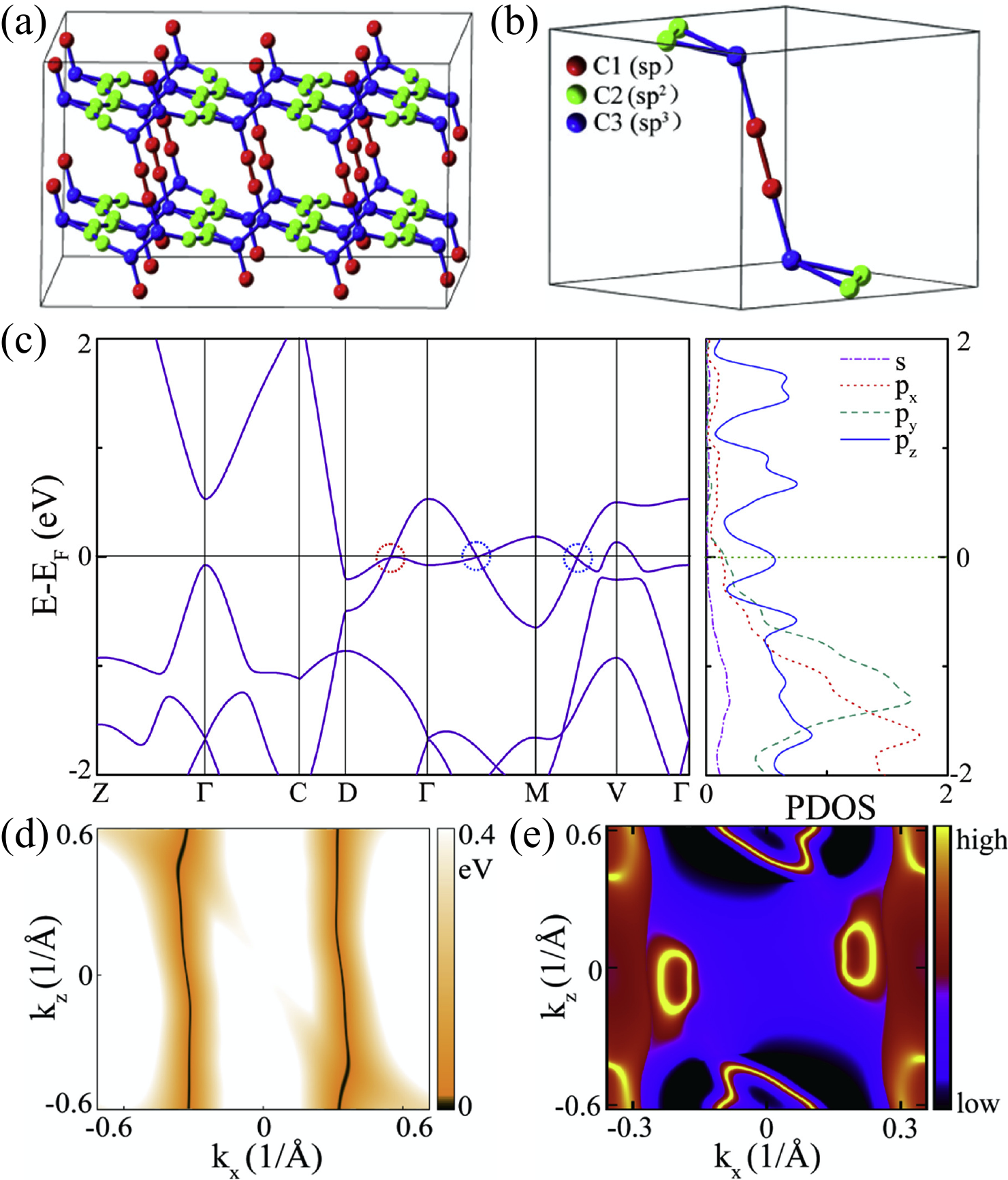}\\
  \caption{(a) The geometrical structure of carboneyane and (b) its primitive cell, where the carbon atoms are
classified into three types C1, C2, C3 by bonding nature, and the red, green and blue colored bonds represent triple, double and single bonds, respectively. (c) Electronic band structures and partial density of states (PDOS) of carboneyane. The blue and red colored circles represent type-I and type-II band dispersions, respectively. (d) Shapes of the two nodal lines obtained from DFT calculations. (e) A constant energy slice at 0.16 eV, where the bright lines showing the drumhead surface states. Copyright (2019) Elsevier. Reprinted (adapted) with permission from ref~\cite{You2019}.}\label{fig2-5}
\end{figure}
In 2019, the first carbon allotrope with simultaneously all three kind of hybrid bonds ($sp$, $sp^2$ and $sp^3$) was proposed, which is coined as carboneyane, as shown in Fig.~\ref{fig2-5}~\cite{You2019}. It has the space group of $C2/m$ with a low equilibrium density of 1.43 $g/cm^3$. There are eight carbon atoms in one primitive cell and the ratio of $sp$, $sp^2$, and $sp^3$ hybrid atoms is 1:2:1. Carboneyane is found to be feasible in experiments due to its energetically, dynamically and kinetically stable structure and much lower enthalpy than that of T-carbon. The calculations on the mechanical properties of carboneyane show that it has good ductility. Due to the exotic bonding mode and unique distribution of different hybrid bonds, carboneyane shows the anisotropy in electrical transport.

Carboneyane is shown to be a topological metal with a pair of nodal lines appealing in the (010) plane, which are protected by the mirror symmetry. The nodal lines are hybrid with type-I and type-II Weyl points and are robust against the negligibly weak SOC in carboneyane. The drumhead surface states can be obtained in carbonyane as the feature of a nodal line semimetal. Owing to the high specific capacity of Li, K, Mg atoms (558 mAh/g) and low energy barrier of Li$^+$ migration (0.094 eV) in carboneyane, it may exhibit great potential applications in the storage and absorption of atoms as well as the battery anode.

\subsection{Cycligraphyne and Cyclicgraphdiyne}
\begin{figure}[!htbp]
  \centering
  \includegraphics[scale=0.33,angle=0]{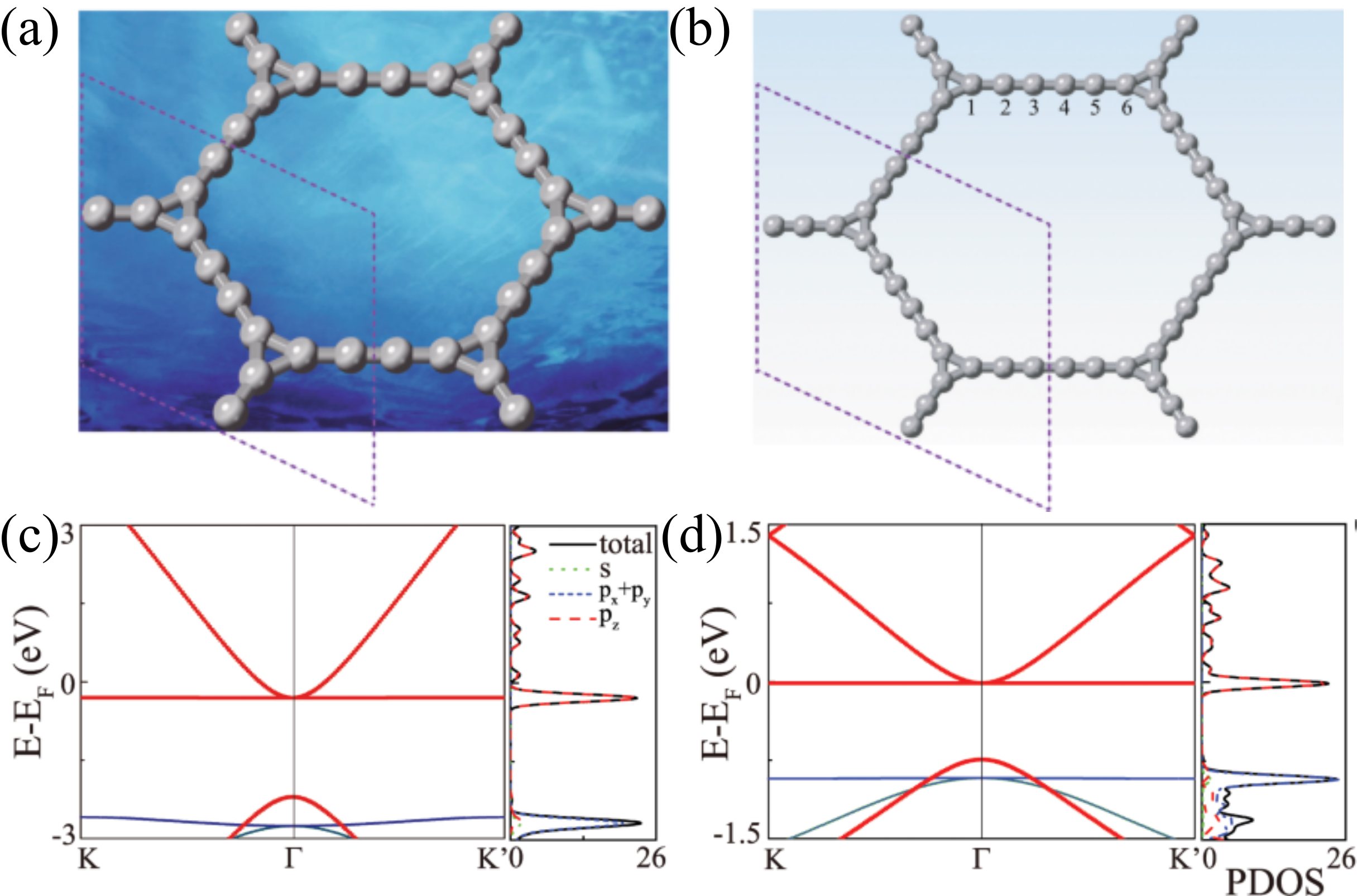}\\
  \caption{The crystal structures of cyclicgraphyne and (b) cyclicgraphdiyne, and (c) and (d) their corresponding band structures and partial DOS. Reprinted (adapted) with permission from ref~\cite{You2019a}.}\label{fig2-6}
\end{figure}

The view from [110] direction of T-carbon shows a kagome-like lattice with the triangles separated compared to kagome lattice [Fig.~\ref{fig2-1}(c)], which leads to the flat band above Fermi level in the path $L$-$\Gamma$ [Fig.~\ref{fig2-2}]. If one and two acetylene bonds are inserted between adjacent carbon triangles Fig.~\ref{fig2-1}(c), two new stable structures, coined as cyclicgraphyne and cyclicgraphdiyne~\cite{You2019a}, can be obtained, respectively, as shown in Fig.~\ref{fig2-6}, which have the space group of $P6/mmm$. The calculated results show a flat band with a band width of about 2 and 145 meV near Fermi level in the above two structures, respectively, which mainly comes from the valley part of the total electron charge density. By doping holes into cyclicgraphdiyne to make its flat band partially occupied, a phase transition from non-magnetic to ferromagnetic (half-metal) states can be obtained.

Both cyclicgraphyne and cyclicgraphdiyne have a crossing point between the flat band and another parabolic band, in which the flat band is called singular flat band (SFB)~\cite{Rhim2019}. The system with SFB can develop anomalous Landau level spectrum as the response of the magnetic field~\cite{Rhim2020}. Since the two structures only contain light carbon atoms, the spin-orbit coupling has little effect on the band structure just like graphene. In addition, both the ferromagnetism caused by the instability of the flat band and electron correlation effect as a result of effective Coulomb interaction have negligible effects on the bandwidth of the flat band in cyclicgraphdiyne~\cite{Rhim2020}. The perfect flat band structure of these two carbon allotropes makes them promising candidates for the experimental realization of SFBs.

\section{Prospects and outlook}

\begin{figure}[!htbp]
  \centering
  \includegraphics[scale=0.43,angle=0]{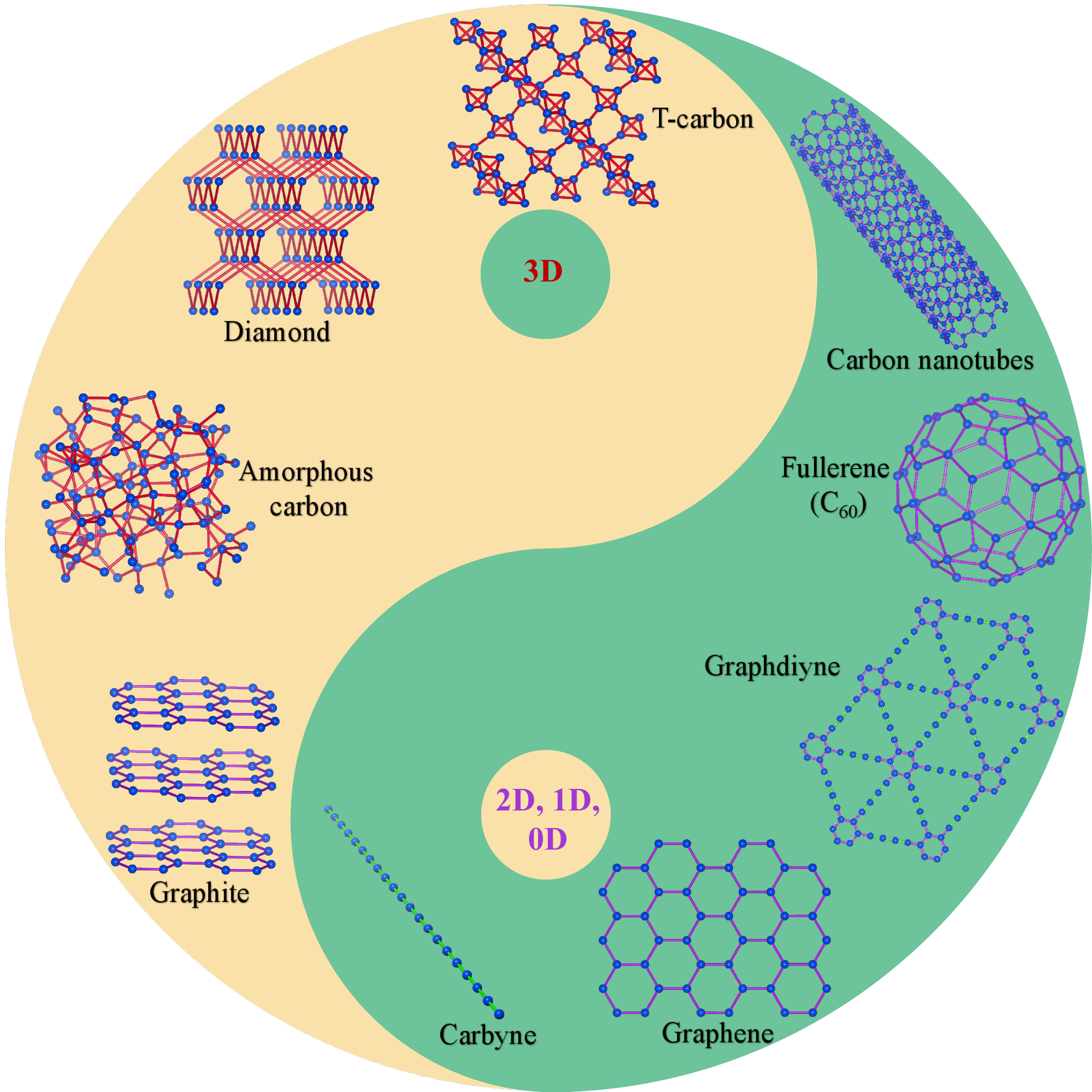}\\
  \caption{Members of carbon family.}\label{fig18-1}
\end{figure}

In the past decade since the proposal of T-carbon, various properties of T-carbon have been widely investigated. However, besides these studies, there are still many interesting properties of T-carbon that need to explore. The future works on T-carbon may focus on the following aspects: how to promote its application in the field of energy; how to find new T-carbon based superconductors; and how to use the existing results on T-carbon to inspire the research of other classical carbon structures or vice versa. Furthermore, the experimental realization of T-carbon adds a new member to the carbon family (see Fig.~\ref{fig18-1}). As the massive production of T-carbon is feasible, one may expect that T-carbon may have wide practical applications in near future, which calls for more experimental explorations.

The possibility of improving the thermoelectric properties of T-carbon remains to be further studied. The characteristics of a good thermoelectric material can be loosely said as "phonon-glass $\&$ electron-crystal", which indicates the materials with a low thermal conductivity as in a glass, and a high electrical conductivity as in crystals~\cite{Rowe2018}. To achieve this purpose, doping elements in semiconductors is a wise and feasible method~\cite{Chen2003}. Precise control and optimization of doping in the hollow structure of T-carbon should be studied in depth in future. In addition, one can also improve the thermoelectric performance of T-carbon by introducing nanotwin structures~\cite{Zhou2017}, applying external field~\cite{Qin2016a} or bond nanodesigning~\cite{Qin2018,Qin2016b}.

Although T-carbon has a high hydrogen storage capacity, it still needs to be improved. The effects of different surface areas, pore shapes, and pore volumes on hydrogen storage in T-carbon should be investigated. It was reported that carbon materials could induce chemisorption to enhance their hydrogen storage capacity by dispersing metal catalysts onto the surface~\cite{Adams2010}. Moreover, the ability of carbon materials to bind hydrogen molecules can also be enhanced by doping with appropriate elements~\cite{Kim2006}. These methods could be applied to promote the hydrogen storage capacity in T-carbon. 

In addition to the applications in thermoelectrics and hydrogen storage, the porous structure of T-carbon makes it an alternative cathode material in lithium-ion battery, which requires further experimental attempts including its cycle stability, power density, and reversible capacity. On the other hand, the voids in T-carbon can be considered as molecular sieves for filtering small molecules. The potential applications of T-carbon in other energy fields, such as electrochemistry, adsorption, energy storage, supercapacitors, aerospace, electronics and so on, are worthy of further study.

Beyond the applications of T-carbon in energy fields, new T-carbon based superconductors are also highly expected. Besides Na doped T-carbon superconductor, are there any other superconductors with much higher superconducting transition temperature in T-carbon doped with other elements, such  as IIIA group elements as well as in double or triple doped T-carbon? Topological superconductivity is a hot topic for topological quantum computation. Whether it is possible to achieve both topological states and superconductivity in T-carbon-based materials so that to realize Majorana fermions is worthy of further study.

The peculiar global graphitization of T-carbon with the tensile strain was rarely reported in other carbon materials, which provided a strategy to improve the ductility and failure strain of other carbon materials. The reason for the high ductility of T-carbon is attributed to the $sp^2$ dominated networks, which exhibit a comparable or even higher failure stress compared to the original $sp^3$ networks of T-carbon above the critical temperature. Therefore, appropriate methods should be implemented to improve the mechanical strength of $sp^2$ networks or reduce the mechanical strength of $sp^3$. 

The studies on other carbon materials also enlightens the further researches of T-carbon. Nanodiamonds inherit many properties of bulk diamonds, but as the size decreases, quantum effects become significant, which makes nanodiamonds exhibit many unique characteristics, such as strong absorption in the visible region. The physical properties of nanodiamonds can be tuned by impurities on the surface or in cores of nanodiamonds~\cite{Baidakova2007}. Similarly, the nanostructures of T-carbon may exhibit many novel characters beyond bulk T-carbon, which requires further study. In the past decades, the autoelectron emission has attracted much attention in various carbon nanostructures, aiming at making a field-emission cathode. Carbon nanotubes, carbon fibers, the nanodiamond powder-based coatings were revealed to have good applications in the field-emission cathode~\cite{Egorov2017}, which can be extended to T-carbon.

To achieve practical applications of T-carbon, the most essential issue is to massively produce T-carbon. How to improve the current experimental routes or adopt new experimental methods to synthesize massive T-carbon should be the focus of future experiments. The CVD method is widely used in the mass production of carbon allotropes including diamond, carbon nanotubes, graphene and so on~\cite{Schwander2011,Prasek2011,Xin2018}. For example, about 10 tonnes of diamond films are produced annually by CVD~\cite{Kharisov2019}. There are many types of CVD methods, among others the PECVD has been used for the synthesis of T-carbon. More CVD-based T-carbon synthesis schemes can be implemented by attempting different CVD experimental conditions, including energy supply, temperature range, pressure range, heated gas or ionization plasma used as carbon-containing precursors, substrate selection and the pretreatment of the samples~\cite{Schwander2011}. In addition, high-quality T-carbon samples could be grown in a carbon-hydrogen chemical vapor transport system with minimum experimental effort~\cite{Regel2001,Heinemann2020}.

High temperature preparation techniques such as arc discharge or laser irradiation are suitable for the synthesis of T-carbon. For instance, CNTs can be formed by arc discharge deposition and pulsed laser deposition, and diamond can be obtained directly from carbon by nanosecond laser irradiation in air~\cite{Prasek2011,Narayan2015}.
The chemical technologies are also feasible for the synthesis of carbon allotropes. The reaction of potassium with CO$_2$ results in the formation of 3D honeycomb-like graphene~\cite{Wei2017}. The impact of the pressure of 25–45 GPa and temperature of 800–900 ℃ was found to decompose ankerite Ca(Fe$^{2+}$,Mg)(CO$_3$)$_2$ to form diamond~\cite{Chen2018}. Choosing suitable carbon-containing precursors in suitable reaction conditions may produce T-carbon by chemical technologies. Furthermore, one could try if there is a feasible way to obtain a large number of C$_4$ units, and then use them as superatoms to further synthesize T-carbon. Therefore, more experimental attempts to massively prepare T-carbon would come soon, which will lead to more experimental and theoretical works on T-carbon in future.

\section* {Acknowledgments}
This work is supported in part by the National Key R\&D Program of China (Grant No. 2018YFA0305800), the Strategic Priority Research Program of the Chinese Academy of Sciences (Grants No. XDB28000000), the National Natural Science Foundation of China (Grant No.11834014), and Beijing Municipal Science and Technology Commission (Grant No. Z191100007219013).

%

\end{document}